\documentclass[11pt]{article}
\usepackage[utf8]{inputenc}
\usepackage{tensor}
\usepackage{bbm}
\usepackage{amssymb}
\usepackage{authblk}
\usepackage{xcolor}
\usepackage{graphicx}
\usepackage{jheppub}
\usepackage{bm}
\newcommand{\p}{\partial}
\usepackage[multiple]{footmisc}
\usepackage{subcaption}
\usepackage{multicol}
\usepackage{multirow}
\usepackage{rotating}
\usepackage{array}
\usepackage{longtable}
\usepackage{booktabs}
\usepackage[mathscr]{euscript}
\usepackage{tikz}
\usepackage{tikz-cd}
\usepackage{braket}
\usepackage{dsfont}

\tikzset{
    partial ellipse/.style args={#1:#2:#3}{
        insert path={+ (#1:#3) arc (#1:#2:#3)}
    }
}

\usetikzlibrary{calc,decorations.markings}
\usetikzlibrary{decorations.pathmorphing}
\usetikzlibrary{decorations.pathreplacing}
\usetikzlibrary{shapes,backgrounds}
\usetikzlibrary{fadings}
\usetikzlibrary{cd}

\tikzfading
[
  name=fade out,
  inner color=transparent!0,
  outer color=transparent!100
]

\newcommand{\be}{ \begin{equation}}
\newcommand{\ee}{\end{equation}}

\newcommand{\bket}[1]{\| #1 \rangle\hspace{-0.075cm}\rangle}

\makeatletter
\newcommand*\bigcdot{\mathpalette\bigcdot@{.65}}
\newcommand*\bigcdot@[2]{\mathbin{\vcenter{\hbox{\scalebox{#2}{$\m@th#1\bullet$}}}}}
\makeatother

\title{\boldmath D-branes in $\mathrm{AdS}_3\times \mathrm{S}^3\times \mathbb{T}^4$ at $k=1$ and their holographic duals}
\author[a]{Matthias R.~Gaberdiel,} 
\author[a,b]{Bob Knighton,} 
\author[a]{Jakub Vo\v{s}mera
} 
\affiliation[a]{Institut f\"{u}r Theoretische Physik, ETH Z\"{u}rich\\
Wolfgang-Pauli-Straße 27, 8093 Z\"{u}rich, Switzerland}
\affiliation[b]{Kavli Institute for Theoretical Physics,\\
University of California, Santa Barbara, CA 93106}
\emailAdd{gaberdiel@itp.phys.ethz.ch
}
\emailAdd{robejr@ethz.ch}
\emailAdd{jvosmera@phys.ethz.ch} 
\abstract{String theory on $\text{AdS}_3\times \text{S}^3\times \mathbb{T}^4$ with minimal $k=1$ NS-NS flux can be described in terms of a free field worldsheet theory in the hybrid formalism. We construct various D-branes of this string theory and calculate their associated cylinder amplitudes. We find that these amplitudes match with the cylinder correlators of certain boundary states of the dual symmetric orbifold CFT $\text{Sym}(\mathbb{T}^4)$, thus suggesting a direct correspondence between these boundary conditions. We also show that the disk amplitudes of these D-branes localise to those points in the worldsheet moduli space where the worldsheet disk holomorphically covers the spacetime disk.
	
}
\arxivnumber{}

\begin{document}
 \maketitle
\flushbottom
\begingroup\allowdisplaybreaks


\section{Introduction and summary}

It has recently become apparent that string theory on ${\rm AdS}_3\times {\rm S}^3 \times \mathbb{T}^4$ with minimal ($k=1$) NSNS flux is exactly dual to the symmetric orbifold of $\mathbb{T}^4$. The worldsheet description of this string theory is exactly solvable in the hybrid formalism \cite{Berkovits:1999im}, using the free field realisation of the $\mathfrak{psu}(1,1|2)_1$ WZW model at level $k=1$ \cite{Eberhardt:2018ouy,Dei:2020zui}. This allows one to show that the perturbative string spectrum matches exactly that of the symmetric orbifold theory in the large $N$ limit \cite{Eberhardt:2018ouy}, see also \cite{Gaberdiel:2018rqv,Giribet:2018ada} for earlier indications. Furthermore, it was shown in \cite{Eberhardt:2019ywk,Dei:2020zui,Eberhardt:2020akk,Knighton:2020kuh} that the correlation functions of the symmetric orbifold theory can be reproduced from this worldsheet perspective. 

Given this perturbative correspondence, one may wonder what can be said about the non-perturbative states of string theory, and whether they also match with some suitable states in the symmetric orbifold. One relatively accessible class of non-perturbative contributions are described by D-branes, and they can be efficiently analysed using boundary CFT techniques on the worldsheet. In fact, since the relevant background is  far from `geometrical' --- at level $k=1$, the size of the AdS space is of the order of the string length --- this is essentially the only approach in this context. 

In this paper we will undertake first steps  towards characterising and analysing the possible boundary conditions of the $\mathfrak{psu}(1,1|2)_1$ worldsheet theory. We will concentrate on the `maximally symmetric' D-branes, i.e.\ the D-branes that preserve the full $\mathfrak{psu}(1,1|2)$ symmetry, up to an automorphism. If we restrict to the $\mathfrak{sl}(2,\mathds{R})$ algebra that describes the ${\rm AdS}_3$ factor following \cite{Maldacena:2000hw}, the branes we consider are closely related to those constructed already some time ago in \cite{Bachas:2000fr,Giveon:2001uq,Petropoulos:2001qu,Lee:2001xe,Hikida:2001yi,Rajaraman:2001cr,Lee:2001gh,Ponsot:2001gt}. However, since at $k=1$ the closed string spectrum is much simpler --- it only contains one continuous representation, as well as its spectrally flowed images --- our analysis differs a bit from what was done before. 

As was already explained in \cite{Bachas:2000fr,Giveon:2001uq,Petropoulos:2001qu,Lee:2001xe,Hikida:2001yi,Rajaraman:2001cr,Lee:2001gh,Ponsot:2001gt} (and as we review in Appendix~\ref{app:geometry}), there are two families of branes in ${\rm AdS}_3$ that preserve the maximal ${\rm SL}(2,\mathds{R})$ symmetry. They are the ${\rm AdS}_2$ and the spherical branes, and their respective geometry is sketched in Figure~\ref{fig:geometry}. In this paper we shall mainly consider the spherical branes whose worldvolumes are located at a fixed time, and hence describe `instantonic' branes in ${\rm AdS}_3$. They extend all the way to the boundary, and create a one-dimensional circular `defect' at the 2d boundary, see the left panel of Figure~\ref{fig:geometry}. Our main result is that this 2d `defect' is to be identified with a certain D-brane of the symmetric orbifold theory. This therefore furnishes a concrete realisation of the scenario proposed in \cite{Takayanagi:2011zk}, see also \cite{Fujita:2011fp}.\footnote{This is to be contrasted with another proposal for AdS/BCFT \cite{Karch:2000gx} where AdS D-branes are related to defects in the dual CFT.}

\begin{figure}
	\centering
	\begin{tikzpicture}[scale = 0.75]
	\begin{scope}
	\draw[thick] (0,0) [partial ellipse = 180:360:3 and 1];
	\draw[thick, dashed] (0,0) [partial ellipse = 0:180:3 and 1];
	\draw[thick, red] (0,4) [partial ellipse = 180:360:3 and 1];
	\draw[thick, red, dashed] (0,4) [partial ellipse = 0:180:3 and 1];
	\fill[red, opacity = 0.25] (0,4) ellipse (3 and 1);
	\draw[thick] (-3,0) -- (-3,8);
	\draw[thick] (3,0) -- (3,8);
	\draw[thick] (0,8) ellipse (3 and 1);
	\node at (0,4) {\footnotesize spherical brane};
	\end{scope}
	\begin{scope}[xshift = 10 cm]
	\draw[thick, dashed] (0,0) [partial ellipse = 0:180:3 and 1];
	\draw[thick, red, dashed] (1,8+0.9428) -- (1,0.9428);
	\draw[thick, red, dashed] (1,0.9428) -- (-1,-0.9428);
	\fill[red, opacity = 0.2] (-1,-0.9428) -- (-1,8-0.9428) -- (1,8+0.9428) -- (1,0.9428) -- (-1,-0.9428);
	\draw[thick, red] (-1,-0.9428) -- (-1,8-0.9428);
	\draw[thick, red] (-1,8-0.9428) -- (1,8+0.9428);
	\draw[thick] (0,0) [partial ellipse = 180:360:3 and 1];
	\draw[thick] (-3,0) -- (-3,8);
	\draw[thick] (3,0) -- (3,8);
	\draw[thick] (0,8) ellipse (3 and 1);
	\node at (0,4.15) {\footnotesize $\text{AdS}_2$};
	\node at (0,3.75) {\footnotesize brane};
	\end{scope}
	\end{tikzpicture}
	\caption{Geometry of the spherical and ${\rm AdS}_2$ branes.}
	\label{fig:geometry}
\end{figure}
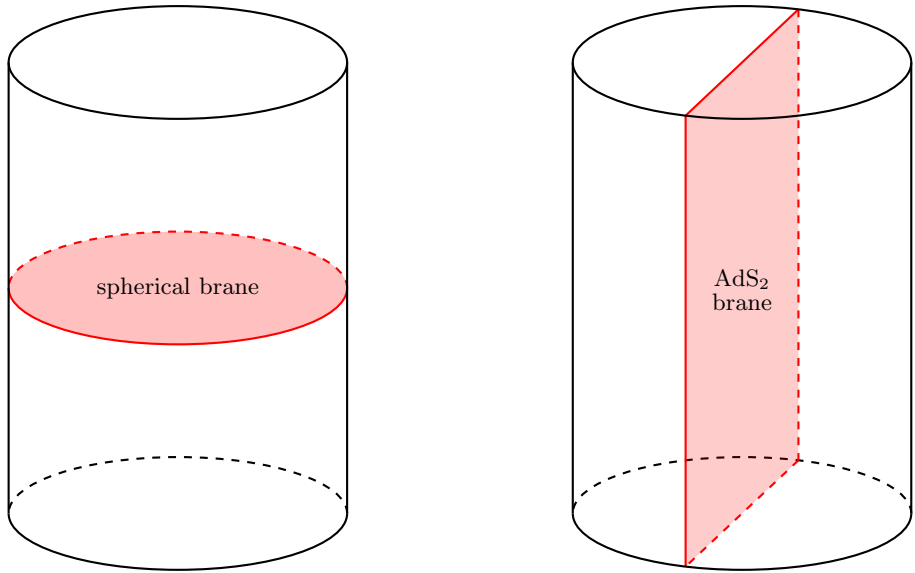

As evidence in favour of this identification we calculate the cylinder diagram (or more precisely the superstring amplitude that describes the exchange of  closed strings between two such branes), where we separate the branes by a certain distance $\hat{t}$ along the time direction, see Figure \ref{fig:cylinder}. From the worldsheet perspective the corresponding cylinder correlator turns out to be of the form (for the sake of simplicity we are suppressing here various chemical potentials, see Section~\ref{sec:calculation} below for more details)
\be\label{Zuv}
\hat{Z}_{u|v}(\hat{t};\hat{\tau})  = 
\sum_{w\in\mathbb{Z}}\int_0^1 d\lambda\,  
\sum_{r\in \mathbb{Z}+\lambda}e^{2\pi i r(\hat{t}-w\hat{\tau})} \hat{Z}_{u|v}^{\mathbb{T}^4}(\hat{t};\hat{\tau}) \ .
\ee
Here $\hat{\tau}$ is the worldsheet modulus, while $\hat{t}$ is the $\mathfrak{sl}(2,\mathds{R})$ chemical potential, respectively. The parameters $u$ and $v$ describe the different D-branes in the $\mathbb{T}^4$ sector, and $\hat{Z}_{u|v}^{\mathbb{T}^4}(\hat{t};\hat{\tau})$ is the corresponding cylinder amplitude.
We should mention that its $\hat{t}$ dependence comes from the ghost contribution since initially the torus degrees of freedom are uncharged with respect to the $\mathfrak{sl}(2,\mathds{R})$ subalgebra of $\mathfrak{psu}(1,1|2)$, see \cite{Gaberdiel:2021njm} and the discussion around eq.~(\ref{eq:Pghost}) below. 

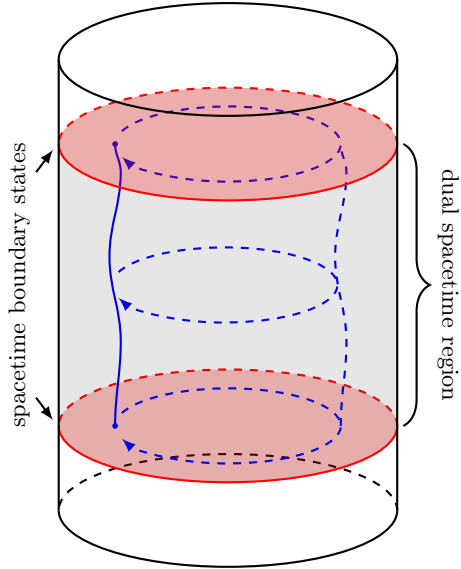
\begin{figure}
	\centering
	\begin{tikzpicture}[scale = 0.75]
	\begin{scope}
	\draw[thick] (0,0) [partial ellipse = 180:360:3 and 1];
	\draw[thick, dashed] (0,0) [partial ellipse = 0:180:3 and 1];
	\draw[thick, red, dashed] (0,1.5) [partial ellipse = 0:180:3 and 1];
	\fill[red, opacity = 0.25] (0,1.5) ellipse (3 and 1);
	\begin{scope}[xshift = -1 cm]
	\fill[blue] (-1,1.5) circle (0.05);
	\fill[blue] (-1,6.5) circle (0.05);
	\draw[thick, blue] (-1,1.5) to[out = 90, in = -90] (-0.9,3) to[out = 90, in = -90] (-1.1,4.5) to[out = 90, in = -90] (-0.9,6) to[out = 90, in = -90] (-1,6.5);
	\end{scope}
	\begin{scope}[xshift = 1 cm]
	\draw[thick, blue, dashed] (1,1.5) to[out = 90, in = -90] (1.1,3) to[out = 90, in = -90] (0.9,4.5) to[out = 90, in = -90] (1.1,6) to[out = 90, in = -90] (1,6.5);
	\end{scope}
	\draw[thick, blue, -latex, dashed] (0,1.5) [partial ellipse = 160:-160:2 and 2/3];
	\draw[thick, blue, -latex, dashed] (0,6.5) [partial ellipse = 160:-160:2 and 2/3];
	\draw[thick, blue, -latex, dashed] (-0.06,4) [partial ellipse = 160:-160:2 and 2/3];
	\fill[black, opacity = 0.1] (0,1.5) [partial ellipse = 180:360:3 and 1] to (3,1.5) to (3,6.5) to (0,6.5) [partial ellipse = 0:180:3 and 1] to (-3,6.5) to (-3,1.5);
	\draw[thick, red] (0,1.5) [partial ellipse = 180:360:3 and 1];
	\draw[thick, red] (0,6.5) [partial ellipse = 180:360:3 and 1];
	\draw[thick, red, dashed] (0,6.5) [partial ellipse = 0:180:3 and 1];
	\fill[red, opacity = 0.25] (0,6.5) ellipse (3 and 1);
	\draw[thick] (-3,0) -- (-3,8);
	\draw[thick] (3,0) -- (3,8);
	\draw[thick] (0,8) ellipse (3 and 1);
	\draw[thick, decorate,decoration={brace,amplitude=10pt},xshift=-4pt,yshift=0pt] (3.25,6.5) -- (3.25,1.5) node [black, midway, xshift=0.6cm, rotate = -90] {\footnotesize dual spacetime region};
	\node[rotate = 90] at (-3.7,4) {\footnotesize spacetime boundary states};
	\draw[thick, -latex] (-3.4,6) -- (-3.1,6.4);
	\draw[thick, -latex] (-3.4,2) -- (-3.1,1.6);
	\end{scope}
	\end{tikzpicture}
	\caption{Cylinder string amplitude for two spherical branes in $\mathrm{AdS}_3$. At $k=1$ all closed strings are essentially `long strings', i.e.\ are close to the asymptotic boundary of  $\mathrm{AdS}_3$.}
	\label{fig:cylinder}
\end{figure}

From a worldsheet perspective this correlator needs to be integrated over $\hat{\tau}$, but because of the integral over $\lambda$ and the sum over $r$, the exponential factor in (\ref{Zuv}) becomes a $\delta$-function, 
\be
\int_0^1 d\lambda\,  \sum_{r\in \mathbb{Z}+\lambda}\, e^{2\pi i r(\hat{t}-w\hat{\tau})}  = \delta\bigl(\hat{t}-w\hat{\tau} \bigr) \ .
\ee
Thus the $\hat\tau$ integral is effectively trivial, and the integrated amplitude simplifies to 
\be
\sum_{w=1}^\infty \frac{1}{w}  \hat{Z}^{\mathbb{T}^4}_{u|v}(\hat{t};\tfrac{\hat{t}}{w})\ . 
\ee
The key observation is that this is precisely the singe particle contribution to the cylinder diagram in the symmetric orbifold theory, where we choose the boundary condition $u$ (resp.\ $v$) in all factors of the symmetric orbifold. (The relevant boundary states in the symmetric orbifold theory are constructed in detail in Section~\ref{sec:symorb}.)

We have also performed the corresponding open string analysis, both in the worldsheet theory as well as in the symmetric orbifold, and they also match by a similar mechanism. The details of the delta function condition are a little bit different, see eq.~(\ref{eq:ZWZW}) below, but in essence the calculation works the same, reflecting that both the boundary states on the worldsheet and in the symmetric orbifold satisfy the Cardy condition; this is explained in Section~\ref{sec:open}. 

Finally, we have confirmed that the correlation functions in the presence of this boundary condition reproduce those of the symmetric orbifold theory, see Section~\ref{sec:correlators}. This follows from that fact that both the correlators on the worldsheet as well as those in the symmetric orbifold can be expressed in terms of covering maps on manifolds with boundaries, mirroring naturally the arguments of \cite{Eberhardt:2019ywk,Dei:2020zui}. 
\bigskip

The reason why we concentrate on the spherical branes in this paper is that their worldsheet boundary states involve all spectrally flowed sectors. As explained in Appendix~\ref{app:geometry}, see also e.g.\ \cite{Hikida:2001yi}, this is simple to understand geometrically: since spectral flow corresponds to the winding number along the circular direction of the ${\rm AdS}_3$ boundary cylinder, only a brane that satisfies a Neumann boundary condition in the angular $\phi$ coordinate can couple to these `winding' sectors, see again Fig.~\ref{fig:geometry}. This translates into the statement that the corresponding symmetric orbifold boundary state involves all twisted sectors states, and hence is `maximally fractional'. This makes its construction relatively straightforward. 

On the other hand, the ${\rm AdS}_2$ branes that satisfy a Dirichlet boundary condition in $\phi$ cannot couple to any spectrally flowed sector on the worldsheet. Their interpretation from the symmetric orbifold perspective is therefore more delicate since all the symmetric orbifold states come from sectors with non-trivial spectral flow on the worldsheet \cite{Eberhardt:2018ouy}. We should mention that the two sets of branes are inequivalent from the point of view of global ${\rm AdS}_3$, i.e.\ in the worldsheet theory based on the WZW model of ${\rm SL}(2,\mathds{R})$, see the comments around eq.~(\ref{t3t3}) in Appendix~\ref{app:geometry}. On the other hand, if we were to consider thermal ${\rm AdS}_3$ whose boundary is a torus, the two gluing conditions are related to one another by the modular $S$ matrix that exchanges the roles of the two torus directions. Then we can apply similar arguments also for the ${\rm AdS}_2$ branes, as was already noticed in \cite{Lee:2001gh}. However, this then involves a different worldsheet theory that has, in particular, also spectrally flowed sectors with $w\neq \bar{w}$, and it is not obvious how this theory is related to the symmetric orbifold theory. 
\medskip

The paper is organised as follows. In Section~\ref{sec:psu} we give a brief review of the $\mathrm{PSU}(1,1|2)$  WZW model at level $k=1$, and in Section~\ref{sec:psubound} we construct the symmetric boundary states of this model. Section~\ref{sec:symorb} summarises the key features of the symmetric orbifold theory and constructs the maximally fractional boundary states that are relevant for the comparison with the spherical branes in ${\rm AdS}_3$. We also sketch there the construction of another natural family of branes whose ${\rm AdS}_3$ interpretation, however, is not yet clear. 
The details of the cylinder amplitude calculation are described in Section~\ref{sec:calculation}. In Section~\ref{sec:correlators} we discuss correlation functions in the presence of these D-branes, and argue that they reproduce the expected symmetric orbifold results. We conclude in Section~\ref{sec:discussion}, and there are a number of appendices where some of the more technical material is described.
\bigskip

\noindent{\bf Note added:} We understand that related work has also been done independently in \cite{Alex}. We have thus coordinated the release of our papers.


\section{\boldmath The \texorpdfstring{$\mathrm{PSU}(1,1|2)_1$}{PSU(1,1|2)1} WZW model}\label{sec:psu}

In this section we shall fix our conventions and remind the reader about the structure of the  WZW model based on the supergroup $\mathrm{PSU}(1,1|2)$ at level $k=1$. 

\subsection[Representations of \texorpdfstring{$\mathfrak{psu}(1,1|2)_1$}{psu(1,1|2)1}]{\boldmath Representations of \texorpdfstring{$\mathfrak{psu}(1,1|2)_1$}{psu(1,1|2)1}}

Let us begin by reviewing the structure of the superalgebra $\mathfrak{psu}(1,1|2)_1$. Its maximal bosonic subalgebra is $\mathfrak{sl}(2,\mathds{R})_1\oplus \mathfrak{su}(2)_1$, and we denote the corresponding generators by $J_m^a$ and $K_m^a$, with $a\in\{\pm, 3\}$, respectively. In addition there are the fermionic generators $S_m^{\alpha\beta\gamma}$ where $\alpha,\beta,\gamma\in \{\pm\}$ denote spinor indices with respect to suitable $\mathfrak{su}(2)$ algebras. At level $k=1$, these modes satisfy the (anti-)commutation relations
\begin{subequations}\label{eq:psu-relations}
\begin{align}
    [J_m^3,J_n^3]&= -\tfrac{1}{2}m\delta_{m+n,0}\ ,\\
    [J_m^3,J_n^\pm]&=\pm J^\pm_{m+n}\ ,\\
    [J_m^+,J_n^-]&=m\delta_{m+n,0}-2J_{m+n}^3\ ,\\
    [K_m^3,K_n^3]&=+\tfrac{1}{2}m\delta_{m+n,0}\ ,\\
    [K_m^3,K_n^\pm]&=\pm K_{m+n}^\pm\ ,\\
    [K_m^+,K_n^-]&=m\delta_{m+n,0}+2K_{m+n}^3\ ,\\
    [J_m^a,S_n^{\alpha\beta\gamma}]&=\tfrac{1}{2}c_a\tensor{(\sigma^a)}{^\alpha_\mu}S^{\mu\beta\gamma}_{m+n}\ ,\label{eq:JS}\\
    [K_m^a,S_n^{\alpha\beta\gamma}]&=\tfrac{1}{2}\tensor{(\sigma^a)}{^\beta_\nu}S^{\alpha\nu\gamma}_{m+n}\ ,\label{eq:KS}\\
    \{S^{\alpha\beta\gamma}_m,S^{\mu\nu\rho}_n\}&=m\epsilon^{\alpha\mu}\epsilon^{\beta\nu}\epsilon^{\gamma\rho}\delta_{m+n,0}-\epsilon^{\beta\nu}\epsilon^{\gamma\rho}c_a(\sigma_a)^{\alpha\mu}J^a_{m+n}+\epsilon^{\alpha\mu}\epsilon^{\gamma\rho}(\sigma_a)^{\beta\nu}K^a_{m+n}\ .\label{eq:AntiComm}
\end{align}
\end{subequations}
Here $\sigma^a$ are the Pauli matrices, while $c_a$ equals $-1$ for $a=-$, and $+1$ otherwise, and the $a$ indices are raised and lowered by the standard $\mathfrak{su}(2)$ invariant form, see \cite{Eberhardt:2018ouy} for our conventions. More specifically, we have 
\begin{subequations}
\begin{align}
   [J^3_0,S^{\pm\beta\gamma}_m] & = \pm \tfrac{1}{2}S^{\pm \beta\gamma}_m\ , \qquad     
& [J^\pm_0,S^{\mp\beta\gamma}_m]  = \pm S^{\pm \beta\gamma}_m\ ,\\
    [K^3_0,S^{\alpha\pm\gamma}_m] & = \pm \tfrac{1}{2}S^{\alpha\pm\gamma}_m\ , \qquad 
&  [K^\pm_0,S^{\alpha\mp\gamma}_m]  = S^{\alpha\pm\gamma}_m\ .
\end{align}
\end{subequations}
The highest-weight states of a highest weight representation form a Clifford representation with respect to the fermionic zero modes, whose different summands sit in representations of the bosonic zero mode algebra $\mathfrak{sl}(2,\mathds{R})\oplus \mathfrak{su}(2)$. At level $k=1$, only the short supermultiplets of the form 
\begin{align}
\label{eq:Short}
    \begin{array}{ccc}
         & (\mathscr{C}^\frac{1}{2}_{\lambda},\mathbf{2}) &\\
        (\mathscr{C}^1_{\lambda+\frac{1}{2}},\mathbf{1}) &  & (\mathscr{C}^0_{\lambda+\frac{1}{2}},\mathbf{1})
    \end{array}
\end{align}
are allowed \cite{Eberhardt:2018ouy}, and we shall denote the corresponding affine representation by $\mathscr{F}_\lambda$. Here $\mathscr{C}_\lambda^j$ is the continuous representation  of $\mathfrak{sl}(2,\mathds{R})$ where $j\in \mathds{R}\cup(\frac{1}{2}+i\mathds{R})$ parametrises the quadratic Casimir as
\begin{align}
\mathcal{C}^{\mathfrak{sl}(2,\mathds{R})}=-j(j-1) \ , 
\end{align}
and $\lambda\in [0,1)\cong \mathds{R}/\mathbb{Z}$ denotes the fractional part of the $J_0^3$ eigenvalue $m$. 
Furthermore, the representations of  $\mathfrak{su}(2)$ are the standard finite-dimensional representations labelled by their dimension $\mathbf{m}$.  
The states in (\ref{eq:Short}) span then the highest weight space from which the full representation $\mathscr{F}_\lambda$ is generated by the action of the negative  modes of $\mathfrak{psu}(1,1|2)_1$. 

\subsection{Spectral flow}

Additional representations can be generated by applying the spectral flow automorphism of the current superalgebra
\begin{subequations}
\label{eq:SpecFlow}
\begin{align}
    \sigma^w(J_m^3) &= J_m^3+ \tfrac{w}{2}\delta_{m,0}\,,\\
    \sigma^w(J_m^\pm) &= J^\pm_{m\mp w}\,,\\
    \sigma^w(K_m^3) &= K_m^3+ \tfrac{w}{2}\delta_{m,0}\,,\\
    \sigma^w(K_m^\pm) &= K^\pm_{m\pm w}\,,\\[1mm]
    \sigma^w(S_m^{\alpha\beta\gamma})&=S^{\alpha\beta\gamma}_{m+\frac{1}{2}w(\beta-\alpha)}\ , 
\end{align}
\end{subequations}
where $w\in\mathbb{Z}$. On the Virasoro modes, spectral flow then acts as 
\begin{align}
    \sigma^w(L_n) = L_n +w(K_n^3-J_n^3)\,.
\end{align}
Spectral flow can be thought of as being implemented by an  operator $S_w$ on the vector space of states, i.e.\ 
\begin{align}
    \sigma^w(\mathscr{F}_\lambda) \equiv S_w(\mathscr{F}_\lambda) \qquad \hbox{with} \qquad 
    \sigma^w(W_n)\equiv S_w^{-1}W_n S_w 
\end{align}
for any mode $W_n$.  If $\Phi\in\mathscr{F}_\lambda$ is a state in any (unflowed) highest-weight representation of $\mathfrak{psu}(1,1|2)_1$, we will denote its image under spectral flow by $w$ units as 
\begin{align}
    [\Phi]^w \equiv S_w \, \Phi\in \sigma^w(\mathscr{F}_\lambda)\ , 
\end{align}
so that 
\begin{align}
  W_n\, [\Phi]^w  = S_w \, S_w^{-1}\, W_n \, S_w \Phi =   [\sigma^w(W_n)\Phi]^w \ , \label{eq:bracket}
\end{align}
for any mode $W_n$. Note that in contrast to the case of the $\mathfrak{su}(2)_1$ subalgebra where each unit of spectral flow simply exchanges the two highest weight representations, spectral flow for $\mathfrak{sl}(2;\mathds{R})_1$ genuinely leads to new representations; in particular, the resulting representations are typically not highest weight.

\subsection{Characters and their modular properties}

For the calculation of the cylinder diagram we will need the (unspecialised) supercharacters corresponding to $\sigma^w(\mathscr{F}_\lambda)$,
\begin{subequations}
\begin{align}
    \widetilde{\mathrm{ch}}[\sigma^w(\mathscr{F}_\lambda)](t,z;\tau)
    &\equiv \mathrm{tr}_{\sigma^w(\mathscr{F}_\lambda)} \big[(-1)^F q^{L_0 - \frac{c}{24}}x^{J_0^3}y^{K_0^3}\big]\\
    &=(-1)^w q^\frac{w^2}{2}\sum_{r\in \mathbb{Z}+\lambda} x^r q^{-rw}\frac{\theta_1(\frac{t+z}{2};\tau)\theta_1(\frac{t-z}{2};\tau)}{\eta(\tau)^4} \ ,\label{eq:char}
\end{align}
\end{subequations}
where $q=e^{2\pi i \tau}$ is the usual modular parameter, and $x = e^{2\pi i t}$ and $y=e^{2\pi i z}$ are the $\mathfrak{sl}(2;\mathds{R})$ and $\mathfrak{su}(2)$ fugacities, respectively. (The tilde indicates that we are considering the character with the insertion of $(-1)^F$.)
Relative to the expression used in \cite{Eberhardt:2020bgq} we have introduced the factor $(-1)^w$ which accounts for the (natural) fermion number of the ground states upon spectral flow (see also \cite{Ferreira:2017pgt} and \cite{Giribet:2007wp}). 
Under the modular S-transformation, we have
\begin{align}
e^{\frac{\pi i}{2\tau}(t^2-z^2)}\widetilde{\mathrm{ch}}[\sigma^w(\mathscr{F}_\lambda)](\tfrac{t}{\tau},\tfrac{z}{\tau};-\tfrac{1}{\tau})=\sum_{w'\in\mathbb{Z}}\int_0^1 d\lambda'\, S_{(w,\lambda),(w',\lambda')}\,\widetilde{\mathrm{ch}}[\sigma^{w'}(\mathscr{F}_{\lambda'})](t,z;\tau)\,,\label{eq:SMod}
\end{align}
where we have introduced the modular S-matrix
\begin{align}
    S_{(w,\lambda),(w',\lambda')}=\frac{|\tau|}{-i\tau}\,e^{2\pi i [w'(\lambda-\frac{1}{2})+w(\lambda'-\frac{1}{2})]}\,.\label{eq:SMat}
\end{align}
Note that $S_{(w,\lambda),(w',\lambda')}$ depends on $\tau$, as is typical for logarithmic CFTs. This dependence will however drop out of all physical quantities such as the fusion coefficients that can be determined via the Verlinde formula as in \cite{Eberhardt:2018ouy}.

\subsection{Bulk spectrum}

We will also need the precise form of the closed string spectrum. Following Maldacena and Ooguri \cite{Maldacena:2000hw}, see also \cite{Eberhardt:2018ouy}, it equals  
\begin{align}
    \mathcal{H} = \bigoplus_{w\in\mathbb{Z}}\int_{\lambda\in [0,1)}d\lambda\, \sigma^w(\mathscr{F}_\lambda)\otimes\overline{\sigma^w(\mathscr{F}_\lambda)}\ ,\label{eq:BulkSpec}
\end{align}
which gives rise to a modular-invariant partition function. We should mention that for $\lambda=\frac{1}{2}$, the representation $\mathscr{F}_{1/2}$ is actually indecomposable and needs to be treated with some care, see \cite{Eberhardt:2018ouy} for details. However, as in that paper, this subtlety will not play an important role in the following.

\subsection{The free field realisation}\label{sec:free-fields}

The $\mathfrak{psu}(1,1|2)_1$ WZW model has another important property which we will later employ to calculate correlation functions in the presence of these D-branes, see Section~\ref{sec:correlators}. Namely, the worldsheet theory admits a realisation in terms of free fields \cite{Eberhardt:2018ouy}, which is analogous to the standard free field construction of the $\mathfrak{su}(N)_1$ WZW models in terms of free fermions. For a more detailed treatment of this construction, see \cite{Dei:2020zui}.

Given four spin-$\frac{1}{2}$ chiral bosons $\xi^{\pm},\eta^{\pm}$ (which, by convention, are called `symplectic' bosons) and four spin-$\frac{1}{2}$ chiral fermions $\psi^{\pm},\chi^{\pm}$ satisfying the OPEs
\begin{equation}\label{eq:symplectic-ope}
\xi^{\alpha}(z)\,\eta^{\beta}(w)\sim\frac{\varepsilon^{\alpha\beta}}{z-w}\ , \qquad \qquad
\psi^{\alpha}(z)\chi^{\beta}(w)\sim\frac{\varepsilon^{\alpha\beta}}{z-w}\ ,
\end{equation}
we can construct spin-$1$ currents as the bilinears
\begin{equation}\label{eq:freefield}
\begin{array}{ll}
J^3=-\frac{1}{2}\left(\xi^+\eta^-+\xi^-\eta^+\right)\ ,\qquad \qquad &\quad K^3=\frac{1}{2}(\psi^+\chi^-+\psi^-\chi^+)\ ,\\
J^{\pm}=\xi^{\pm}\eta^{\pm}\ , \qquad &\quad K^{\pm}=\mp\psi^{\pm}\chi^{\pm}\ ,  \\
S^{\alpha\beta+}=\xi^{\alpha}\chi^{\beta}\ , \qquad &\quad S^{\alpha\beta-}=-\eta^{\alpha}\psi^{\beta}\ ,\\
U=\frac{1}{2}\left(\xi^+\eta^--\xi^-\eta^+\right)\ , \qquad &\quad V=\frac{1}{2}\left(\psi^-\chi^+-\psi^+\chi^-\right)\ ,
\end{array}
\end{equation}
where the products are taken to be normal-ordered. As their names suggest, the currents $J^a$ and $K^a$ generate the bosonic subalgebras $\mathfrak{sl}(2,\mathbb{R})_1$ and $\mathfrak{su}(2)_1$ respectively of $\mathfrak{psu}(1,1|2)_1$, and satisfy the commutation relations \eqref{eq:psu-relations}. The supercurrents, on the other hand, satisfy the anti-commutation relations
\begin{equation}
\begin{split}
\{S^{\alpha\beta\gamma}_m,S^{\mu\nu\rho}_n\}&=m\epsilon^{\alpha\mu}\epsilon^{\beta\nu}\epsilon^{\gamma\rho}\delta_{m+n,0}-\epsilon^{\beta\nu}\epsilon^{\gamma\rho}c_a(\sigma_a)^{\alpha\mu}J^a_{m+n}\\
&\hspace{2cm}+\epsilon^{\alpha\mu}\epsilon^{\gamma\rho}(\sigma_a)^{\beta\nu}K^a_{m+n}+\epsilon^{\alpha\mu}\epsilon^{\beta\nu}\delta^{\gamma,-\delta}Z_{m+n}\ ,
\end{split}
\end{equation}
where we have defined
\begin{equation}
Y=U-V\ ,\qquad \qquad Z=U+V\ ,
\end{equation}
which satisfy the commutation relations
\begin{equation}
[Z_m,Z_n]=[Y_m,Y_n]=0\ , \qquad \qquad[Z_m,Y_n]=-m\delta_{m+n,0}\ .
\end{equation}
Finally, note that the current $Z$ is central, other than its nontrivial commutator with $Y$, and so if we take $Z_n=0$, we would recover the usual $\mathfrak{psu}(1,1|2)_1$ algebra. As pointed out in \cite{Dei:2020zui}, getting rid of the $Z$ and $Y$ modes amounts to taking the coset with respect to the $\mathfrak{u}(1)$ generated by $Z$ (that is, we consider only the subspace of states with $Z_n\ket{\psi}=0$ for all $n\geq 0$), and then quotienting by descendant states $Z_{-n}\ket{\psi}$ for $n>0$, which are all null in the coset. By abuse of notation, we will refer to this process as `gauging' the $Z$ current. The end result of this process is a worldsheet theory which is equivalent to the $\mathfrak{psu}(1,1|2)_1$ theory.

States in the free field theory are constructed from highest-weight representations of the algebra \eqref{eq:symplectic-ope} in the Ramond sector. The highest weight states $\ket{m,j}$ in this sector are labelled by their $J^3_0$ eigenvalue $m$ and their $U_0$ eigenvalue $j-\frac{1}{2}$, and the symplectic boson zero modes act on them as
\begin{equation}\label{eq:symplectic-representation}
\begin{array}{ll}
\xi^{+}_0\ket{m,j}=\ket{m+\tfrac{1}{2},j-\tfrac{1}{2}}\ , \qquad \qquad & \eta^+_0\ket{m,j}=(m+j)\ket{m+\tfrac{1}{2},j+\tfrac{1}{2}}\ ,\\
\xi^{-}_0\ket{m,j}=-\ket{m-\tfrac{1}{2},j-\tfrac{1}{2}}\ , \qquad \qquad & \eta^-_0\ket{m,j}=-(m-j)\ket{m-\tfrac{1}{2},j+\tfrac{1}{2}}\ .
\end{array}
\end{equation}
On these states, $Z_n=U_n$, and so the coset condition $Z_n=0$ with $n=0$ imposes $U_0=0$ or $j=\frac{1}{2}$. The resulting representation is labelled by the fractional part $\lambda$ of the $J^3_0$ eigenvalue.

With respect to the fermions, we take the states in \eqref{eq:symplectic-representation} to be highest-weight with respect to $\psi^+$ and $\chi^+$. Allowing the remaining two fermions to act on these states furnishes a Clifford representation. In particular, the states
\begin{equation}
\ket{m,j}\ , \qquad \qquad\psi^-_0\chi^-_0\ket{m,j}
\end{equation}
together furnish the $\textbf{2}$ of $\mathfrak{su}(2)$, while the states
\begin{equation}
\psi^-_0\ket{m,j}\ , \qquad \qquad\chi^-_0\ket{m,j}
\end{equation}
individually furnish the $\textbf{1}$ of $\mathfrak{su}(2)$. Imposing the `gauging' of the $\mathfrak{u}(1)$ current $Z$ forces the states transforming in the $\textbf{2}$ of $\mathfrak{su}(2)$ to have $j=\frac{1}{2}$, while the states transforming in the two $\textbf{1}$s of $\mathfrak{su}(2)$ have $j=0$ and $j=1$, respectively. This is how the supermultiplets \eqref{eq:Short} of $\mathfrak{psu}(1,1|2)_1$ arise in the free field construction.

Finally, spectral flow is realised on the free fields via the automorphism
\begin{equation}
\begin{array}{ll}
\sigma^w(\xi^{\pm}_r)=\xi^{\pm}_{r\mp\frac{w}{2}}\ , \qquad &\qquad\sigma^w(\eta^{\pm}_r)=\eta^{\pm}_{r\mp\frac{w}{2}}\,,\\
\sigma^w(\psi^{\pm}_r)=\psi^{\pm}_{r\pm\frac{w}{2}}\ , \qquad &\qquad\sigma^w(\chi^{\pm}_r)=\chi^{\pm}_{r\pm\frac{w}{2}}\,,
\end{array}
\end{equation}
which exactly reproduces the spectral flow of the $\mathfrak{psu}(1,1|2)_1$ currents, while acting trivially on the currents $Y$ and $Z$.

\section{\boldmath The boundary states of \texorpdfstring{$\mathrm{PSU}(1,1|2)_1$}{PSU(1,1|2)1}}\label{sec:psubound}

In this section we shall construct (some) boundary states of the ${\rm PSU}(1,1|2)_1$ WZW model. We shall first discuss the relevant gluing conditions and then construct the corresponding Ishibashi states. We assume that the reader has some basic knowledge about the boundary states of a CFT \cite{Cardy:1989ir};  suitable introductory texts are for example  \cite{Recknagel:2013uja,Gaberdiel:2002my}.

In the following we shall concentrate on the boundary conditions that preserve the full $\mathfrak{psu}(1,1|2)$ symmetry, up to an automorphism. Following \cite{Bachas:2000fr} there are two interesting cases, corresponding to whether the automorphism is trivial or a specific order two transformation.  In this paper we shall mainly focus on the case with a ``trivial'' gluing automorphism.\footnote{Here we have put the label ``trivial'' in quotation marks, because in the conventions of \cite{Maldacena:2000hw}, which we are adopting, they seem to involve a non-trivial automorphism. This is due to the fact that \cite{Maldacena:2000hw} implicitly apply an additional automorphism $\bar{J}^3\to -\bar{J}^3$, $\bar{J}^\pm\to \bar{J}^\mp$ in the anti-holomorphic sector.} The corresponding D-branes describe the so-called {\em spherical branes} of \cite{Ponsot:2001gt,Bachas:2000fr}, and as we shall see they are compatible with spectral flow. On the other hand, the $\text{AdS}_2$ branes of \cite{Bachas:2000fr} correspond to a gluing condition with a non-trivial (order $2$) automorphism of $\mathfrak{sl}(2;\mathds{R})_1$, and, as we shall explain, they are not compatible with spectral flow. As a consequence their interpretation in the dual CFT is more delicate. 

\subsection{Spherical branes}

As explained above, we take the ``trivial" gluing conditions to be 
	\begin{align}
	J^{\, 3}(z)=-\bar{J}^{\, 3}(\bar{z})\ , \qquad 
	 J^{\, \pm}(z)=\bar{J}^{\, \mp}(\bar{z}) \label{eq:GluJIdp}
	\end{align}
at $z=\bar{z}$, where the right-movers are denoted by a bar. 
On the other hand, since $\mathfrak{su}(2)$ does not admit any outer automorphisms, we may always, without loss of generality, assume that the $\mathfrak{su}(2)$ gluing conditions are trivial
	\begin{align}
	K^a(z)=\bar{K}^a(\bar{z})\,.\label{eq:GluKIdp}
	\end{align}
Finally, we need to impose suitable gluing conditions on the fermionic currents so as to preserve the full $\mathfrak{psu}(1,1|2)_1$ algebra. It is easy to see that these are
\begin{align}
S^{\alpha\beta\gamma}(z)=\varepsilon\tensor{(i\sigma^2)}{^\alpha_\mu}\, \bar{S}^{\mu\beta\gamma}(\bar{z})\ ,\label{eq:GluSId}
\end{align}
where $\varepsilon=\pm$ plays the role of the worldsheet spin structure. At $k=1$ where the radii of both the $\mathrm{AdS}_3$ and $\mathrm{S}^3$ are comparable to the string length, we expect the worldvolumes of these branes to be fuzzy, but generally localised around a particular value in time (since formally they satisfy a Dirichlet boundary condition along  the $\mathrm{AdS}_3$ time direction). Adopting the terminology of \cite{Ponsot:2001gt}, we will call them spherical branes. 

Translating the above gluing conditions into the closed string language we are thus looking for Ishibashi states \cite{Ishibashi:1988kg} that are characterised by 
\begin{subequations}
\label{eq:IshibashiSpherical}
\begin{align}
    (J_n^3-\bar{J}_{-n}^3)|w,\lambda,\varepsilon\rangle\!\rangle_{\mathrm{S}} &=0\ , \label{J3cond}\\
    (J_n^\pm+\bar{J}_{-n}^\mp)|w,\lambda,\varepsilon\rangle\!\rangle_{\mathrm{S}} &=0\ ,\\
    (K_n^a+\bar{K}_{-n}^a)|w,\lambda,\varepsilon\rangle\!\rangle_{\mathrm{S}} &=0\ ,\\
    (S_n^{\alpha\beta\gamma}+\varepsilon\tensor{(i\sigma^2)}{^\alpha_\mu}\bar{S}_{-n}^{\mu\beta\gamma})|w,\lambda,\varepsilon\rangle\!\rangle_{\mathrm{S}} &=0\ .
\end{align}
\end{subequations}
The label ``$\mathrm{S}$'' here stands for ``spherical'' and is there to distinguish these states from the Ishibashi states for the $\mathrm{AdS}_2$ branes, which we introduce below and label by ``$\mathrm{AdS}$''. The corresponding boundary conditions in the free field construction of $\mathfrak{psu}(1,1|2)_1$ take the form
\begin{equation}\label{eq:free-field-spherical}
\begin{split}
(\xi^{\alpha}_{r}+e^{i\varphi}(\sigma^2)\indices{^\alpha_{\beta}}\, \bar{\xi}^{\beta}_{-r})|w,\lambda,\varepsilon\rangle\!\rangle_{\mathrm{S}}&=0\ ,\\
(\eta^{\alpha}_{r}+e^{-i\varphi}(\sigma^2)\indices{^\alpha_{\beta}}\, \bar{\eta}^{\beta}_{-r})|w,\lambda,\varepsilon\rangle\!\rangle_{\mathrm{S}}&=0\ ,\\
(\psi^{\alpha}_r-i\varepsilon\,e^{i\varphi}\, \bar{\psi}^{\alpha}_{-r})|w,\lambda,\varepsilon\rangle\!\rangle_{\mathrm{S}}&=0\ ,\\
(\chi^{\alpha}_r-i\varepsilon\,e^{-i\varphi}\, \bar{\chi}^{\alpha}_{-r})|w,\lambda,\varepsilon\rangle\!\rangle_{\mathrm{S}}&=0\ ,
\end{split}
\end{equation}
where $e^{i\varphi}$ is some arbitrary phase, which can be gauged away by a global $\text{U}(1)$ transformation. It is easy to see that this then leads to the gluing conditions (\ref{eq:IshibashiSpherical}) using the free field relations of eq.~(\ref{eq:freefield}). Furthermore, on the auxiliary $\mathfrak{u}(1)$ fields $Y$ and $Z$, the resulting boundary conditions are trivial and hence compatible with the gauging procedure. 
\smallskip

Let us first consider the unflowed representations. The condition (\ref{J3cond}) with $n=0$ implies that the left- and right-moving $J^3_0$ eigenvalues must be equal.\footnote{The following analysis will be done on the level of $\mathfrak{psu}(1,1|2)$; we could also arrive at the same results using the free field gluing conditions (\ref{eq:free-field-spherical}).}  In order for this to be possible we need $\lambda=\bar{\lambda}$, and this is indeed the case for all representations appearing in the 
bulk spectrum \eqref{eq:BulkSpec}. Thus in the unflowed sector all ${\cal F}_\lambda$ sectors have an Ishibashi state which we shall denote by $|0,\lambda,\varepsilon\rangle\!\rangle_{\mathrm{S}}$. 

Next we observe that the Ishibashi state in the $w$-spectrally flowed sector can be simply obtained from the unflowed Ishibashi state via 
\begin{align}
 |w,\lambda,\varepsilon\rangle\!\rangle_{\mathrm{S}}=\big[|0,\lambda,\varepsilon\rangle\!\rangle_{\mathrm{S}}\big]^w  \ . 
\end{align}
Thus there exists a (non-trivial) Ishibashi state in each sector, and we shall label it by 
\begin{align}
|w,\lambda,\varepsilon\rangle\!\rangle_{\mathrm{S}}\qquad \text{for all $w\in\mathbb{Z}$ and $\lambda\in [0,1)$.}
\end{align}
For the following we shall mainly need the elementary overlaps
\begin{align}
    _{\mathrm{S}}\langle\!\langle w',\lambda',\mp| \hat{q}^{\frac{1}{2}(L_0+\bar{L}_0-\frac{c}{12})}\hat{x}^{\frac{1}{2}(J_0^3+\bar{J}_0^3)}|w,\lambda,\pm\rangle\!\rangle_{\mathrm{S}} &= 
        \delta_{w,w'}\, \delta_{\lambda,\lambda'}\,\widetilde{\mathrm{ch}}[\sigma^{w}(\mathscr{F}_\lambda)](\hat{t};\hat{\tau}) \ , 
    \label{eq:overlap}
\end{align}
where the character appearing on the right-hand-side involves $(-1)^F$, see eq.~(\ref{eq:char}), since 
\begin{align}
    (-1)^F|w,\lambda,\varepsilon\rangle\!\rangle_{\mathrm{S}}=|w,\lambda,-\varepsilon\rangle\!\rangle_{\mathrm{S}}\ .
\end{align}
Here $\hat{q}$ and $\hat{x}$ are defined by
\be
\hat{q}=e^{2\pi i \hat{\tau}}   \ , \qquad 
\hat{x}=e^{2\pi i \hat{t}} \ ,  
\ee
where $\hat{\tau}$ and $\hat{t}$ are real. Identifying $\hat{q}^{\frac{1}{2}(L_0+\bar{L}_0-\frac{c}{12})}$ with the worldsheet propagator, the overlap \eqref{eq:overlap} computes the zero-point worldsheet conformal block on a cylinder with length $\hat{\tau}$. The $\mathfrak{sl}(2;\mathds{R})$ chemical potential $\hat{t}$, on the other hand, will be interpreted as the length of the corresponding spacetime cylinder, which is cut out on the boundary of $\mathrm{AdS}_3$ by the two spherical branes. Similarly, we could have introduced a chemical potential $\hat{\zeta}$ for the $\mathfrak{su}(2)$ currents $K_n^a$ by inserting
$\hat{y}^{\frac{1}{2}(K_0^3+\bar{K}_0^3)}$ into the overlap with $\hat{y}=e^{2\pi i \hat{\zeta}}$ --- we will suppress this in order not to clutter the formulae.

The inclusion of the spacetime propagator $\hat{x}^{\frac{1}{2}(J_0^3+\bar{J}_0^3)}$ can be thought of either as a way of `unspecialising' the cylinder conformal blocks, or, as computing a twisted
overlap between Ishibashi states of the form 
\begin{align}
|w,\lambda,\varepsilon;\hat{t}\, \rangle\!\rangle_{\mathrm{S}}=\hat{x}^{\frac{1}{2}(J_0^3+\bar{J}_0^3)}|w,\lambda,\varepsilon\rangle\!\rangle_{\mathrm{S}} = 
\hat{x}^{J_0^3}|w,\lambda,\varepsilon\rangle\!\rangle_{\mathrm{S}} \ , 
\label{eq:shifted}
\end{align}
shifted by a distance $\hat{t}$ along the spacetime cylinder. (Here we have used the gluing condition (\ref{J3cond}) in the final equation.) These shifted Ishibashi states then satisfy the gluing conditions 
\begin{subequations}
\label{eq:IshibashiSphericalTw}
\begin{align}
    (J_n^3-\bar{J}_{-n}^3)|w,\lambda,\varepsilon;\hat{t}\rangle\!\rangle_{\mathrm{S}} &=0\ ,\\
    (e^{\pm \pi i \hat{t}}J_n^\pm+e^{\mp \pi i \hat{t}}\bar{J}_{-n}^\mp)|w,\lambda,\varepsilon;\hat{t}\rangle\!\rangle_{\mathrm{S}} &=0\ ,\\[1mm]
    (K_n^a+\bar{K}_{-n}^a)|w,\lambda,\varepsilon;\hat{t}\rangle\!\rangle_{\mathrm{S}} &=0\ ,\\
    (e^{\pm \frac{1}{2}\alpha \pi i \hat{t}} S_n^{\alpha\beta\gamma}+\varepsilon
    e^{ \frac{1}{2}\beta \pi i \hat{t}}
    \tensor{(i\sigma^2)}{^\alpha_\mu}\bar{S}_{-n}^{\beta\gamma})|w,\lambda,\varepsilon;\hat{t}\rangle\!\rangle_{\mathrm{S}} &=0\ .
\end{align}
\end{subequations}
We note that for  $\hat{t}=m\in \mathbb{Z}$, the shifted Ishibashi states \eqref{eq:shifted} again obey the Ishibashi conditions \eqref{eq:IshibashiSpherical}, with $\varepsilon \mapsto (-1)^m\varepsilon $, i.e.\ 
\begin{align}
    |w,\lambda,\varepsilon;m\rangle\!\rangle_{\mathrm{S}} 
    = (-1)^{mw} e^{2\pi i m\lambda}  |w,\lambda,(-1)^m\varepsilon \rangle\!\rangle_{\mathrm{S}}
    \ .
\end{align}
This reflects the periodicity $\hat{t}\to \hat{t}+2$ of the ${\rm SL}(2;\mathds{R})$ group manifold, which was promoted to global $\mathrm{AdS}_3$ by considering its universal covering.

\subsubsection{Consistent boundary states}

Next we want to assemble these Ishibashi states into consistent boundary states. Since we have an Ishibashi state from each closed string sector, we would expect that there exists a D-brane whose open string spectrum will just consist of the `vacuum' representation  $\mathscr{F}_{1/2}$. In fact, this property seems to hold for any boundary state of the form 
\begin{align}
\| W,\Lambda,\varepsilon\rangle\!\rangle_\mathrm{S} = \sum_{w\in \mathbb{Z}}\int_0^1 d\lambda\, e^{2\pi i [w(\Lambda-\frac{1}{2})+(\lambda-\frac{1}{2}) W]}|w,\lambda,\varepsilon\rangle\!\rangle_\mathrm{S} \ , \label{eq:BSspherical}
\end{align}
where $W$ and $\Lambda$ are some parameters whose 
physical interpretation will be discussed shortly. Indeed, the unspecialised $\mathfrak{psu}(1,1|2)_1$ overlap equals
\begin{subequations}
\label{eq:psuoverlap}
\begin{align}
\hat{Z}^{\text{S}}_{(W_1,\Lambda_1)|(W_2,\Lambda_2)}(\hat{t};\hat{\tau})&\equiv\,_{\text{S}}\langle\!\langle \Lambda_2,W_2,\mp\| \hat{q}^{\frac{1}{2}(L_0+\bar{L}_0-\frac{c}{12})}\hat{x}^{\frac{1}{2}(J_0^3+\bar{J}_0^3)}\| \Lambda_1,W_1,\pm\rangle\!\rangle_{\text{S}}\\
&=\sum_{w\in\mathbb{Z}}\int_0^1 d\lambda \, e^{2\pi i w(\Lambda_1-\Lambda_2)} \, 
e^{2\pi i (\lambda-\frac{1}{2})(W_1-W_2)}\widetilde{\text{ch}}[\sigma^w(\mathscr{F}_\lambda)](\hat{t};\hat{\tau})\ , \label{3.14b}
\end{align}
\end{subequations}
where, as before, taking the opposite values of the parameter $\varepsilon$ for the two boundary states effectively inserts $(-1)^F$ into the overlap, and thus yields the $\mathfrak{psu}(1,1|2)_1$ supercharacter $\widetilde{\mathrm{ch}}$, as appropriate when discussing boundary states of supergroups.
Performing the modular S-transformation to the open-string channel, we obtain the corresponding boundary superpartition function
	\begin{align}
{Z}^{\text{S}}_{(W_1,\Lambda_1)|(W_2,\Lambda_2)}({t};{\tau})
	&= e^{-\frac{\pi i\tau}{2t^2}}\, \widetilde{\mathrm{ch}}[\sigma^{W_2-W_1}(\mathscr{F}_{\frac{1}{2}-\Lambda_1+\Lambda_2})](-\tfrac{\tau}{t};\tau)\ ,\label{eq:SphericalOpen}
	\end{align}
where we define the open-string variables $t$ and $\tau$ via 
\be
    \hat{t}=-\frac{1}{t}\ , \qquad     \hat{\tau}=-\frac{1}{\tau}\ . 
\ee
The prefactor $e^{-\frac{\pi i\tau}{2t^2}}$ is familiar from the modular transformation of Jacobi forms; it reflects the fact that the `open string' spectrum is twisted by the insertion of the chemical potential in the closed string overlap, see eq.~(\ref{eq:shifted}). For $W_1=W_2$ and $\Lambda_1=\Lambda_2$ the open string spectral flow is trivial, and we just get $\mathscr{F}_{1/2}$ as anticipated. 

In order to understand the interpretation of $W$ and $\Lambda$ we observe that 
\begin{align}
(-1)^{W(F+1)}e^{\pi i W(J_0^3+\bar{J}_0^3)}\| 0,\Lambda-\tfrac{W}{2},\varepsilon\rangle\!\rangle_\mathrm{S} = \| W,\Lambda,\varepsilon\rangle\!\rangle_\mathrm{S} \ .
\end{align}
Thus up to twisting by $(-1)^F$ and suitably adjusting $\Lambda$, the parameter $W$ can be identified with the shift of the boundary state in the time direction of $\mathrm{AdS}_3$, see eq.~(\ref{eq:IshibashiSphericalTw}). On the other hand, $\Lambda$ describes the Wilson line in the angular direction of $\mathrm{AdS}_3$ along which our boundary state satisfies a Neumann boundary condition; indeed, the relative Wilson lines $\Lambda_1$ and $\Lambda_2$ introduce the factors $e^{2\pi i w(\Lambda_1-\Lambda_2)}$ into the $w$-flowed sectors of the boundary state overlap, see 
eq.~(\ref{3.14b}). For the following these parameters will not play an important role, and in order to simplify our expressions we shall set them to $W=0$ and $\Lambda=\frac{1}{2}$ from now on; then the boundary state simplifies to 
\begin{align}
\|\varepsilon\rangle\!\rangle_{\text{S}} \equiv \|W=0,\Lambda=\tfrac{1}{2},\varepsilon\rangle\!\rangle_{\text{S}}= \sum_{w\in \mathbb{Z}}\int_0^1 d\lambda\, |w,\lambda,\varepsilon\rangle\!\rangle_\mathrm{S}\ .\label{eq:BLambda}
\end{align}

\subsection[\texorpdfstring{$\mathrm{AdS}_2$}{AdS2} branes]{\boldmath \texorpdfstring{$\mathrm{AdS}_2$}{AdS2} branes}\label{sec:AdS2}

For completeness let us also discuss the boundary states describing the $\mathrm{AdS}_2$ branes. Their gluing conditions are `twined' relative to the gluing conditions for the above spherical branes, and hence look trivial in our conventions, i.e.\ we have at $z=\bar{z}$ 
\begin{subequations}
\label{eq:GluJp}
\begin{align}
	J^a(z)&=\bar{J}^a(\bar{z})\ ,\\
	K^a(z)&=\bar{K}^a(\bar{z})\ ,\\
	S^{\alpha\beta\gamma}(z)&=\varepsilon\, \bar{S}^{\alpha\beta\gamma}(\bar{z})\ ,
	\end{align}
\end{subequations}
where $\varepsilon$ is again a sign. This time, the corresponding Ishibashi states are characterised by 
\begin{subequations}
\label{eq:IshibashiAdS}
\begin{align}
    (J_n^a+\bar{J}_{-n}^a)|0,\lambda,\varepsilon\rangle\!\rangle_{\mathrm{AdS}}&=0\ , \label{J31}\\[0.5mm]
    (K_n^a+\bar{K}_{-n}^a)|0,\lambda,\varepsilon\rangle\!\rangle_{\mathrm{AdS}}&=0\ ,\\
    (S_n^{\alpha\beta\gamma}+ \varepsilon\, \bar{S}_{-n}^{\alpha\beta\gamma})|0,\lambda,\varepsilon\rangle\!\rangle_{\mathrm{AdS}}&=0\ ,
\end{align}
\end{subequations}
or, in terms of the free fields,
\begin{equation}\label{eq:ads2-free-fields}
\begin{split}
\left(\xi^{\alpha}_r-i\,e^{i\varphi}\bar{\xi}^{\alpha}_{-r}\right)|0,\lambda,\varepsilon\rangle\!\rangle_{\mathrm{AdS}}&=0\,,\\
\left(\eta^{\alpha}_r-i\,e^{-i\varphi}\bar{\eta}^{\alpha}_{-r}\right)|0,\lambda,\varepsilon\rangle\!\rangle_{\mathrm{AdS}}&=0\,,\\
\left(\psi^{\alpha}_r-i\varepsilon\,e^{i\varphi}\bar{\psi}^{\alpha}_{-r}\right)|0,\lambda,\varepsilon\rangle\!\rangle_{\mathrm{AdS}}&=0\,,\\
\left(\chi^{\alpha}_r-i\varepsilon\,e^{-i\varphi}\bar{\chi}^{\alpha}_{-r}\right)|0,\lambda,\varepsilon\rangle\!\rangle_{\mathrm{AdS}}&=0\,,
\end{split}
\end{equation}
where, again $e^{i\varphi}$ is some arbitrary phase which plays no physical role. Just as in the case of the spherical branes, the free field boundary conditions also impose the trivial boundary conditions on $Y_n$ and $Z_n$.

Now, the first line of \eqref{eq:IshibashiAdS} implies $m=-\bar{m}$, and hence only has a solution in the unflowed sector if $\lambda = - \bar{\lambda}\ \hbox{(mod $1$)}$. Given the structure of the bulk spectrum \eqref{eq:BulkSpec}, this is only possible for $\lambda\in \{0,\tfrac{1}{2}\}$, and thus we only have the two Ishibashi states 
\begin{align}
|0,\lambda,\varepsilon\rangle\!\rangle_{\mathrm{AdS}}\qquad \text{with $\lambda\in \{0,\tfrac{1}{2}\}$} \ . 
\end{align}
It is also not difficult to convince oneself that no Ishibashi states exist for non-trivial spectral flow; a geometric argument for this is given in Appendix~\ref{app:A.3}, but one can also arrive at this conclusion algebraically. 

The elementary overlap relevant for the case of the $\text{AdS}_2$ branes now reads 
\begin{align}
    _{\mathrm{AdS}}\langle\!\langle 0,\lambda',\mp | \hat{q}^{\frac{1}{2}(L_0+\bar{L}_0-\frac{c}{12})}\hat{\xi}^{\frac{1}{2}(J_0^3-\bar{J}_0^3)}|0,\lambda,\pm\rangle\!\rangle_{\mathrm{AdS}} &= 
    \delta_{\lambda,\lambda'}\,\widetilde{\mathrm{ch}}[\mathscr{F}_\lambda](\hat{\theta};\hat{\tau}) \ ,    
 \label{eq:overlapAdS}
\end{align}
where $\hat{\xi}=e^{2\pi i\hat{\theta}}$ 
implements the translation by $\hat{\theta}$ in the angular direction of the $\mathrm{AdS}_3$ boundary cylinder. (This is again the direction in which the branes satisfy a Dirichlet boundary condition on the $\mathrm{AdS}_3$ boundary, see the discussion in Appendix~\ref{app:A.3}.) Alternatively, we may describe this translation in terms of shifted Ishibashi states (c.f.\ the analysis of eq.~(\ref{eq:IshibashiSphericalTw}))
\begin{align}\label{AdS2rot}
    |0,\lambda,\varepsilon;\hat{\theta}\rangle\!\rangle_{\mathrm{AdS}}=\hat{\xi}^{\frac{1}{2}(J_0^3-\bar{J}_0^3)}|0,\lambda,\varepsilon\rangle\!\rangle_{\mathrm{AdS}} =\hat{\xi}^{J_0^3}|0,\lambda,\varepsilon\rangle\!\rangle_{\mathrm{AdS}} \ ,
\end{align}
which satisfy the twisted Ishibashi conditions
\begin{subequations}
\label{eq:IshibashiAdSTw}
\begin{align}
(J_n^3+\bar{J}_{-n}^3)|0,\lambda,\varepsilon;\hat{\theta}\rangle\!\rangle_{\mathrm{AdS}} &=0\ ,\\
(e^{\pm \pi i \hat{\theta}}J_n^\pm+e^{\mp \pi i \hat{\theta}}\bar{J}_{-n}^\pm)|0,\lambda,\varepsilon;\hat{\theta}\rangle\!\rangle_{\mathrm{AdS}} &=0\ ,\\[1mm]
(K_n^a+\bar{K}_{-n}^a)|0,\lambda,\varepsilon;\hat{\theta}\rangle\!\rangle_{\mathrm{AdS}} &=0\ ,\\
(e^{+\frac{1}{2}\alpha \pi i \hat{\theta}} S_n^{\alpha\beta\gamma}+\varepsilon
    e^{-\frac{1}{2}\alpha \pi i \hat{\theta}}
   \bar{S}_{-n}^{\alpha\beta\gamma})\, |0,\lambda,\varepsilon;\hat{\theta}\rangle\!\rangle_{\mathrm{AdS}} &=0\ .
\end{align}
\end{subequations}  
For $\hat{\theta}=m\in\mathbb{Z}$, these again reduce to the Ishibashi conditions \eqref{eq:IshibashiAdS} with the sign in the supercurrent condition being $(-1)^m\varepsilon$; explicitly, we have
\begin{align}\label{AdS2brane}
    |0,\lambda,\varepsilon;m\rangle\!\rangle_{\mathrm{AdS}} 
    =e^{2\pi i m\lambda}  |0,\lambda,(-1)^m\varepsilon\rangle\!\rangle_{\mathrm{AdS}}
    \ .
\end{align}
For $\hat{\theta}\to\hat{\theta}+2$, we recover the periodicity of the $\mathrm{AdS}_3$ bulk in its angular coordinate, where the Ishibashi states pick up a phase of $e^{4\pi i\lambda}=+1$ (since $\lambda\in \{0,\frac{1}{2}\}$) upon fully encircling the center of $\mathrm{AdS}_3$. On the other hand, when translating the Ishibashi states to an antipodal position by shifting $\hat{\theta}\to\hat{\theta}+1$, they acquire a factor of $(-1)^F$, apart from picking up a phase of $e^{2\pi i \lambda}=\pm 1$.

\subsubsection{Consistent boundary states}

For the $\mathrm{AdS}_2$ branes the situation is in a sense opposite to what we had for the spherical branes: now we only have two Ishibashi states, and thus we expect to find a relatively `big' open string spectrum. In any case, given that there are only two Ishibashi states, there is essentially only one ansatz we can make for the boundary states, namely
%
\begin{align}
    \|\Theta,\varepsilon\rangle\!\rangle_{\text{AdS}}&= \frac{1}{\sqrt{2}}\sum_{\lambda\in\{0,\frac{1}{2}\}}e^{2\pi i (\lambda-\frac{1}{2})\Theta}|0,\lambda,\varepsilon\rangle\!\rangle_\text{AdS} \ ,\label{eq:BSAdS}
\end{align}
where $\Theta$ is an integer, which can be restricted to $\Theta\in \{0,1\}$. As in the case of the spherical branes, calculating the overlaps of these boundary states reveals the spectrum of boundary operators, and we find in this case
\begin{align}\label{AdS2self}
    \mathcal{H}_{\Theta_1|\Theta_2} = \bigoplus_{w\in 2\mathbb{Z}+\Theta_1-\Theta_2}\int_0^1 d\lambda\, \sigma^w(\mathscr{F}_\lambda)\ . 
\end{align}
Specifically, the string beginning and ending on the same brane consists of all $w$-even flowed representations, while the string stretching between branes differing in $\Theta$ consists of all the $w$-odd flowed representations, see also \cite{Petropoulos:2001qu,Lee:2001xe}. Since the two boundary states are related  via 
\begin{align}
    (-1)^{\Theta (F+1)}e^{\pi i \Theta (J_0^3-\bar{J}_0^3)}\|0,\varepsilon\rangle\!\rangle_\mathrm{AdS}=\|\Theta,\varepsilon\rangle\!\rangle_{\text{AdS}}\ ,
\end{align}
the two $\mathrm{AdS}_2$ branes are mapped into one another upon a rotation by $\pi$ in the angular direction. That is to say, they stretch between two fixed antipodal points on the  $\mathrm{AdS}_3$-boundary, but have opposite orientation. (The branes that start and end at a different pair of antipodal points can be obtained as in eq.~(\ref{AdS2rot});  they then satisfy the modified boundary conditions of eq.~(\ref{eq:IshibashiAdSTw}), see also \cite{Bachas:2000fr}.)

\section{Boundary states in the symmetric orbifold}
\label{sec:symorb}

In the previous section we have constructed the symmetry preserving boundary states of the ${\rm PSU}(1,1|2)$ WZW model. They can be combined with suitable boundary states from the torus factor to describe the D-branes of the full worldsheet theory, see the discussion in Section~\ref{sec:calculation} below. As we will see, at least for the spherical branes, these D-branes can be directly identified with suitable D-branes in the dual symmetric orbifold theory. In order to make this identification more precise we now need to construct the corresponding boundary states of the symmetric orbifold theory. This will be the subject of this section. 

We shall first briefly review the bulk spectrum of a symmetric orbifold theory. We shall then consider  the boundary states which satisfy the same gluing condition in each individual tensor factor of the symmetric orbifold. The corresponding Ishibashi states exist in each twisted sector, and they give rise to what one may call the maximally-fractional boundary states. In the main part we shall explain the construction for a generic bosonic symmetric orbifold; the modifictations that arise for the supersymmetric $\mathbb{T}^4$ theory are described in Appendix~\ref{app:T4symorb}.

\subsection{Bulk spectrum}

Let us start by reviewing briefly the (bulk) spectrum of a symmetric orbifold CFT; for simplicity we shall here focus on the case where the seed theory ${\rm X}$ is a (diagonal) quasirational bosonic  CFT with spectrum 
\be
{\cal H}^{\rm X} =  \bigoplus_{\alpha} {\cal H}_\alpha \otimes \bar{\cal H}_\alpha \ , 
\ee
where $\alpha$ runs over the different irreducible representations of the chiral algebra of ${\rm X}$.\footnote{If ${\rm X}$ is rational, the sum over $\alpha$ is finite; in the quasirational case, there are countably many representations that appear. For example, the torus theory ${\rm X} = \mathbb{T}^4$ is quasirational.} 

We are interested in the torus amplitude (partition function) $\mathcal{Z}^{S_N}(t,\bar{t}\,)$ of the symmetric orbifold theory ${\rm X}^{\otimes N}/S_N$. Let us denote the holomorphic and anti-holomorphic torus moduli by $t$ and $\bar{t}$, with $x=e^{2\pi i t}$ and $\bar{x}=e^{-2\pi i \bar{t}}$, and let $Z(t,\bar{t}\,)$ be the partition function of the seed theory ${\rm X}$. The simplest method to compute the symmetric orbifold partition function is by going to the grand canonical ensemble \cite{Dijkgraaf:1996xw} since the generating function (the grand canonical partition function) can be expressed in terms of the partition function $Z(t,\bar{t}\,)$ of the seed theory as
\begin{align}\label{eq:bulk-sym-partition}
\mathfrak{Z}(p,t,\bar{t}\,) = \sum_{N=1}^\infty p^N \mathcal{Z}^{S_N}(t,\bar{t}\,)= \exp \bigg(\sum_{k=1}^\infty p^k  T_k Z(t,\bar{t}\,)\bigg)\ ,
\end{align}
where $T_k$ is the Hecke operator
\begin{align}\label{eq:Hecke-operator}
T_k Z(t,\bar{t}\,) = \frac{1}{k}\sum_{w|k}\sum_{\kappa=0}^{w-1} Z\big(\tfrac{(k/w)t+\kappa}{w},\tfrac{(k/w)\bar{t}+\kappa}{w}\big)\ .
\end{align}
Here $w$ denotes the length of any individual cycle by which we are twisting, while the sum over $\kappa$ projects onto $\mathbb{Z}_w$-invariant states. (Recall that the twisted sectors of the symmetric orbifold theory are labelled by the conjugacy classes of $S_N$, which in turn are described by the cycle shapes, i.e.\ the partitions of $N$.) 

Denoting by $d(h,\bar{h})$ the non-negative integer multiplicities of the states in the seed CFT,
it is possible to rewrite the grand canonical partition function in a manifestly multiparticle form \cite{Dijkgraaf:1996xw}
\begin{align}
\mathfrak{Z}(p,t,\bar{t}\,) = \prod_{w=1}^\infty \prod_{h,\bar{h}}\Big[1-p^w x^{\frac{h}{w}+\frac{c}{24}(w-\frac{1}{w})-\frac{cw}{24}}\bar{x}^{\frac{\bar{h}}{w}+\frac{c}{24}(w-\frac{1}{w})-\frac{cw}{24}}\Big]^{-d(h,\bar{h})}\bigg|_{h-\bar{h}\in w\mathbb{Z}}\ ,\label{eq:multiparticle}
\end{align}
where $c$ denotes the central charge of the seed CFT. In particular, the ground-state conformal dimension $\Delta_w$ in $w$-twisted sector can be read off to equal 
\begin{align}
    \Delta_w = \frac{c}{24}\Big(w-\frac{1}{w}\Big)\ .
\end{align}
We can also see from \eqref{eq:multiparticle} that the single-particle states come only from conjugacy classes containing just one cycle, while generic permutations yield multiparticle states as per their cycle shape. Furthermore, the actual single-particle partition function equals
\begin{align}
\mathcal{Z}_\text{s.p.}(t,\bar{t}\,) = \sum_{w=1}^\infty Z(\tfrac{t}{w},\tfrac{\bar{t}}{w})\Big|_{h-\bar{h}\in w\mathbb{Z}}\ .
\end{align}
Finally, the bulk grand canonical partition function $\mathfrak{Z}(p,t,\bar{t}\,)$ admits a geometric interpretation in terms of (unramified) covering spaces of the base torus \cite{Bantay:1997ek}. Indeed, if we think of \eqref{eq:bulk-sym-partition} as a diagrammatic expansion, then the `connected part' would be
\begin{equation}
\log\mathfrak{Z}(p,t,\bar{t}\,)=\sum_{k=1}^{\infty}p^kT_kZ(t,\bar{t}\,)\ .
\end{equation}
Now, the Hecke operator can be thought of as summing the partition function $Z(t,\bar{t}\,)$ over each distinct torus which holomorphically covers the base torus $k$ times. The modulus of each covering torus is the argument appearing in \eqref{eq:Hecke-operator}, and the factor of $1/k$ corrects for the size of the automorphism group $\mathbb{Z}_{w}\times\mathbb{Z}_{k/w}$ of the corresponding covering, as is typical in diagrammatic expansions. Since the only (unramified) covering space of a torus is a disjoint union of tori, equation \eqref{eq:bulk-sym-partition} then tells us that the partition function of the symmetric orbifold ${\rm X}^{\otimes N}/S_N$ is calculated as a sum over the ${\rm X}$ partition function on all disjoint products of tori which cover the base torus $N$ times. Holographically, this sum over covering spaces arises from a localisation principle in which the integral over the worldsheet torus modulus $\tau$ localises to only those values for which the worldsheet holomorphically covers the boundary of thermal $\text{AdS}_3$, as was shown in \cite{Eberhardt:2018ouy,Eberhardt:2020bgq}, see also \cite{Maldacena:2000kv}.

\subsection{Boundary states of the seed theory}

In order to describe the boundary states of the symmetric orbifold theory, we first need to understand those of the seed theory itself. We shall always assume that these boundary states have already been constructed (and e.g.\ for the torus theory we have primarily in mind, this is the case). The main aim of our analysis here is to explain how these (known) boundary conditions of the seed theory can be  `lifted' to the symmetric orbifold theory. 

We shall only be interested in boundary states of the seed theory that preserve the full chiral algebra of ${\rm X}$. Let us denote the chiral fields of ${\rm X}$ collectively by $W(z)$, then the branes of interest will satisfy the boundary condition 
\be
W(z) = (\Omega \bar{W})(\bar{z}) \ , \qquad \hbox{at $z = \bar{z}$,}
\ee
where $\Omega$ is an automorphism of the chiral algebra of ${\rm X}$. In terms of boundary states we then consider the Ishibashi states 
\be
|\beta\rangle\!\rangle \in {\cal H}_\beta \otimes \bar{\cal H}_\beta \ , 
\ee
that are characterised by 
\be
\Bigl(W_n + (-1)^h (\Omega \bar{W})_{-n} \Bigr) \, | \beta\rangle\!\rangle = 0 \ , 
\ee
where $h$ is the conformal dimension of $W$, and $\beta$ runs over those representations for which $\Omega(\beta)=\beta$. (If $\Omega$ is inner, then this will be the case for all representations, but if $\Omega$ is outer, the $\beta$'s will only run over some subset of  representations.)  The corresponding boundary states will then be of the form 
\be\label{seedbound}
\| u \rangle\!\rangle = \sum_\beta B_\beta(u) \, | \beta\rangle\!\rangle \ , 
\ee
where $B_\beta(u)$ are some suitable coupling constants that depend on the structure of ${\rm X}$ (and the choice of $\Omega$ --- if $\Omega$ is trivial, then they can be expressed in terms of the usual $S$-matrix elements). 

In particular, we shall assume that they satisfy the Cardy condition \cite{Cardy:1989ir}, which means that the overlap takes the form 
\be\label{Cardyseed}
\langle\!\langle u \| e^{\pi i \hat{t}(L_0+\bar{L}_0-\frac{c}{12})}\, \| v \rangle\!\rangle  = 
\sum_\beta \bar{B}_\beta(u)  B_\beta(v) \, \chi_\beta(\hat{t}\,) =  \sum_\alpha n_{u|v}^{\alpha} \, \chi_\alpha(t) \ , 
\ee
where in the last step we have performed the S-modular transformation to the open string, writing $\hat{t}=-1/t$. In particular,  the relative open string spectrum consists of the representations $\alpha$ appearing with multiplicity $n_{u|v}^{\alpha} \in \mathbb{N}_0$. 

\subsection{Lifting to the symmetric orbifold}

We can lift these seed theory branes directly to the tensor product theory $\mathrm{X}^{\otimes N}$ by imposing the  `factorised' gluing conditions 
\be\label{factglue}
    W^{(i)}(z) = (\Omega \bar{W})^{(i)}(\bar{z}) \ , \qquad i=1,\ldots, N
\ee
at $z=\bar{z}$. Here the $W^{(i)}(z)$ are the chiral fields in the $i$'th copy of $\mathrm{X}^{\otimes N}$, and $\Omega$ is  the same for all factors, i.e.\ does not depend on $i$. The corresponding boundary states are then essentially just the tensor products of the above seed theory branes, 
\be
\|{\bf u} \rangle\!\rangle = \sum_{\underline{\beta}} B_{\underline{\beta}}({\bf u}) \, | {\underline{\beta}}\rangle\!\rangle \ , \qquad B_{\underline{\beta}}({\bf u}) = \prod_{i=1}^{N} B_{\beta_i}(u_i) \ , \label{tensorbranes}
\ee
where ${\bf u} = (u_1,\ldots,u_N)$ and ${\underline{\beta}}=(\beta_1,\ldots,\beta_N)$ are the obvious multi-indices, and the $B_{\beta_i}(u_i)$ are the coefficients appearing in eq.~(\ref{seedbound}). 

In the next step we now need to impose the $S_N$ orbifold projection. For general ${\bf u}$, the boundary states are not orbifold invariant, and we need to sum over the images $\sigma({\bf u})$, where $\sigma$ acts on the multi-index ${\bf u}$ by permuting the entries. The simplest branes, however, arise provided we choose an $S_N$ invariant ${\bf u}$, i.e.\ we take ${\bf u}=(u,\ldots,u)$. Then the above boundary state is by itself orbifold invariant, and it will give rise to a `maximally-fractional' D-brane once we have added in the corresponding twisted sector Ishibashi states, as we are about to describe.\footnote{Obviously there are also in-between possibilities, e.g.\ the boundary state is invariant under some subgroup of $S_N$, and these branes can be constructed similarly. In this paper we shall concentrate on the `maximally-fractional' case that is directly relevant for what we have in mind.}

In order to explain the structure of the twisted sector Ishibashi states we need to introduce some notation. Let us consider the conjugacy class labelled by $[\sigma]$, where $\sigma$ has the cycle shape corresponding to the partition 
\be
N = \sum_{j=1}^{r} l_j \ , \label{eq:partition}
\ee
i.e.\ it consists of $r$ cycles of length $l_j$, $j=1,\ldots, r$. (Here we include also cycles of length one.) In the $[\sigma]$ twisted sector, the modes are then of the form 
\be
W^{[j]}_p \ , \qquad p \in \tfrac{1}{l_j} \mathbb{Z} \ , \ \ (j=1,\ldots,r) \ . 
\ee
(This is to say, for each cycle we have one set of modes, but their mode numbers are now not in general integer, but rather run over $\tfrac{1}{l_j} \mathbb{Z}$.) The Ishibashi states that corresponds to the gluing condition (\ref{factglue}) are then characterised by 
\be
\Bigl( W^{[j]}_p - (-1)^h (\Omega \bar{W}^{[j]})_{-p} \Bigr) \, | {\underline{\beta}} \rangle \!\rangle_{[\sigma]} = 0 \ , \qquad p \in \tfrac{1}{l_j} \mathbb{Z} \ , \ \ (j=1,\ldots,r)  \ , 
\ee
and they are now labelled by $\underline{\beta}=(\beta_1,\ldots,\beta_r)$, i.e.\ there is one $\beta$ parameter for each cycle. The relevant overlap between two such Ishibashi states is 
\be
{}_{[\sigma]}\langle\!\langle {\underline{\beta}}|\, e^{\pi i \hat{t}(L_0+\bar{L}_0-\frac{Nc}{12})}\,| {\underline{\beta}}\rangle\!\rangle_{[\sigma]} =
\prod_{j=1}^{r} \, \chi_{\beta_j} (\tfrac{\hat{t}}{l_j}) \label{eq:BasicOverlap}
\ee
provided that the multi-indices (and the twisted sectors) agree, and zero otherwise. 
\smallskip

With these preparations in place we can now write down the maximally fractional symmetric orbifold D-branes: they are labelled by a pair $(u,\rho)$, where $u$ labels the D-branes of the seed theory as in (\ref{seedbound}), while $\rho$ is a representation of $S_N$, and they are explicitly given by\footnote{The parameter $\Lambda$ that appeared in the worldsheet description corresponds to multiplying the contribution from the $l_j$-cycle  by $e^{2\pi i \Lambda l_j}$. Since $N=\sum_j l_j$ the total effect of this is to multiply the whole boundary state by the phase $e^{2\pi i \Lambda N}$. Consistency then requires $\Lambda \in \frac{1}{N}\, \mathbb{Z}$, which is invisible from the worldsheet perspective.}
\be\label{maxfracbrane}
\| u , \rho \rangle\!\rangle = \sum_{(\underline{\beta},[\sigma])} B_{(\underline{\beta},[\sigma])}(u,\rho)\, 
|  {\underline{\beta}} \rangle \!\rangle_{[\sigma]} \ , \qquad 
B_{(\underline{\beta},[\sigma])}(u,\rho) = \Bigl( \frac{|[\sigma]|}{N!} \Bigr)^{\frac{1}{2}}\, \chi_\rho([\sigma]) \, \prod_{j=1}^{r}  B_{\beta_j}(u) \ ,
\ee
where $|[\sigma]|$ denotes the number of elements in the conjugacy class $[\sigma]$, while $\chi_\rho([\sigma])$ is the character of the conjugacy class $[\sigma]$ in the representation $\rho$. Indeed, the overlap between two such boundary states equals then 
\begin{align}
\langle\!\langle u,\rho_1\| \, e^{\pi i \hat{t}(L_0+\bar{L}_0-\frac{|\omega|c}{12})}\,\| v,\rho_2\rangle\!\rangle & = \frac{1}{N!}\sum_{\sigma\in S_N} \bar{\chi}_{\rho_1}(\sigma) \, \chi_{\rho_2}(\sigma) 
\prod_{j=1}^{r} \sum_{\beta_j} \bar{B}_{\beta_j}(u) \, B_{\beta_j}(v) \chi_{\beta_j} (\tfrac{\hat{t}}{l_j}) \nonumber \\
& = \frac{1}{N!}\sum_{\sigma\in S_N} \bar{\chi}_{\rho_1}(\sigma) \, \chi_{\rho_2}(\sigma)  \prod_{j=1}^{r} \sum_{\alpha} n_{u|v}^{\alpha} \, \chi_\alpha( l_j t)  \label{symorbopen}\\
& = \sum_{\alpha} n_{u|v}^{\alpha}  \Bigl[ \frac{1}{N!}\sum_{\sigma\in S_N} \bar{\chi}_{\rho_1}(\sigma) \, \chi_{\rho_2}(\sigma) \, {\rm Tr}_{\alpha^{\otimes N}} \bigl(\sigma e^{2\pi i t (L_0 - \frac{Nc}{24})} \bigr) \Bigr] \ , \nonumber 
\end{align}
where in going to the middle line we have used (\ref{Cardyseed}) for each $j$. Finally, we note that the square bracket in the last line simply projects onto those states in $\alpha^{\otimes N}$ that transform in the representation $\rho_1 \otimes  \rho_2^\ast$ of $S_N$; in particular, our D-branes therefore satisfy the Cardy condition \cite{Cardy:1989ir}.

\subsubsection{The grand canonical ensemble}\label{sec:opengrand}

In our application to holography we will be interested in the D-branes for which $\rho_1=\rho_2={\rm id}$ is the trivial representation. In this case the group projection in (\ref{symorbopen}) is to the $S_N$ invariant states. It is then convenient not to work with a fixed $S_N$ orbifold, but rather to introduce a fugacity $p$, and work in the grand canonical ensemble, for which the boundary partition function is defined to be
\begin{align}
{\mathfrak{Z}}_{u|v}(p,{t}) = \sum_{N=1}^\infty p^N {\mathcal{Z}}^{S_N}_{(u,\mathrm{id})|(v,\mathrm{id})}({t})\ .
\end{align}
In order to massage this into a simpler form, we consider eq.~\eqref{symorbopen} with $\rho_1=\rho_2=\text{id}$, which reads
\begin{equation}
Z^{S_N}_{(u,\text{id})|(v,\text{id})}(t)=\frac{1}{N!}\sum_{\sigma\in S_N}\prod_{j=1}^{r}Z_{u|v}(l_jt)\ .
\end{equation}
The sum over elements $\sigma\in S_N$ depends only on the conjugacy class of the permutation, characterised by the partition $\sum_{j}l_j=N$. It is more convenient to write such a partition as a sum $\sum_{j}k_jn_j$, where $k_j$ are \textit{distinct} integers, and $n_j$ are their multiplicities. In terms of these partitions, the size of the corresponding conjugacy class is given by
\begin{equation}
|C_{k_j,n_j}|=\frac{N!}{\prod_{j}k_j^{n_j}n_j!}\ ,
\end{equation}
and so the symmetric orbifold partition function takes the form
\begin{equation}
Z^{S_N}_{(u,\text{id})|(v,\text{id})}(t)=\sum_{\text{partitions of }N}\prod_{j}\frac{1}{k_j^{n_j}n_j!}Z_{u|v}(k_jt)^{n_j}\ .
\end{equation}
From this, the grand canonical partition function is immediately calculated, and we find
\begin{equation}
\begin{split}
\mathfrak{Z}_{u|v}(p,{t})&=\sum_{N=1}^{\infty}p^N\sum_{\text{partitions of }N}\prod_{j}\frac{1}{k_j^{n_j}n_j!}Z_{u|v}(k_jt)^{n_j}\\
&=\prod_{k=1}^{\infty}\left(\sum_{n=1}^{\infty}\frac{p^{nk}}{k^nn!}Z_{u|v}(kt)^n\right)=\exp\left(\sum_{k=1}^{\infty}\frac{p^k}{k}Z_{u|v}(kt)\right)\ .\label{eq:ZGCclosed}
\end{split}
\end{equation}
Moreover, if we expand the seed CFT partition function in terms of multiplicities, i.e.
\begin{equation}
Z_{u|v}(t)=\sum_{h}d_{u|v}(h)x^{h-\frac{c}{24}}\ ,
\end{equation}
then we can write $\mathfrak{Z}_{u|v}(p,t)$ in a manifestly multiparticle form analogous to \eqref{eq:multiparticle}. Indeed,
\begin{equation}
\begin{split}
\mathcal{Z}_{u|v}(t)&=\exp\left(\sum_{h}d_{u|v}(h)\sum_{k=1}^{\infty}\frac{p^k}{k}x^{k(h-\frac{c}{24})}\right)\\
&=\exp\left(-\sum_{h}d_{u|v}(h)\log\left(1-px^{h-\frac{c}{24}}\right)\right)\\
&=\prod_{h}\left(1-p\,x^{h-\frac{c}{24}}\right)^{-d_{u|v}(h)}\ .
\end{split}
\end{equation}

Just as in the bulk case, the grand canonical partition function computed between maximally fractional branes also admits a nice geometric interpretation in terms of covering spaces of the cylinder. Since a cylinder has vanishing Euler characteristic, its only (unramified) covering spaces are disjoint unions of cylinders whose moduli are integer multiplies of the modulus $t$. Considering only the connected component of \eqref{eq:ZGCclosed}, namely
\begin{equation}
\log\mathfrak{Z}_{u|v}(p,t)=\sum_{k=1}^{\infty}\frac{p^k}{k}Z_{u|v}(kt)\ ,
\end{equation}
we see that the coefficient of $p^k$ in this expansion is then the seed-CFT partition function evaluated on an cylinder with modulus $kt$, which is indeed the only connected covering space of the base cylinder. Furthermore, the factor of $1/k$ again accounts for the size of the automorphism group $\mathbb{Z}_k$ of the covering. In analogy with the bulk calculation, one would expect that the mechanism through which \eqref{eq:ZGCclosed} is reproduced in $\text{AdS}_3$ is via the worldsheet modulus integral localising to only those worldsheets which holomorphically cover the cylinder on the boundary of $\text{AdS}_3$. This is indeed the case, as we will see in Section \ref{sec:calculation}.

\subsection{Symmetrised permutation branes}\label{sec:perm}

There is another natural class of branes for the symmetric orbifold theory we can construct, and they may also play a role in this AdS/CFT duality. They do not directly arise from the branes of the seed theory, but are instead associated to permutation branes of the ${\rm X}^{\otimes N}$ theory \cite{Recknagel:2002qq}. Here we will only briefly describe the main points behind their construction and leave the details for future work.

The permutation branes of the covering tensor product theory are characterised by the gluing conditions
\be\label{permglu}
\bigl( W^{(i)}_n - (-1)^h (\Omega\bar{W})^{(\sigma(i))}_{-n} \bigr) \, \|\sigma; a \rangle\!\rangle = 0 \ , 
\ee
where $\sigma$ is a (fixed) element of $\sigma\in S_N$, and $a$ labels the different such branes. 
These boundary states are, however, not individually invariant under the full permutation group since the action of $\pi\in S_N$ maps
\be\label{eq:piaction}
\pi \Bigl( \|\sigma;a\rangle\!\rangle \Bigr) = \|\pi \sigma \pi^{-1};a \rangle\!\rangle \ . 
\ee
In order to obtain a consistent symmetric orbifold brane we therefore have to sum over all representatives of the conjugacy class associated to $\sigma$. On the other hand, each of these constituent branes is invariant under the action of the corresponding centraliser, and thus the resulting symmetric orbifold brane will also couple to the twisted sectors associated to the elements in the centraliser. While this general structure is familiar from the usual boundary state construction for orbifold theories, the details are likely to be somewhat different since in our case the gluing conditions \eqref{permglu} are not invariant under the action of the orbifold group $S_N$, as is apparent from \eqref{eq:piaction}.

The situation is simplest if $\sigma$ is a cyclic permutation of order $N$, since the centraliser then just consists of the cyclic group generated by $\sigma$. (More general choices of $\sigma$ can be dealt with using analogous techniques but the formulae become rather cumbersome.) Moreover, if we fix $N$ to be prime (as we shall do from now on), all elements of the centraliser, except for the identity, are again $N$-cycles, i.e.\ they lie in the same conjugacy class as $\sigma$ itself. In the symmetric orbifold theory, the corresponding D-brane will therefore only couple to the $N$-cycle twisted sector, which effectively `decouples' in the large $N$ limit.\footnote{For non-prime $N$, lower-order cycles appear, but they will always be of multi-cycle form. Thus the only single-cycle twisted sector that contributes is again the $N$-cycle twisted sector.}

Let us describe this construction in a bit more detail. Before going to the symmetric orbifold, i.e.\ in the tensor product theory ${\rm X}^{\otimes N}$, the $\sigma = (1\cdots N)$ permutation boundary states are labelled by a single parameter $u$, see eq.~(\ref{seedbound}), and are explicitly of the form \cite{Recknagel:2002qq}
\be
\| u \rangle\!\rangle^{\sigma} =  \sum_\beta \frac{B_\beta(u)}{(S_{\beta 0})^{(N-1)/2}} \, | \beta^{\otimes N}\rangle\!\rangle^{\sigma} \ , 
\ee
where the Ishibashi states on the right live in the sector labelled by $\beta^{\otimes N}$, and $S_{\beta 0}$ is the usual $S$-matrix element.\footnote{This is the correct ansatz for $\Omega={\rm id}$, and we are assuming here that $\Omega$ is such that it does not allow for any additional permutation Ishibashi states.} The corresponding open string partition function then equals 
\be
Z_{u|v} (t) =  \sum_{\alpha} n^{\alpha}_{u|v}  \, \sum_{\alpha_i}  N^{\alpha_1 \cdots \alpha_N}_{\alpha}  \, \chi_{\alpha_1}(t)\cdots  \chi_{\alpha_N}(t)\ , 
\ee
where $N^{\alpha_1 \cdots \alpha_N}_{\alpha}$ is the multiplicity with which $\alpha$ appears in the fusion of $\alpha_1 \otimes_{\rm f} \cdots \otimes_{\rm f} \alpha_N$. 

In going to the symmetric orbifold, we need to make the boundary state $S_N$ invariant, i.e.\ we need to 
sum over the different $N$-cycle permutation branes. We also need to add in a contribution from the $N$-cycle twisted sector. Thus the symmetric orbifold boundary states will be of the form
\be\label{permsym}
\| u,\ell\rangle\!\rangle^{(S_N)} =  \frac{1}{\sqrt{N!}} \sum_{\pi\in S_N}\hspace{-0.1cm}{}^{'}  \, \| u \rangle\!\rangle^{\pi\sigma \pi^{-1}} +\|u,\ell\rangle\!\rangle^{(S_N)}_\mathrm{tw} \ ,
\ee
where the sum over $\pi$ only runs over those $(N-1)!$ permutations for which the $\pi \sigma\pi^{-1} $ are pairwise different, and $\ell=0,\ldots,N-1$ labels the $\mathbb{Z}_N$ twisted charge. The contribution from the $N$-cycle twisted sector $\|u,\ell\rangle\!\rangle^{(S_N)}_\mathrm{tw}$ consists of a linear combination of states $| \beta \rangle\!\rangle_{[(1\cdots N)]}^{(r)}$ for $r=0,\ldots,N-1$, which are characterised by the gluing condition
\be
\bigl(W_p -   e^{2\pi i rp} (-1)^h (\Omega \bar{W})_{-p} \bigr) | \beta \rangle\!\rangle_{[(1\cdots N)]}^{(r)} = 0 \ \,,
\ee
where $p\in \frac{1}{N} \mathbb{Z}$. One finds that the open string spectrum between the permutation boundary states of eq.~(\ref{permsym}) and the maximally fractional branes of eq.~(\ref{maxfracbrane}) consists of $N$-cycle twisted states that are suitably projected by $\mathbb{Z}_N$. On the other hand, the open string spectrum between two permutation boundary states of the form of eq.~(\ref{permsym}) consists of even-permutation twisted states that are again suitably $\mathbb{Z}_N$ projected.\footnote{This is somewhat reminiscent of the structure of the ${\rm AdS}_2$ brane overlaps, see the discussion around eq.~(\ref{AdS2self}).} The details will be described elsewhere.

\section{The cylinder amplitude}
\label{sec:calculation}

In this section we return to the worldsheet perspective. We begin by constructing the full D-branes of the worldsheet theory. This is to say  we combine a brane from $\mathfrak{psu}(1,1|2)_1$ with a boundary state for the $\mathbb{T}^4$ sector, as well as one for the ghosts. We then determine their cylinder amplitudes. On the worldsheet we need to integrate over the modular parameter $\tau$, and as we shall see, this integral can actually be done explicitly. For the case of the spherical branes on ${\rm AdS}_3$, the resulting cylinder amplitude agrees then precisely with a cylinder amplitude in the dual symmetric orbifold CFT. This suggests that there is a direct correspondence between the spherical D-branes of ${\rm AdS}_3\times {\rm S}^3 \times \mathbb{T}^4$ and the maximally fractional D-branes in the symmetric orbifold of $\mathbb{T}^4$. 

We also do a similar computation in the open string channel (both on the worldsheet and in the dual CFT), and not surprisingly, but somewhat nontrivially, the result also agrees. This seems to suggest that the open-closed duality (Cardy condition) on the worldsheet is essentially equivalent to that in the dual CFT, at least for these spherical branes.

\subsection{Closed-string channel calculation}

Let us consider the worldsheet boundary states of the form 
\begin{align}
    \|u,\varepsilon\rangle\!\rangle_\mathrm{S} 
     = \|\varepsilon\rangle\!\rangle_\mathrm{S}\, \|u,\mathrm{R},\varepsilon\rangle\!\rangle_{\mathbb{T}^4}\,\|\text{ghost},\varepsilon\rangle\!\rangle\ ,\label{eq:BS}
\end{align}
where $\|\varepsilon\rangle\!\rangle_\mathrm{S}$ is the $\mathfrak{psu}(1,1|2)_1$ boundary state \eqref{eq:BLambda}, while $\|u,\mathrm{R},\varepsilon\rangle\!\rangle_{\mathbb{T}^4}$ and $\|\text{ghost},\varepsilon\rangle\!\rangle$ are boundary states for the $\mathbb{T}^4$ and the $\rho\sigma$ ghost system, respectively. Since the $\mathbb{T}^4$ is topologically twisted, the torus boundary states just consist of the RR part of a usual supersymmetric $\mathbb{T}^4$ boundary state. Note that, as usual, we have aligned the spin structure $\varepsilon$ across the three factors.

We are interested in the worldsheet cylinder amplitude, 
\begin{align}
\hat{Z}_{u|v}^\text{w}(\hat{t},\hat{\zeta};\hat{\tau})=\,_\mathrm{S}\langle\!\langle v,\mp\|\hat{q}^{\frac{1}{2}(L_0+\bar{L}_0-\frac{c}{12})}\hat{x}^{\frac{1}{2}(J_0^3+\bar{J}_0^3)}\hat{y}^{\frac{1}{2}(K_0^3-\bar{K}_0^3)}
\|u,\pm\rangle\!\rangle_\mathrm{S}\ ,\label{eq:P}
\end{align}
where we have inserted a $(-1)^F$ factor (i.e.\ considered opposite spin structures for the two boundary states), as is convenient for supergroup CFTs. Here $\hat{q}=e^{2\pi i \hat{\tau}}$ describes the propagation along the worldsheet time, while $\hat{x} = e^{2\pi i \hat{t}}$ with $\hat{t}$ real measures the separation between the two branes along the boundary of $\mathrm{AdS}_3$, i.e.\ it separates the two spherical branes by the distance $\hat{t}$ along the boundary of $\mathrm{AdS}_3$, see eq.~(\ref{eq:shifted}). 

Given that the boundary state in (\ref{eq:BS}) factorises, the same will be true for the cylinder diagram (\ref{eq:P}).  The explicit form of the $\mathfrak{psu}(1,1|2)_1$ overlap was already derived in Section~\ref{sec:psu}, except that we now also want to introduce the $\mathfrak{su}(2)$ chemical potential parametrised by $\hat{\zeta}$. Using eq.~(\ref{eq:char}), this modifies (\ref{3.14b}) to
\be\label{eq:P1}
\hat{Z}^{\rm S}(\hat{t},\hat{\zeta};\hat{\tau}) = \sum_{w\in\mathbb{Z}} \int_0^1 d\lambda 
(-1)^w \hat{q}^{\frac{w^2}{2}} \sum_{r\in\mathbb{Z} + \lambda} \hat{x}^{r} \hat{q}^{-rw} \, 
\frac{\theta_1(\frac{\hat{t}+\hat{\zeta}}{2};\hat{\tau})\theta_1(\frac{\hat{t}-\hat{\zeta}}{2};\hat{\tau})}{\eta(\hat{\tau})^4} \ , 
\ee
where relative to eq.~(\ref{3.14b}) we have  set $W_1=W_2=0$ and $\Lambda_1=\Lambda_2=\frac{1}{2}$. Next we use 
\be
  \int_0^1 d\lambda \sum_{r\in \mathbb{Z}+\lambda} e^{2\pi i r(\hat{t}-w\hat{\tau})} = \delta(\hat{t}-w\hat{\tau}) = \frac{1}{w} \delta( \tfrac{\hat{t}}{w} - \hat{\tau}) \ ,
\ee
as well as the theta function identity 
\begin{align}
    \theta_1(\tfrac{w\hat{\tau}\pm \hat{\zeta}}{2};\hat{\tau}) = \bigg\{\begin{array}{ll}
        \hat{q}^{- \frac{ w^2}{8}}e^{\frac{\pi i w}{2}}\theta_1(\pm \tfrac{\hat{\zeta}}{2};\hat{\tau}) & \text{if $w\in 2\mathbb{Z}$} \\
        \hat{q}^{- \frac{ w^2}{8}}e^{\frac{\pi i w}{2}}\theta_4(\pm \tfrac{\hat{\zeta}}{2};\hat{\tau}) & \text{if $w\in 2\mathbb{Z}+1$,}
    \end{array}
\end{align}
to rewrite (\ref{eq:P1}) as 
\begin{align}
\hat{Z}^{\text{S}}(\hat{t},\hat{\zeta};\hat{\tau})
&=\sum_{\substack{w=1\\ \text{$w$ even}}}^\infty \frac{1}{w}
 \hat{x}^\frac{w}{4}
\delta(\tfrac{\hat{t}}{w}-\hat{\tau})\frac{\theta_1(+\frac{\hat{\zeta}}{2};\hat{\tau})\theta_1(-\frac{\hat{\zeta}}{2};\hat{\tau})}{\eta(\hat{\tau})^4}+\nonumber\\
&\hspace{3cm}+\sum_{\substack{w=1\\ \text{$w$ odd}}}^\infty \frac{1}{w}
 \hat{x}^\frac{w}{4}
\delta(\tfrac{\hat{t}}{w}-\hat{\tau})\frac{\theta_4(+\frac{\hat{\zeta}}{2};\hat{\tau})\theta_4(-\frac{\hat{\zeta}}{2};\hat{\tau})}{\eta(\hat{\tau})^4}
\ .\label{eq:PWZW}
\end{align}
Here we have used that, because of the delta function $\delta(\hat{t}-w\hat{\tau})$ in the $\mathfrak{psu}(1,1|2)_1$ overlap, only positive values of $w$ contribute. Geometrically, the delta function means that 
the spacetime cylinder modulus $\hat{t}$ localises to those values $w\hat{\tau}$, for which a holomorphic covering map from the worldsheet cylinder exists, see the discussion in Section~\ref{sec:opengrand}; this is obviously the analogue of what was found in \cite{Eberhardt:2018ouy,Eberhardt:2019ywk,Dei:2020zui}.

As regards the ghost part overlap, we postulate that it leads to
\begin{align}
    \hat{Z}_{\text{ghost}}(\hat{\tau})= \frac{\eta(\hat{\tau})^4}{\theta_1(0;\hat{\tau})\theta_1(0;\hat{\tau})}\ ,\label{eq:Pghost}
\end{align}
since the ghosts do not carry any chemical potential.\footnote{Strictly speaking the denominator of eq.~(\ref{eq:Pghost}) is divergent because of the zero modes in the $\theta_1$ factor. This will cancel against a similar contribution from the numerator in eq.~(\ref{eq:PT4}) below.} This is somewhat different than what was proposed in \cite{Eberhardt:2018ouy}, but it ties in naturally with \cite{Gaberdiel:2021njm}; in particular, the $\mathfrak{su}(2)$ quantum numbers are already correctly accounted for in terms of the $\mathfrak{su}(2)$ subalgebra of $\mathfrak{psu}(1,1|2)$ (and there is no need to involve any chemical potential from the $\mathbb{T}^4$ or the ghost sector). Finally, the $\mathbb{T}^4$ part gives 
\begin{align}
   \hat{Z}_{u|v,\tilde{\mathrm{R}}}^{\mathbb{T}^4}(\hat{\tau}) = \frac{\theta_1(0;\hat{\tau})
    \theta_1(0;\hat{\tau})
    }{\eta(\hat{\tau})^6}\hat{\Theta}_{u|v}^{\mathbb{T}^4}(\hat{\tau})\ ,\label{eq:PT4}
\end{align}
where the $\hat{\Theta}$ function describes the winding and momentum modes. Now the numerator of eq.~(\ref{eq:PT4}) cancels precisely with the denominator of eq.~(\ref{eq:Pghost}), while the numerator of 
eq.~(\ref{eq:Pghost}) cancels, say the bosonic oscillators in eq.~(\ref{eq:PWZW}). Finally, putting all of these pieces together and performing the $\hat{\tau}$ integral we find 
\begin{align}
\int_0^{\infty} {d\hat{\tau}}\, \hat{Z}^\text{w}_{u|v}(\hat{t},\hat{\zeta};\hat{\tau})
&=\sum_{\substack{w=1\\ \text{$w$ even}}}^\infty \frac{1}{w}\hat{x}^\frac{w}{4}\hat{Z}^{\mathbb{T}^4}_{u|v,\tilde{\mathrm{R}}}(\hat{\zeta},\tfrac{\hat{t}}{w})+\sum_{\substack{w=1\\ \text{$w$ odd}}}^\infty \frac{1}{w}\hat{x}^\frac{w}{4}\hat{Z}^{\mathbb{T}^4}_{u|v,\tilde{\mathrm{NS}}}(\hat{\zeta},\tfrac{\hat{t}}{w})\ , 
\end{align}
where 
\begin{align}
\hat{Z}^{\mathbb{T}^4}_{u|v,\tilde{\mathrm{R}}}(\hat{\zeta},\tfrac{\hat{t}}{w}) & = \frac{\theta_1(+\frac{\hat{\zeta}}{2};\tfrac{\hat{t}}{w})\theta_1(-\frac{\hat{\zeta}}{2};\tfrac{\hat{t}}{w})}{\eta(\tfrac{\hat{t}}{w})^6} \, \hat{\Theta}_{u|v}^{\mathbb{T}^4}(\tfrac{\hat{t}}{w}) \ , \\
\hat{Z}^{\mathbb{T}^4}_{u|v,\tilde{\mathrm{NS}}}(\hat{\zeta},\tfrac{\hat{t}}{w}) & = 
\frac{\theta_4(+\frac{\hat{\zeta}}{2};\tfrac{\hat{t}}{w})\theta_4(-\frac{\hat{\zeta}}{2};\tfrac{\hat{t}}{w})}{\eta(\tfrac{\hat{t}}{w})^6} \, \hat{\Theta}_{u|v}^{\mathbb{T}^4}(\tfrac{\hat{t}}{w}) \ . 
\end{align}

In order to compare this to the dual CFT answer we now need to recall that the worldsheet theory only sees the single particle sector of the symmetric orbifold theory. Thus in order to match with the dual CFT we need to include string field theoretic multi-worldsheet states. This is most conveniently done as in 
\cite{Eberhardt:2020bgq}, i.e.\ by introducing a chemical potential $\sigma$ (with the associated fugacity $p=e^{2\pi i \sigma}$) for the number of F-strings wound around the bulk of the $\mathrm{AdS}_3$, and working in a grand canonical ensemble,\footnote{In analogy with the sphere contribution to the closed string partition function of \cite{Eberhardt:2018ouy}, it is natural to assume that the contribution to the path integral due to two disconnected discs accounts for the ground-state shift $\hat{x}^{-\frac{w}{4}}$ in the individual $w$-twisted sectors.}
\begin{align}
    \hat{\mathfrak{Z}}_{u|v}(\hat{\zeta};p,\hat{t}\,)=\exp\bigg(\sum_{\substack{w=1\\ \text{$w$ even}}}^\infty \frac{p^w}{w}\hat{Z}^{\mathbb{T}^4}_{u|v,\tilde{\mathrm{R}}}(\hat{\zeta},\tfrac{\hat{t}}{w})+\sum_{\substack{w=1\\ \text{$w$ odd}}}^\infty \frac{p^w}{w}\hat{Z}^{\mathbb{T}^4}_{u|v,\tilde{\mathrm{NS}}}(\hat{\zeta},\tfrac{\hat{t}}{w})\bigg)\ .\label{eq:WorldsheetClosed}
\end{align}
This then matches precisely with the symmetric orbifold answer, see eq.~\eqref{eq:ZT4hat}. 

Two comments are in order at this point. First, note that the difference $\Lambda_1-\Lambda_2$ of the worldsheet $\mathfrak{psu}(1,1|2)_1$ Wilson lines would introduce the phase factor $e^{2\pi i w(\Lambda_1-\Lambda_2)}$ into the sum over $w$ in \eqref{eq:WorldsheetClosed}. This is therefore equivalent to multiplying the $w$-twisted part of the $\mathrm{Sym}(\mathbb{T}^4)$ overlap by the same phase, which in turn can be reproduced by multiplying the $\mathrm{Sym}_N(\mathbb{T}^4)$ boundary states at fixed $N$ by the global phase $e^{2\pi i \Lambda_i}$. This is therefore invisible at fixed $N$ --- multiplying all the boundary states by a fixed phase is invisible in all cylinder diagrams.

Second, we note that the whole analysis would also go through (with minor modifications) had we not twisted the worldsheet overlaps with $(-1)^F$. In that case, we would instead end up with an $\mathrm{NS}$ cylinder correlator in $\mathrm{Sym}(\mathbb{T}^4)$.

\subsection{Open-string channel}
\label{sec:open}

We can also redo the calculation of the cylinder string amplitude in the open-string channel, where, by virtue of the open-closed duality, we expect to obtain the same result as in the closed-string channel. The key ingredient here is the $(-1)^F$-twisted worldsheet boundary partition function  
\begin{align}
    {Z}^\mathrm{w}_{u|v}(t,\zeta;\tau)=\mathrm{Tr}_{\mathcal{H}_{u|v}} \Bigl( (-1)^F \, q^{L_0-\frac{c}{24}}\, e^{-2\pi i \frac{\tau}{t}J_0^3} \, e^{2\pi i\frac{\tau}{t}  \zeta K_0^3} \Bigr)\ ,\label{eq:Z}
\end{align}
which can be obtained by modular S-transforming the boundary states overlap \eqref{eq:P} with the relations
\be 
\hat{\tau} = -\tfrac{1}{\tau}\ , \qquad \hat{t} = -\tfrac{1}{t}\ , \qquad \hat{\zeta} = -\hat{t}\, \zeta 
\ee
between the open- and closed-string channel parameters.
The trace \eqref{eq:Z} again factorises into the $\mathfrak{psu}(1,1|2)_1$ WZW part, the ghost part and the $\mathbb{T}^4$ part. First, starting from \eqref{eq:SphericalOpen} with $\Lambda_1=\Lambda_2=\frac{1}{2}$, $W_1=W_2=0$ and substituting the explicit form \eqref{eq:char} for the $\mathscr{F}_{1/2}$ character, we obtain the $\mathfrak{psu}(1,1|2)_1$ part of the boundary partition function
\begin{align}
	{Z}^{\text{S}}(t,\zeta;\tau)
	&=  \tau\,\sum_{\substack{k=1\\
	\text{$k$ even}}}^\infty \frac{1}{k} \hat{x}^\frac{k}{4}e^{\frac{\pi i k\zeta^2}{2t}}  \delta({\tau}-kt)\, \frac{\theta_1(+\frac{k\zeta}{2};\tau)\theta_1(-\frac{k\zeta}{2};\tau)}{\eta(\tau)^4}\nonumber\\
	&\hspace{2cm}-\tau\,\sum_{\substack{k=1\\
	\text{$k$ odd}}}^\infty \frac{1}{k} \hat{x}^\frac{k}{4}e^{\frac{\pi i k\zeta^2}{2t}}  \delta({\tau}-kt) \, \frac{\theta_2(+\frac{k\zeta}{2};\tau)\theta_2(-\frac{k\zeta}{2};\tau)}{\eta(\tau)^4}\ , \label{eq:ZWZW}
\end{align}
where the sum over $k$ corresponds to the sum over $w$ in \eqref{eq:PWZW}. In the process, we have also used the $\theta$-function identity
\begin{align}
    \theta_1(-\tfrac{k}{2};\tau) =\bigg\{\begin{array}{ll}
       (-1)^{\frac{k}{2}}\theta_1(0;\tau)  &  \text{if $k\in 2\mathbb{Z}$}\\
        (-1)^{\frac{k-1}{2}}\theta_2(0;\tau) &  \text{if $k\in 2\mathbb{Z}+1$\,.}
    \end{array}
\end{align}
Note that as for the 
closed-string overlap \eqref{eq:PWZW}, the boundary partition function \eqref{eq:ZWZW} is $\delta$-function localised at $\tau=kt$ for $k\in \mathbb{Z}$; this corresponds precisely to those values of the spacetime cylinder modulus $t$ where a holomorphic covering map from the worldsheet cylinder exists. Furthermore, since $t,\tau\geqslant 0$, we can restrict again to $k\geqslant 1$. The modular transformation of the ghost and $\mathbb{T}^4$ factors, see eq.~(\ref{eq:Pghost}) and (\ref{eq:PT4}) respectively, are standard, 
\be
    {Z}^{\text{ghost}}({\tau}) = -\frac{\eta( {\tau})^4}{\theta_1(0; {\tau})\theta_1(0; {\tau})}\ , \qquad 
    {Z}_{u|v,\tilde{\mathrm{R}}}^{\mathbb{T}^4}(\tau)  = -\frac{\theta_1(0; {\tau})\theta_1(+0; {\tau})}{\eta( {\tau})^6} {\Theta}_{u|v}^{\mathbb{T}^4}( {\tau})\ ,\label{eq:ZT}
\ee
where the overall minus signs are included to compensate for the properties of the $\theta_1$ functions under the modular S-transformation.

Altogether, multiplying \eqref{eq:ZWZW}  and \eqref{eq:ZT} and integrating over the period modulus of the worldsheet cylinder, we obtain the amplitude\footnote{Here we have also included the usual factor of $\tau$ in the worldsheet measure that comes from the modular transformation of the ghosts.}
\begin{align}
    	\int_0^{\infty} \frac{d\tau}{\tau}\,{Z}^\mathrm{w}_{u|v}( t,\zeta;\tau)=\sum_{\substack{k=1\\ \text{$k$ even}}}^\infty \frac{1}{k}\hat{x}^\frac{k}{4} e^{\frac{\pi i k\zeta^2}{2t}} 
	 Z_{u|v,\tilde{\mathrm{R}}}^{\mathbb{T}^4}(k\zeta; kt)+
	 \sum_{\substack{k=1\\ \text{$k$ odd}}}^\infty \frac{1}{k}\hat{x}^\frac{k}{4} e^{\frac{\pi i k\zeta^2}{2t}} 
	 Z_{u|v,{\mathrm{R}}}^{\mathbb{T}^4}(k\zeta; kt)
	 \,,
\end{align}
where the $\mathbb{T}^4$ boundary partition function $Z_{i|j,{\mathrm{R}}}^{\mathbb{T}^4}$ can be obtained from \eqref{eq:ZT} by replacing $\theta_1\to i\theta_2$.

As before, the worldsheet analysis only captures the single string spectrum, and in order to relate this to the full dual CFT we need to go to the grand canonical ensemble by introducing a chemical potential $p$ for the number of times $k$ the worldsheet cylinder covers the boundary spacetime cylinder. Eliminating the $\hat{x}^{k/4}$ factor as before by taking into account the contribution to the string amplitude due to two disconnected discs, we obtain the 1-loop open-string partition function
\begin{align}
  {\mathfrak{Z}}_{u|v}(\zeta;p,t)=\exp\bigg(  \sum_{\substack{k=1\\ \text{$k$ even}}}^\infty \frac{p^k}{k} e^{\frac{\pi i k\zeta^2}{2t}} 
	 Z_{u|v,\tilde{\mathrm{R}}}^{\mathbb{T}^4}(k\zeta; kt)+
	 \sum_{\substack{k=1\\ \text{$k$ odd}}}^\infty \frac{p^k}{k} e^{\frac{\pi i k\zeta^2}{2t}} 
	 Z_{u|v,{\mathrm{R}}}^{\mathbb{T}^4}(k\zeta; kt)\bigg)\,.
\end{align}
This is exactly equal to the partition function \eqref{eq:ZT4} for the Ramond boundary spectrum of the $\mathrm{Sym}(\mathbb{T}^4)$ maximally-fractional boundary states.

\section{Disk correlation functions}\label{sec:correlators}

Up to now we have analysed the partition functions or cylinder amplitudes of our symmetric orbifold boundary conditions from the worldsheet perspective. It is also instructive to consider the correlation functions of (bulk) fields in the presence of these boundary conditions, and it is the aim of this section to study the corresponding correlators. 

In the absence of a boundary, i.e.\ for the case where the symmetric orbifold correlators are evaluated on the full boundary of $\text{AdS}_3$ (which we can take to be the Riemann sphere $\mathbb{CP}^1$),\footnote{Strictly speaking, the Riemann sphere is the compactification of the boundary of $\text{AdS}_3$.} the relevant correlation functions are determined by viewing the worldsheet $\Sigma$ as a covering space of the boundary of $\text{AdS}_3$. The correlation functions are then expressible in terms of the data of the holomorphic map $\Gamma:\Sigma\to\mathbb{CP}^1$ which implements this covering \cite{Eberhardt:2019ywk}. These covering maps are in turn directly related to the computation of correlation functions in the symmetric orbifold CFT \cite{Lunin:2001ew} (see also \cite{Dei:2019iym}), and this perspective makes the equivalence of the $\text{AdS}_3$ string theory correlators and those of the symmetric product orbifold essentially manifest \cite{Eberhardt:2019ywk}. 

In the presence of a boundary, i.e.\ for symmetric orbifold correlators on the disk $\mathbb{D}$, it is then natural to expect that the stringy correlation functions are controlled by a map $\Gamma:\Sigma\to\mathbb{D}$ which covers the unit disk $\mathbb{D}$ holomorphically by the worldsheet $\Sigma$.\footnote{Incidentally, this then requires that the worldsheet also has a boundary, which  ties in naturally with what we have argued above, namely that boundary conditions on the worldsheet correspond to boundary conditions in the symmetric orbifold.} In the following we shall show that this is expectation is indeed borne out. We shall concentrate on the genus zero case where the worldsheet has no handles, and thus the appropriate covering maps will be of the form $\Gamma:\mathbb{D}\to\mathbb{D}$; we review the properties of these maps in Appendix \ref{sec:covering-maps}.  However, as for the analysis of  higher genus closed worldsheets \cite{Eberhardt:2020akk,Knighton:2020kuh}, we expect this to remain true also at higher genus. As was shown in \cite{Dei:2020zui}, the relationship between stringy closed string correlators and these covering maps is most readily seen using the free field realisation of the $\mathfrak{psu}(1,1|2)_1$ model, which we introduced in Section \ref{sec:free-fields}. This will also remain true in the situation with boundaries.

\subsection{Review of the bulk argument}

Before we analyse the correlators in the presence of boundaries, let us briefly review, following \cite{Dei:2020zui,Knighton:2020kuh},  the argument for the closed string case without boundaries.
We work in the free field realisation of $\mathfrak{psu}(1,1|2)_1$ which we introduced in Section~\ref{sec:free-fields}. In terms of these fields, the highest-weight states $\ket{m,j}$ lie in the representation defined by \eqref{eq:symplectic-representation}. The spectrally flowed states $\ket{m,j}^w$ are then defined by the relations
\begin{equation}
A\ket{m,j}^{w}=\left(\sigma^w(A)\ket{m,j}\right)^w\ .
\end{equation}
Promoting $\ket{m,j}^{w}$ to a vertex operator $V_{m,j}^{w}(z)$ then yields the OPEs
\begin{equation}
\begin{split}
\xi^{+}(z)V^{w}_{m,j}(0)&\sim z^{-\frac{w+1}{2}}V_{m+\frac{1}{2},j-\frac{1}{2}}^{w}(0)+\cdots\ ,\\
\xi^-(z)V_{m,j}^{w}(0)&\sim -z^{\frac{w-1}{2}}V_{m-\frac{1}{2},j-\frac{1}{2}}^{w}(0)+\cdots\ ,
\end{split}
\end{equation}
and similarly for $\eta^{\pm}$. That is, $\xi^+$ has a highly singular OPE with $V_{m,j}^{w}$, while $\xi^-$ has a highly regular one.

In addition to the quantum numbers $m$, $j$, and $w$, one can also introduce a dependence with respect the coordinate $x$ on the $\mathrm{AdS}_3$ boundary by using $J^+_0$ as the dual CFT translation operator. Explicitly, we define the vertex operator $V_{m,j}^{w}(x,z)$ via
\begin{equation}
V_{m,j}^{w}(x,z)=e^{xJ^+_0}V_{m,j}^{w}(z)e^{-xJ^+_0}\ .
\end{equation}
The correlation functions of these vertex operators then have a natural interpretation in terms of the dual spacetime CFT. Because of the translation via $J^+_0$, the vertex operators $V_{m,j}^{w}(x,z)$ have slightly modified OPEs with respect to the free fields. In particular, if $x\neq 0$, both $\xi^{\pm}$ (respectively $\eta^{\pm}$) have singular OPEs with $V_{m,j}^{w}(x,z)$
\begin{equation}\label{eq:ope-poles}
\begin{split}
\xi^{+}(z)V^{w}_{m,j}(x,0)&\sim z^{-\frac{w+1}{2}}V_{m+\frac{1}{2},j-\frac{1}{2}}^{w}(x,0)+\cdots\ ,\\
\xi^{-}(z)V^{w}_{m,j}(x,0)&\sim -x\,z^{-\frac{w+1}{2}}V_{m+\frac{1}{2},j-\frac{1}{2}}^{w}(x,0)+\cdots\ ,
\end{split}
\end{equation}
while the linear combination $\xi^-+x\,\xi^+$ (respectively $\eta^-+x\,\eta^+$) has the regular OPE
\begin{equation}\label{eq:ope-regular}
\Bigl(\xi^-(z)+x\,\xi^+(z)\Bigr)\, V^{w}_{m,j}(x,0)\sim -z^{\frac{w-1}{2}}V_{m-\frac{1}{2},j-\frac{1}{2}}^{w}(x,0)+\cdots\ .
\end{equation}
Now, naively one would like to calculate correlation functions of the vertex operators $V_{m,j}^{w}(x,z)$, i.e.\ correlators of the form
\begin{equation}
\Bigl\langle\prod_{i=1}^{n}V_{m_i,j_i}^{w_i}(x_i,z_i)\Bigr\rangle\ .
\end{equation}
However, due to subtleties in defining correlation functions in the hybrid formalism, this is not quite the correct amplitude to consider. In particular, one needs to account for modifications to the correlation function which are analogous to picture-changing in the RNS string (see Section~3 of \cite{Dei:2020zui}). While this process is somewhat subtle, for correlation functions of spectrally-flowed primaries, it essentially amounts to considering the correlation function
\begin{equation}\label{eq:bulk-correlator}
\Bigl\langle\prod_{a=1}^{n-\chi(\Sigma)}W(u_a)\prod_{i=1}^{n}V_{m_i,j_i}^{w_i}(x_i,z_i)\Bigr\rangle\ ,
\end{equation}
where the new fields $W(u)$ obey the OPEs
\begin{equation}\label{eq:ope-xi-w}
\xi^{\pm}(z)W(0)\sim\mathcal{O}(z)\ ,\qquad\eta^{\pm}(z)W(0)\sim\mathcal{O}(1/z)\ .
\end{equation}
Furthermore, the number $n-\chi(\Sigma)=n+2g-2$ of these fields corresponds to the number of picture changing operators which are typically inserted in a genus $g$ superstring amplitude in the RNS formalism. Here we will only consider the planar case $g=0$ for simplicity.

The correlation function \eqref{eq:bulk-correlator} enjoys a remarkable property: it localises on the worldsheet moduli space to those configurations $z_i$ for which a holomorphic map $\Gamma:\mathbb{CP}^1\to\mathbb{CP}^1$ exists. Here $\Gamma$ is characterised by the property that it has exactly $n$ critical points $z=z_i$, near which it has the local behaviour
\begin{equation}\label{eq:bulk-covering-condition}
\Gamma(z)\sim x_i+\mathcal{O}\bigl((z-z_i)^{w_i}\bigr)\ .
\end{equation}
The proof of this property is readily shown utilising the free fields. Indeed, we can define two meromorphic functions on the worldsheet by
\begin{equation}
\omega^{\pm}(z)=\Bigl\langle\xi^{\pm}(z)\prod_{a=1}^{n-2}W(u_a)\prod_{i=1}^{n}V_{m_i,j_i}^{w_i}(x_i,z_i)\Bigr\rangle\ .
\end{equation}
Because of the OPEs eqs.~\eqref{eq:ope-poles} and \eqref{eq:ope-xi-w}, we see that the $\omega^{\pm}$ have poles of order $\frac{w_i+1}{2}$ at $z=z_i$, and simple zeroes at $z=u_a$. Thus, we can define new functions
\begin{equation}
P^{\pm}(z)=\frac{\prod_{i=1}^{n}(z-z_i)^{\frac{w_i+1}{2}}}{\prod_{a=1}^{n-2}(z-u_a)}\omega^{\pm}(z)\ ,
\end{equation}
which then have no poles (away from $z=\infty$), and no longer have zeroes at $z=u_a$. Thus, they must be polynomials, and their degree is readily calculated to be
\begin{equation}
N=1+\sum_{i=1}^{n}\frac{w_i-1}{2}\ .
\end{equation}
We then notice that the ratio $\Gamma(z)=-P^-(z)/P^+(z)$ (when it is defined) satisfies eq.~\eqref{eq:bulk-covering-condition}. Furthermore, the degree $N$ of the polynomials $P^{\pm}$ is exactly the degree of the covering map calculated from the Riemann-Hurwitz formula. One is therefore led to conclude that this ratio is precisely the covering map of equation \eqref{eq:bulk-covering-condition}. 

An equivalent way of stating the relationship between the covering map and the correlators is implicitly through the so-called `incidence relation'
\begin{equation}\label{eq:bulk-incidence-relation}
\omega^-(z)+\Gamma(z)\,\omega^+(z)=0\ .
\end{equation}
Note that \eqref{eq:bulk-incidence-relation} can only be satisfied when such a covering map exists. However, this only happens at discrete points in the worldsheet moduli space. If such a map does not exist, then some part of the above argument must fail. Indeed, if such a $\Gamma$ does not exist, we must conclude that $\omega^+(z)=0$. However, by eq.~\eqref{eq:ope-poles}, we see that the near $z=z_i$, the leading behaviour of $\omega^+$ is given by
\begin{equation}
\omega^+(z)\sim(z-z_i)^{\frac{w_i+1}{2}}\Bigl\langle\prod_{a=1}^{n-2}W(u_a)V_{m_i+\frac{1}{2},j_i-\frac{1}{2}}^{w_i}(x_i,z_i)\prod_{k\neq i}V_{m_k,j_k}^{w_k}(x_k,z_k)\Bigr\rangle\ .
\end{equation}
Thus, if $\omega^+$ identically vanishes, the correlation functions themselves must vanish as well. Therefore, the correlation functions \eqref{eq:bulk-correlator} are only nonzero if a covering map exists.
\smallskip

From the symmetric orbifold point of view, the covering maps $\Gamma$ are also used in the calculation of correlation functions. Indeed, a general formula for the correlation function of twisted sector ground states can be found, and it takes the schematic form \cite{Dei:2019iym}
\begin{equation}
\Braket{\mathcal{O}^{w_1}(x_1,\bar{x}_1)\cdots\mathcal{O}^{w_n}(x_n,\bar{x}_n)}\sim\sum_{\Gamma}\prod_{i=1}^{n}|a_i^{\Gamma}|^{-h_i}\ ,
\end{equation}
where $a_i^{\Gamma}$ is the coefficient of $(z-z_i)^{w_i}$ in eq.~\eqref{eq:bulk-covering-condition}. In fact, using the incidence relation \eqref{eq:bulk-incidence-relation} and the OPEs of eqs.~\eqref{eq:ope-poles} and \eqref{eq:ope-regular}, one can show that the string theory correlators take precisely this form, up to some unknown normalisation constants (see Section 4.4 of \cite{Dei:2020zui}). Thus, the above argument shows that the worldsheet correlators reproduce the structure of the symmetric orbifold correlators. As we will see below, these arguments continue to hold when we consider correlators in the presence of a boundary.

\subsection{The worldsheet doubling trick}

We now turn our attention to the correlation function
\begin{equation}\label{eq:basic-correlator}
\Bigl\langle \, \prod_{a=1}^{n-1}W(u_a)\prod_{i=1}^{n}V_{m_i,j_i}^{w_i}(x_i,z_i)\Bigr\rangle_B \equiv 
\Bigl\langle  \, \prod_{a=1}^{n-1}W(u_a)\prod_{i=1}^{n}V_{m_i,j_i}^{w_i}(x_i,z_i)  \, \Bigl|\!\Bigl| B  \Bigr\rangle\!\!\Bigr\rangle \ ,
\end{equation}
where the index $B$ on the left-hand-side indicates that this correlator is evaluated in the presence of the boundary state $|\!| B\rangle\!\rangle$. Since we take the boundary state $\bket{B}$ to be located on the real line, the above correlator is taken on the upper half-plane $\mathbb{H}$, which is conformally equivalent to the unit disk $\mathbb{D}$. Note that, in contrast to the case of correlation functions on the sphere, we have $n-1$ $W$-field insertions since the Euler characteristic of the disk is $\chi(\mathbb{D})=1$. To start with we shall consider the case where the boundary state $B$ corresponds to a spherical D-brane; the situation for the ${\rm AdS}_2$ branes is discussed in Section~\ref{sec:AdS2branesdisk}.

In order to compute the correlation function \eqref{eq:basic-correlator}, we now use the same trick as for the correlators on the sphere. That is, we define the correlator-valued complex functions
\begin{equation}
\omega^{\pm}(z)=\Bigl\langle \,\xi^{\pm}(z)\prod_{a=1}^{n-1}W(u_a)\prod_{i=1}^{n}V_{m_i,j_i}^{w_i}(x_i,z_i)\Bigr\rangle_{(u,\varepsilon)_{\rm S}} \ ,
\end{equation}
where $(u,\varepsilon)_{\rm S}$ indicates that the boundary condition is a spherical brane, see eq.~(\ref{eq:BS}). We now want to explore their complex-analytic properties, and for this we can use a doubling trick to consider a function on the full Riemann sphere instead of on the disk.

To do this, we note that the boundary conditions \eqref{eq:free-field-spherical} obeyed by the boundary state $ \|u,\varepsilon\rangle\!\rangle_\mathrm{S}$ can be expressed in terms of complex variables as
\begin{equation}
\begin{split}
\left(\xi^{\pm}(z)\pm e^{i\varphi}\bar{\xi}^{\pm}(\bar{z})\right) \|u,\varepsilon\rangle\!\rangle_\mathrm{S}&=0\ ,\\
\left(\eta^{\pm}(z)\pm e^{-i\varphi}\bar{\eta}^{\mp}(\bar{z})\right) \|u,\varepsilon\rangle\!\rangle_\mathrm{S} &=0\ ,
\end{split}
\end{equation}
when $z$ is on the real line. These boundary conditions are automatically taken into account if we define an analytically-continued field
\begin{equation}
\Xi^{\pm}(z)=\left\{
\begin{array}{ll}
\xi^{\pm}(z)\ ,\qquad  & \Im(z)\geq 0\\
\mp e^{i\varphi}\,\overline{\xi}^{\mp}(\bar{z})\ , \qquad \qquad & \Im(z)<0
\end{array} \right.
\end{equation}
(and similarly for $\eta^{\pm}$), which is defined on the entire Riemann sphere and not just the upper half-plane, see Figure \ref{fig:doubling-trick}.

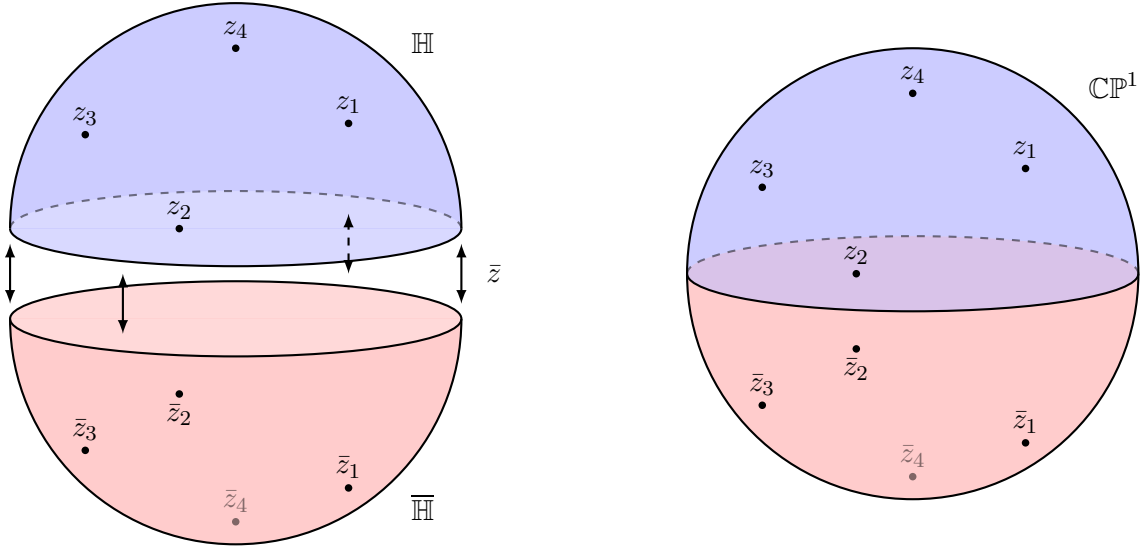
\begin{figure}
\centering
\begin{tikzpicture}
\begin{scope}
\begin{scope}[yshift = 0.6cm]
\fill[blue, opacity = 0.2] (0,0) [partial ellipse = 0:180:3 and 3];
\fill[white] (0,0) [partial ellipse = 0:180:3 and 0.5];
\fill[blue, opacity = 0.15] (0,0) [partial ellipse = 0:360:3 and 0.5];
\draw[thick] (0,0) [partial ellipse = 180:360:3 and 0.5];
\draw[thick, dashed, opacity = 0.5] (0,0) [partial ellipse = 0:180:3 and 0.5];
\draw[thick] (0,0) [partial ellipse = 0:180:3 and 3];
\fill (1.5,1.4) circle (0.05);
\node[above] at (1.5,1.4) {$z_1$};
\fill (-0.75,0) circle (0.05);
\node[above] at (-0.75,0) {$z_2$};
\fill (-2,1.25) circle (0.05);
\node[above] at (-2,1.25) {$z_3$};
\fill (0,2.4) circle (0.05);
\node[above] at (0,2.4) {$z_4$};
\node[above right] at (2.2,2.2) {$\mathbb{H}$};
\end{scope}
\begin{scope}[yshift = -0.6cm]
\fill[red, opacity = 0.2] (0,0) [partial ellipse = 180:360:3 and 3];
\fill[white] (0,0) [partial ellipse = 180:360:3 and 0.5];
\fill[red, opacity = 0.15] (0,0) [partial ellipse = 0:360:3 and 0.5];
\draw[thick] (0,0) ellipse (3 and 0.5);
\draw[thick] (0,0) [partial ellipse = 180:360:3 and 3];
\fill (1.5,-2.25) circle (0.05);
\node[above] at (1.5,-2.25) {$\bar{z}_1$};
\fill (-0.75,-1) circle (0.05);
\node[below] at (-0.75,-1) {$\bar{z}_2$};
\fill (-2,-1.75) circle (0.05);
\node[above] at (-2,-1.75) {$\bar{z}_3$};
\fill[opacity = 0.5] (0,-2.7) circle (0.05);
\node[above, opacity = 0.5] at (0,-2.7) {$\bar{z}_4$};
\node[below right] at (2.2,-2.2) {$\overline{\mathbb{H}}$};
\end{scope}
\draw[thick, latex-latex] (3,0.4) -- (3,-0.4);
\draw[thick, latex-latex] (-3,0.4) -- (-3,-0.4);
\draw[thick, latex-latex] (-1.5,-0) -- (-1.5,-0.8);
\draw[thick, latex-latex, dashed] (1.5,0) -- (1.5,0.8);
\node[right] at (3.2,0) {$\bar{z}$};
\end{scope}
\begin{scope}[xshift = 9cm]
\fill[red, opacity = 0.2] (0,0) [partial ellipse = 180:360:3 and 3];
\fill[blue, opacity = 0.2] (0,0) [partial ellipse = 0:180:3 and 3];
\fill[white] (0,0) [partial ellipse = 0:360:3 and 0.5];
\fill[red, opacity = 0.1] (0,0) [partial ellipse = 0:360:3 and 0.5];
\fill[blue, opacity = 0.15] (0,0) [partial ellipse = 0:360:3 and 0.5];
\draw[thick] (0,0) [partial ellipse = 180:360:3 and 0.5];
\draw[thick, dashed, opacity = 0.5] (0,0) [partial ellipse = 0:180:3 and 0.5];
\draw[thick] (0,0) circle (3);
\node[above right] at (2.2,2.2) {$\mathbb{CP}^1$};
\fill (1.5,1.4) circle (0.05);
\node[above] at (1.5,1.4) {$z_1$};
\fill (-0.75,0) circle (0.05);
\node[above] at (-0.75,0) {$z_2$};
\fill (-2,1.15) circle (0.05);
\node[above] at (-2,1.15) {$z_3$};
\fill (0,2.4) circle (0.05);
\node[above] at (0,2.4) {$z_4$};
\fill (1.5,-2.25) circle (0.05);
\node[above] at (1.5,-2.25) {$\bar{z}_1$};
\fill (-0.75,-1) circle (0.05);
\node[below] at (-0.75,-1) {$\bar{z}_2$};
\fill (-2,-1.75) circle (0.05);
\node[above] at (-2,-1.75) {$\bar{z}_3$};
\fill[opacity = 0.5] (0,-2.7) circle (0.05);
\node[above, opacity = 0.5] at (0,-2.7) {$\bar{z}_4$};
\end{scope}
\end{tikzpicture}
\caption{The doubling trick. We compute correlators on the upper-half-plane $\mathbb{H}$ by considering a correlator on the `doubled' worldsheet, given by the Riemann sphere.}
\label{fig:doubling-trick}
\end{figure}

In terms of this field defined on the full Riemann sphere, we can then define new functions $\Omega^{\pm}(z)$ as
\begin{equation}
\Omega^{\pm}(z)=\Bigl\langle \,\Xi^{\pm}(z)\prod_{a=1}^{n-1}W(u_a)\prod_{i=1}^{n}V_{m_i,j_i}^{w_i}(x_i,z_i)\Bigr\rangle_{(u,\epsilon)_{\rm S}} \ ,
\end{equation}
which satisfy the functional equation
\begin{equation}
\Omega^{\pm}(z)=\mp e^{i\varphi}\,\overline{\Omega^{\mp}(\bar{z})}\ .
\end{equation}
Now, we can use complex analysis on the sphere to determine the form of $\Omega^{\pm}$. Based on the OPEs \eqref{eq:ope-poles} and \eqref{eq:ope-regular} between $\xi^{\pm}$ and $V_{m,j}^{w}$, as well as the OPEs \eqref{eq:ope-xi-w} between $\xi^{\pm}$ and $W$, we find that $\Omega^{\pm}$ has the following properties:

\begin{enumerate}
	\item $\Omega^{\pm}$ have poles of order $\frac{w_i+1}{2}$ at $z=z_i$ and $z=\bar{z}_i$.

	\item $\Omega^{\pm}$ have simple zeroes at $z=u_a$ and $z=\bar{u}_a$.

	\item The linear combination $\Omega^-(z)+x_i\,\Omega^+(z)$ has a zero of order $\frac{w_i-1}{2}$ at $z=z_i$.

	\item The linear combination $\Omega^-(z)+\Omega^+(z)/\bar{x}_i$ has a zero of order $\frac{w_i-1}{2}$ at $z=\bar{z}_i$.
\end{enumerate}

Given these properties of the functions $\Omega^{\pm}(z)$, we can immediately implement the trick of \cite{Dei:2020zui}. We define a function $\Gamma(z)$ implicitly by the relation
\begin{equation}\label{eq:bnd-incidence-relation}
\Omega^-(z)+\Gamma(z)\,\Omega^+(z)=0\ .
\end{equation}
The map $\Gamma$ then satisfies the functional equation $\Gamma(z)=1/\overline{\Gamma(\bar{z})}$, or specifically
\begin{equation}
|\Gamma(z)|^2=1\,,\quad z\in\mathbb{R}\,.
\end{equation}
Moreover, by properties 3.\ and 4., one can show that $\Gamma$ has critical points at $z=z_i$ and $z=\bar{z}_i$, i.e.
\begin{equation}\label{eq:s2coveringlocality}
\begin{array}{ll}
{\displaystyle \Gamma(z)\sim x_i+\mathcal{O}\bigl((z-z_i)^{w_i}\bigr)\ ,} \qquad\quad & z\to z_i\,,\\
{\displaystyle  \Gamma(z)\sim\frac{1}{\bar{x}_i}+\mathcal{O}\bigl(\left(z-\bar{z}_i\right)^{w_i}\bigr)\ ,}\quad & z\to\bar{z}_i\,.
\end{array}
\end{equation}
Finally, by properties 1.\ and 2., one can readily show that $\Gamma$ has degree
\begin{equation}\label{eq:s2coveringdegree}
N=1+\sum_{i=1}^{n}(w_i-1)\ ,
\end{equation}
which is precisely the degree of a disk covering map with critical points of order $w_i$. Together, \eqref{eq:s2coveringlocality} and \eqref{eq:s2coveringdegree} therefore demonstrate that the function $\Gamma$ has the form of a covering map from the upper half-plane to the disk (see Appendix \ref{sec:covering-maps}). Thus, by the same arguments that apply in the bulk case, we see that the correlation functions \eqref{eq:basic-correlator} localise to those points in the moduli space where a covering map $\Gamma:\mathbb{H}\to\mathbb{D}$ exists. Furthermore, as shown there, see eq.~(\ref{eq:rigid}), the corresponding locus has the correct codimension to turn the worldsheet moduli space integral into a finite sum. 

\subsection[The case for \texorpdfstring{$\text{AdS}_2$}{AdS2} branes]{\boldmath The case for \texorpdfstring{$\text{AdS}_2$}{AdS2} branes}\label{sec:AdS2branesdisk}

As an aside we can also preform a similar analysis for the boundary states $\bket{{\rm AdS}_2}$ corresponding to the $\text{AdS}_2$ branes. In this case, the boundary conditions \eqref{eq:ads2-free-fields} become 
\begin{equation}
\begin{split}
\Bigl(\xi^{\pm}(z)-e^{i\varphi}\bar{\xi}^{\pm}(\bar{z})\Bigr) \, \bket{{\rm AdS}_2}&=0\ ,\\
\Bigl(\eta^{\pm}(z)-e^{-i\varphi}\bar{\eta}^{\pm}(\bar{z})\Bigr) \, \bket{{\rm AdS}_2}&=0\ .
\end{split}
\end{equation}
Thus, we can define the analytically continued fields
\begin{equation}
\Xi^{\pm}(z)=
\begin{cases}
\xi^{\pm}(z)\ , \qquad & \Im(z)\geq 0\\
e^{i\varphi}\bar{\xi}^{\pm}(\bar{z})\ , \qquad & \Im(z)<0\ ,
\end{cases}
\end{equation}
so that the complex functions
\begin{equation}
\Omega^{\pm}(z)=\Bigl\langle\,\Xi^{\pm}(z)\prod_{a=1}^{n-1}W(u_a)\prod_{i=1}^{n}V_{m_i,j_i}^{w_i}(x_i,z_i)\Bigr\rangle_{{\rm AdS}_2}
\end{equation}
satisfy the functional equation
\begin{equation}
\Omega^{\pm}(z)=e^{i\varphi}\overline{\Omega^{\pm}(\bar{z})}\ .
\end{equation}
The corresponding covering map constructed on the worldsheet via eq.~\eqref{eq:bnd-incidence-relation} then satisfies the functional equation
\begin{equation}
\Gamma(z)=\overline{\Gamma(\bar{z})}\ ,
\end{equation}
and has appropriate critical points at $z=z_i$ and $z=\bar{z}_i$ with $\Gamma(z_i)=x_i$ and $\Gamma(\bar{z}_i)=\bar{x}_i$. Therefore, by an argument identical to the one for spherical branes, the function $\Gamma$ satisfies precisely the properties of a covering map $\Gamma:\mathbb{H}\to\mathbb{H}$ (see Appendix \ref{sec:covering-maps}). This implies that the correlation functions of $n$ bulk strings in the presence of an $\text{AdS}_2$ brane localise to the points in the moduli space where these covering maps exist. This gives strong support to the idea that also the $\text{AdS}_2$ brane must correspond to a D-brane in the symmetric orbifold theory. However, so far, we have not managed to identify the corresponding D-brane explicitly.

\subsection{Comparison to the symmetric orbifold}

As we have mentioned before, the symmetric orbifold correlation functions on the sphere can be calculated  by employing ramified covering maps $\Gamma:\Sigma\to\mathbb{CP}^1$ to lift the twisted sector states to a surface on which they are single-valued \cite{Lunin:2001ew}. Such correlation functions can be expressed as a sum over such covering maps in a Feynman diagrammatic expansion \cite{Pakman:2009zz}
\begin{equation}\label{eq:symorb-diagram}
\Braket{\mathcal{O}_1^{w_1}(x_1)\cdots\mathcal{O}^{w_n}_n(x_n)}=\sum_{\Gamma}C_{\Gamma}\Braket{\mathcal{O}_1(z_1)\cdots\mathcal{O}_n(z_n)}_{\text{seed}}\,,
\end{equation}
where $C_{\Gamma}$ is the conformal factor that is obtained by pulling back from the base space by $\Gamma$, and the correlators on the right-hand-side are seed theory correlators evaluated on the covering space $\Sigma$.

One would expect that for the symmetric product correlators on a disk a similar construction should work, where now the appropriate covering maps are of the form $\Gamma:\Sigma \to\mathbb{D}$. In particular, for the maximally fractional branes $(u,{\rm id})$ constructed in Section \ref{sec:symorb}, one would expect the natural generalisation of equation \eqref{eq:symorb-diagram} to be
\begin{equation}
\langle \mathcal{O}_1^{w_1}(x_1)\cdots\mathcal{O}^{w_n}_n(x_n)\rangle_{(u,{\rm id})}=\sum_{\Gamma:\Sigma\to\mathbb{D}}C_{\Gamma}\, \langle\mathcal{O}_1(z_1)\cdots\mathcal{O}_n(z_n)\rangle_{u, \text{seed}}\ ,
\end{equation}
where  the right-hand-side is a sum over seed theory correlators evaluated on the covering space $\Sigma$ and with boundary condition $u$ along $\partial\Sigma$. For the genus zero contribution, i.e.\ for the case that $\Sigma$ is a disk, the correlation functions on the right-hand-side can then be calculated as $2n$-point functions on the sphere using the appropriate doubling trick, in complete parallel to the worldsheet computation from above.  In this way, the correlation functions calculated on $\text{AdS}_3\times\text{S}^3\times\mathbb{T}^4$ naturally reproduce the structure that is expected from the symmetric orbifold perspective.

In particular, this allows us to compare the bulk-boundary coefficients of the two sets of branes. In the 
symmetric orbifold CFT, these are given by the second equation of \eqref{maxfracbrane}, where we consider $\rho=\mathrm{id}$, and focus on the contribution of a single $w$-cycle $\sigma$; the combinatorial prefactor $\sqrt{|[\sigma]|/N!}$ equals then 
\begin{align}
\frac{1}{\sqrt{w}}\frac{1}{\sqrt{(N-w)!}}\ .
\end{align}
The second factor, $1/\sqrt{(N-w)!}$, is just the symmetry factor associated with the $N-w$ identical $1$-cycle states, and thus the `single-particle' one-point function from the $w$-cycle twisted sector equals 
\begin{align}
\frac{1}{\sqrt{w}}\, B_\beta(u)\ .
\end{align}
This now agrees precisely with the worldsheet bulk-boundary coefficient associated to the brane of the form (\ref{eq:BS}). In particular, the factor $B_\beta(u)$ just comes from the $\mathbb{T}^4$ boundary state $\|u,\mathrm{R},\varepsilon\rangle\!\rangle_{\mathbb{T}^4}$ in (\ref{eq:BS}),\footnote{The boundary state in (\ref{eq:BS}) is always evaluated in the RR sector (since the $\mathbb{T}^4$ is topologically twisted), while for the symmetric orbifold $B_\beta(u)$ labels the coefficient in the RR sector if $w$ is even, and in the NSNS sector if $w$ is odd. However, because of spectral flow these coefficients are always the same.}
while the  prefactor $1/\sqrt{w}$ reflects the relative normalisation of the worldsheet and spacetime disk correlators, since the worldsheet disk covers the spacetime disk $w$ times.

\section{Discussion}\label{sec:discussion}

In this paper we have constructed the spherical D-branes of string theory on ${\rm AdS}_3\times {\rm S}^3 \times \mathbb{T}^4$ for the specific background that is exactly dual to the symmetric orbifold of $\mathbb{T}^4$ \cite{Eberhardt:2018ouy,Eberhardt:2019ywk,Dei:2020zui}. Since this background is very stringy, we have used the worldsheet description in terms of an $\mathfrak{psu}(1,1|2)_1$ WZW model and constructed the symmetry-preserving boundary states. The geometric picture, see Figure~\ref{fig:geometry}, suggests that the spherical D-branes are dual to a brane in the symmetric orbifold CFT, and this seems indeed to be true. More specifically, the D-brane that is `spherical' on ${\rm AdS}_3 \times {\rm S}^3$ and describes a Dp-brane on $\mathbb{T}^4$, see eq.~(\ref{eq:BS}), is to be identified with the maximally fractional boundary condition of ${\rm Sym}_N (\mathbb{T}^4)$, see eq.~(\ref{maxfracbrane}), where we pick the same Dp-brane boundary for all individual $\mathbb{T}^4$s. We have confirmed this suggestive picture by comparing the relevant cylinder and disk amplitudes. It would be interesting to complete this picture by working out general open-closed amplitudes at arbitrary genus (and involving an arbitrary number of boundaries) on the worldsheet, and comparing them with the $1/N$ expansion of the corresponding symmetric orbifold CFT correlators. It would also be very interesting to understand systematically how the `non-perturbative' nature of the D-branes manifests itself in the dual symmetric orbifold theory. Since the string coupling constant behaves as $g_s\sim 1/\sqrt{N}$, the inclusion of D-branes should lead to effects that go as $e^{-\sqrt{N}}$, and it would be very instructive to see this more explicitly in the symmetric orbifold.

Since these boundary states are instantonic, it would also be interesting to study them using string field theoretic methods as in \cite{Sen:2020cef}. In particular, this should allow one to reproduce the correct ground state shift, see the discussion around eq.~(\ref{eq:WorldsheetClosed}).
\smallskip

The ${\rm AdS}_3$ string background also possesses another symmetry-preserving D-brane which is described by an ${\rm AdS}_2$ geometry in ${\rm AdS}_3$, and we have also constructed the corresponding boundary state in our worldsheet theory, see Section~\ref{sec:AdS2}. However, in this case, the dual CFT interpretation is less clear. Conversely, there are also more general boundary states in the symmetric orbifold theory, in particular, the D-branes that descend from the permutation branes in the tensor product theory, see Section~\ref{sec:perm}, whose ${\rm AdS}_3$ interpretation is also not yet clear. It would be interesting to investigate these boundary conditions further, and to see whether they may in fact be related to one another.

\acknowledgments 

We thank Lorenz Eberhardt, Rajesh Gopakumar and Tom\'{a}\v{s} Proch\'{a}zka for discussions. The work of BK is supported by a grant of the Swiss National Science Foundation, and in part by the Heising-Simons Foundation, the Simons Foundation, and National Science Foundation Grant No.\ NSF PHY-1748958. The work of the group at ETH is furthermore supported more generally by the NCCR SwissMAP which is administered by the Swiss National Science Foundation.

\appendix

\section{\boldmath The geometry of \texorpdfstring{${\rm SL}(2,\mathds{R})$}{SL(2,R)}}\label{app:geometry}

In this appendix we briefly review the geometry of ${\rm SL}(2,\mathds{R})$. Our analysis follows largely \cite{Maldacena:2000hw}. We begin with the Lie algebra of $\mathfrak{sl}(2,\mathrm{R})$, for which we work with the basis \cite{Maldacena:2000hw}
\be\label{generators}
t^1 = \frac{1}{2} \left( \begin{matrix} 1 & 0 \cr 0 & -1 \end{matrix} \right) \ , \qquad 
t^2 = \frac{1}{2} \left( \begin{matrix} 0 & 1 \cr 1 & 0 \end{matrix} \right) \ , \qquad
t^3 = \frac{1}{2} \left( \begin{matrix} 0 & 1 \cr -1 & 0 \end{matrix} \right) \ .
\ee
These generators satisfy the Lie algebra relations 
\be
[t^1,t^2] = t^3 \ , \qquad [t^2,t^3] = - t^1 \ , \qquad [t^3,t^1] = - t^2 \ . 
\ee
The most general Lie group element in ${\rm SL}(2,\mathds{R})$ can then be written as 
\be\label{param}
g =  \begin{pmatrix} \cosh \rho \cos t + \sinh \rho \cos\phi & \cosh\rho \sin t - \sinh\rho \sin\phi \cr
 -\cosh\rho \sin t - \sinh\rho \sin\phi & \cosh \rho \cos t - \sinh \rho \cos\phi \end{pmatrix} = e^{(t+\phi) t^3} e^{2 \rho t^1}  e^{(t-\phi) t^3} \ .
\ee
In terms of ${\rm AdS}_3$, $\rho$ describes the radial coordinate, while $(t,\phi)$ are the coordinates on the boundary cylinder, with $\phi$ being a $2\pi$-periodic variable.\footnote{Note that the conventions of \cite{Maldacena:2000hw} for the normalisation of the time coordinate $t$, which we use in this appendix, differ by an overall factor of $\pi$ relative to the normalisation of the closed-string $\mathfrak{sl}(2;\mathds{R})$ chemical potential used in the main body of this paper.} In ${\rm SL}(2,\mathds{R})$ also $t$ is $2\pi$-periodic, but in order to describe ${\rm AdS}_3$, we should not identify $t\cong t+2\pi$.
Thus we need to work with the covering space of ${\rm SL}(2,\mathds{R})$, whose coordinates are $(\rho,t,\phi)$, with only $\phi$ being $2\pi$-periodic, but no periodicity conditions on $\rho$ or $t$. 

Next we define the Cartan-Weyl basis for the complexification of $\mathfrak{sl}(2,\mathds{R})$. This is to say, we take $t^3$ to be the Cartan generator and define $t^\pm$ via 
\be\label{generators}
t^\pm =  t^1 \pm i\,  t^2 \ ,  \qquad 
t^+ = \frac{1}{2}\, \begin{pmatrix} 1 & i \cr i & -1 \end{pmatrix} \ , \qquad
t^- = \frac{1}{2}\, \begin{pmatrix} 1 & -i \cr -i & -1 \end{pmatrix} \ , 
\ee
so that 
\be\label{sl2}
{}[t^3,t^\pm] = \pm\, i\, t^\pm \ , \qquad [t^+,t^-] = - 2\, i \,  t^3 \ . 
\ee
The most general Lie group element in ${\rm SL}(2,\mathds{R})$ is 
\be
g = \left( \begin{matrix} a & b \cr c & d \end{matrix} \right) \ , \qquad \hbox{with inverse} \qquad 
g^{-1} = \left( \begin{matrix} d & -b \cr -c & a \end{matrix} \right)  \ , 
\ee
where 
\be
a,b,c,d \in \mathds{R} \ , \qquad ad - bc = 1 \ . 
\ee
Under conjugation by $g$, the generators in (\ref{generators}) transform as $\rho_g (t^a) = g t^a g^{-1}$, and we find in particular 
\be
\rho_g(t^3) =  \frac{1}{2}\, \left( \begin{matrix}  - (ac + bd) & a^2+ b^2  \cr  -c^2 - d^2  & (ac+bd) \end{matrix} \right)  \ .
\ee
Thus, within ${\rm SL}(2,\mathds{R})$, there is no inner automorphism that maps $t^3 \mapsto - t^3$. However, within ${\rm SL}(2,\mathbb{C})$ we can find 
\be
g_0 = \begin{pmatrix} i & 0 \cr 0 & -i \end{pmatrix} 
\ee
so that $\rho_{g_0}(t^3) = - t^3$. (Then we also have $\rho_{g_0}(t^\pm)  = t^\mp$.)
The fact that within ${\rm SL}(2,\mathds{R})$ there is no inner automorphism that maps $t^3\mapsto - t^3$ hinges on the fact that $t^3$ is the {\em timelike} direction. More specifically, we have 
\be\label{t3t3}
{\rm tr}( t^3 t^3) = - \frac{1}{2} \ , \qquad {\rm tr}(t^1 t^1) = {\rm tr}(t^2 t^2) = \frac{1}{2} \ , 
\ee
and thus $t^3$ is timelike, while $t^1$ and $t^2$ are spacelike. Thus we conclude that the two gluing conditions corresponding to the spherical and the ${\rm AdS}_2$ branes, see eqs.~(\ref{eq:GluJIdp}) and (\ref{eq:GluJp}), respectively, are not related by an inner automorphism in ${\rm SL}(2,\mathds{R})$. 


\subsection{Currents}

Following  \cite{Maldacena:2000hw} we define the currents via 
\be
J^a = k\, {\rm tr} \bigl( t^a \partial_+ g \, g^{-1} \bigr) \ , \qquad
\tilde{J}^a = k\, {\rm tr} \bigl( (t^a)^\ast\, g^{-1}\, \partial_- g  \bigr) \ , 
\ee
where $t^a$ are our Lie algebra generators from above. Here 
\be
x^\pm = \tau \pm \sigma \ , \qquad \partial_\pm = \frac{\partial}{\partial x^\pm} \ . 
\ee
Since $J^a \equiv J^a(x^+)$ and $\tilde{J}^a \equiv \tilde{J}^a(x^-)$, the most general solution is then 
\be\label{sol}
g (\tau,\sigma) = g_+(x^+) \, g_-(x^-) \ ,
\ee
and we have 
\be\label{currents}
J^a = k\, {\rm tr} \bigl( t^a \partial_+ g_+ \, g_+^{-1} \bigr) \ , \qquad
\tilde{J}^a = k\, {\rm tr} \bigl( (t^a)^\ast\, g_-^{-1}\, \partial_- g_- \bigr) \ .
\ee

\subsection{Spectral flow}

Given a solution of the form (\ref{sol}), a new solution is obtained by multiplying 
\be\label{specf}
g(\tau,\sigma) \rightarrow \tilde{g}(\tau,\sigma) \equiv e^{ w x^+ t^3} \, g(\tau,\sigma) \, e^{ \bar{w} x^- t^3} \ , 
\ee
i.e.\ replacing 
\be
g_+(x^+) \mapsto e^{ w x^+ t^3}\, g_+(x^+) \ , \qquad 
g_-(x^-) \mapsto g_-(x^-)\, e^{ \bar{w} x^- t^3}  \ . 
\ee
Since the original solution $g(\tau,\sigma)$ is periodic in $\sigma$ (with period $2\pi$), $\tilde{g}$ has this periodicity property provided that 
\be
\tilde{g}(\tau,\sigma+2\pi) = e^{2 \pi w t^3} \, \tilde{g}(\tau,\sigma) \, e^{- 2\pi \bar{w} t^3} \equiv \tilde{g}(\tau,\sigma) \ .
\ee
In terms of the parametrisation (\ref{param}) the exponentials in the expression in the middle equation shift $t$ and $\phi$ as 
\be
(t+\phi) \mapsto (t+\phi) + 2\pi w \ , \qquad (t-\phi) \mapsto (t-\phi) - 2\pi \bar{w} \ , 
\ee
i.e.\ correspond to 
\be
t \mapsto t + \pi (w-\bar{w}) \ , \qquad \phi \mapsto \phi + \pi (w + \bar{w}) \ . 
\ee
Since in ${\rm AdS}_3$ we do not have any periodicity in $t$, we need that $w=\bar{w}$. Furthermore, since $\phi$ is $2\pi$-periodic, we need to require that $w=\bar{w}\in\mathbb{Z}$. This reproduces the usual spectral flow. 

Let us next understand what this does to the left- and right-moving currents defined by (\ref{currents}). It follows from an explicit computation that 
\be
J^a \mapsto \frac{ k w}{2} \delta^{a,3} + k\, {\rm tr} \bigl(e^{ - w x^+ t^3}\,  t^a\,   e^{ w x^+ t^3}\ (\partial_+ g_+ ) g_+^{-1} \bigr) \ . 
\ee
In terms of the (complex) Cartan-Weyl basis this then implies 
\be
J^3_n \mapsto J^3_n +  \frac{ k w}{2}\, \delta_{n,0} \ , \qquad 
J^\pm_n \mapsto J^\pm_{n \mp w} \ , 
\ee
where we have expanded the currents in Fourier modes as 
\be
J^a(x^+) = \sum_n J^a_n e^{i n x^+} 
\ee
and used that 
\be
e^{ - w x^+ t^3}\,  t^\pm \, e^{  w x^+ t^3} = e^{\mp i w x^+} \, t^\pm \ . 
\ee
For the right-moving currents the analysis is essentially identical, and using that $(t^\pm)^\ast = t^\mp$ we find 
\be
\tilde{J}^3_n \mapsto \tilde{J}^3_n +  \frac{ k \bar{w}}{2}\, \delta_{n,0} \ , \qquad 
\tilde{J}^\pm_n \mapsto \tilde{J}^\pm_{n \mp \bar{w}} \ .
\ee
Thus spectral flow acts the same way on left- and right-moving currents.

\subsection{Boundary conditions}\label{app:A.3}
\setcounter{equation}{0}

In order to understand the geometrical meaning of the different boundary conditions it is useful to write the currents $J^a$ and $\tilde{J}^a$ in terms of the $(\rho,t,\phi)$ parametrisation, i.e.\ to evaluate (\ref{currents}) using the parametrisation (\ref{param}). This leads to 
\begin{align}
J^3 & = k \Bigl( (\sinh \rho )^2 \, \partial_+ \phi - (\cosh \rho)^2  \, \partial_+ t \Bigr)  \label{J3}\\
J^+ & = k \, e^{-i(\phi+t)} \Bigl( \partial_+ \rho + \frac{i}{2} \sinh(2\rho) \bigl( \partial_+ t - \partial_+ \phi \bigr) \Bigr)  \\ 
J^- & = k \, e^{i(\phi+t)} \Bigl( \partial_+ \rho - \frac{i}{2} \sinh(2\rho) \bigl( \partial_+ t - \partial_+ \phi \bigr) \Bigr) \ , \label{Jm}
\end{align}
as well as  
\begin{align}
\tilde{J}^3 & = k \Bigl( - (\sinh \rho )^2 \, \partial_- \phi - (\cosh \rho)^2  \, \partial_- t \Bigr) \\
\tilde{J}^+ & = k \, e^{i(\phi-t)} \Bigl( \partial_- \rho + \frac{i}{2} \sinh(2\rho) \bigl( \partial_- t + \partial_- \phi \bigr) \Bigr)   \\ 
\tilde{J}^- & = k \, e^{-i(\phi-t)} \Bigl( \partial_- \rho - \frac{i}{2} \sinh(2\rho) \bigl( \partial_- t + \partial_- \phi \bigr) \Bigr) \ . 
\end{align}
This now let's us understand the geometric meaning of the different boundary conditions. 

\subsubsection[The \texorpdfstring{${\rm AdS}_2$}{AdS2} boundary condition]{\boldmath The \texorpdfstring{${\rm AdS}_2$}{AdS2} boundary condition}

The ``trivial" boundary condition corresponds to imposing that 
\be
J^a(z) = \tilde{J}^a(\bar{z})  \qquad \hbox{for $z=\bar{z}$.}
\ee
This then leads to the equations 
\begin{align}
(\sinh \rho )^2 \bigl( \partial_+ \phi + \partial_- \phi \bigr) & = (\cosh \rho)^2\bigl ( \partial_+ t -  \partial_- t \bigr)  \label{tb1} \\ 
\frac{i}{2} \sinh(2\rho) \Bigl[ e^{i\phi} \bigl( \partial_- t + \partial_- \phi \bigr) - e^{-i\phi} 
\bigl( \partial_+ t - \partial_+ \phi \bigr)\Bigr] & = - e^{i\phi} \partial_- \rho + e^{-i\phi} \partial_+ \rho \ .  \label{tb2} 
\end{align}
In order to get a sense of what this means geometrically, it is useful to consider the limit $\rho\rightarrow \infty$, which describes the boundary cylinder. In that limit (\ref{tb1}) simplifies to 
\be\label{tb1p}
\bigl( \partial_+ t -  \partial_+ \phi \bigr)  = \bigl ( \partial_- t +  \partial_- \phi \bigr) 
\ee
while (\ref{tb2}) becomes 
\be\label{tb2p}
e^{i\phi} \bigl( \partial_- t + \partial_- \phi \bigr)  =  e^{-i\phi} 
\bigl( \partial_+ t - \partial_+ \phi \bigr) \ . 
\ee
Plugging (\ref{tb1p}) into (\ref{tb2p}) then leads to 
\be
e^{i\phi} =  e^{-i\phi}  \qquad \hbox{i.e. $\phi=0,\pi$.}
\ee
Thus the resulting brane must obey a {Dirichlet boundary condition in $\phi$} at the boundary; since this implies that $(\partial_+ \phi + \partial_- \phi)=0$, it follows from (\ref{tb1p}) that $(\partial_+ t - \partial_- t)=0$, i.e.\ that it satisfies a {Neumann boundary condition in $t$}, see the right panel of Fig.~\ref{fig:geometry}. The resulting D-brane is the {stretched static D-string} of Section~4 of \cite{Bachas:2000fr}.

\subsubsection{The spherical boundary condition}\label{sec:42}

The other natural boundary condition is 
\begin{align}
J^3(z) & = - \tilde{J}^3(\bar{z})  \qquad   \hbox{for $z=\bar{z}$} \\ 
J^\pm(z) & =  \tilde{J}^\mp(\bar{z})  \qquad \ \  \hbox{for $z=\bar{z}$} \ . 
\end{align}
This then leads to the equations 
\begin{align}
(\sinh \rho )^2 \bigl( \partial_+ \phi - \partial_- \phi \bigr) & = (\cosh \rho)^2\bigl ( \partial_+ t +  \partial_- t \bigr)  \label{tb1} \\ 
\frac{i}{2} \sinh(2\rho) \Bigl[ e^{it} \bigl( \partial_- t + \partial_- \phi \bigr) + e^{-it} 
\bigl( \partial_+ t - \partial_+ \phi \bigr)\Bigr] & =  e^{it} \partial_- \rho - e^{-it} \partial_+ \rho \ .  \label{tb2} 
\end{align}
Again, in the $\rho\rightarrow\infty$ limit this simplifies to 
\be
\partial_+ \phi - \partial_+ t = \partial_- \phi + \partial_- t \ , 
\ee
and
\be
e^{it} \bigl( \partial_- t + \partial_- \phi \bigr) = -  e^{-it} 
\bigl( \partial_+ t - \partial_+ \phi \bigr)  \ . 
\ee
Combining these two equations now implies that 
\be
e^{it} = e^{-it} \qquad \hbox{i.e. $t \in \mathbb{Z} \pi$.}
\ee
The resulting boundary condition thus satisfies a {Dirichlet boundary condition in $t$}, and a {Neumann boundary condition in $\phi$}, see the left panel of Fig.~\ref{fig:geometry}. It can therefore be identified with the (instantonic) {circular D-string} of Section~3 of \cite{Bachas:2000fr}.  

\subsection{Compatibility with spectral flow}

As we have explained above, in the $w$-spectrally flowed sector, $t(x^+,x^-)$ and $\phi(x^+,x^-)$ take the form 
\begin{align}
t(x^+,x^-) & = t_0(x^+,x^-) + \frac{w}{2} (x^+ + x^-) = t_0(x^+,x^-) + w \tau \\ 
\phi(x^+,x^-) & = \phi_0(x^+,x^-) + \frac{w}{2} (x^+ - x^-) = \phi_0(x^+,x^-) + w \sigma\ , 
\end{align}
where $(t_0,\phi_0)$ are the functions associated to the unflowed solution $g(\tau,\sigma)$ in (\ref{specf}); both of them are strictly periodic under $\sigma\mapsto \sigma + 2\pi$, i.e.\ $x^\pm \mapsto x^\pm \pm 2 \pi$. From the closed string perspective, the boundary occurs for a fixed $\tau=\tau_0$, i.e.\ for $x^-= 2 \tau_0- x^+$, and we therefore have 
\begin{align}
\left. t(x^+,x^-) \right|_{x^-=2 \tau_0 - x^+} & = t_0(x^+,2 \tau_0-x^+)  + w \tau_0   \\
\left. \phi(x^+,x^-) \right|_{x^-=2 \tau_0-x^+} & = \phi_0(x^+,2 \tau_0-x^+) + w (x^+ - \tau_0)\ . 
\end{align}
In particular, for $w\neq 0$ we cannot impose a Dirichlet boundary condition on $\phi$ since $\phi$ winds round $w$ times as $x^+\mapsto x^+ + 2\pi$. Thus only the spherical boundary condition of Section~\ref{sec:42} is compatible with spectral flow, in agreement with the analysis of Section~\ref{sec:psubound}. 

\section{\boldmath The D-branes of the symmetric orbifold of \texorpdfstring{$\mathbb{T}^4$}{T4}}\label{app:T4symorb}

In this appendix we explain how the construction of the maximally-fractional boundary states for a bosonic symmetric orbifold theory, see Section~\ref{sec:symorb}, can be generalised to the situation with fermions. We shall mainly focus on the case where the seed CFT is the superconformal theory of four free bosons and fermions on a 4-torus $\mathbb{T}^4$. We will furthermore restrict ourselves to the NS sector in the closed-string channel. Our results can, however, straightforwardly be extended to more general cases.

The main new ingredient we need to take care of comes from the fact that we may pick up minus signs from permuting fermions (because of Fermi statistics). More specifically, if $\sigma\in S_N$ has a cycle shape as described in \eqref{eq:partition}, we find that, see e.g.\ \cite[Section~4.3]{Brunner:2005fv} \cite[Appendix~A]{Gaberdiel:2018rqv}
\begin{align}
\mathrm{tr}_\mathcal{H}\Big[\sigma e^{2\pi i t(L_0 - \frac{Nc}{24})}\Big] = \prod_{j=1}^r Z^{(\epsilon_j)}(l_j t)\ , \qquad \epsilon_j = (-1)^{(l_j-1)} \ , 
\label{eq:BoundaryTraceSCFT}
\end{align}
where $Z^{(\epsilon)}$ is the trace with the insertion of $(-1)^{\epsilon F}$. 


\subsection[Boundary states in the seed \texorpdfstring{$\mathbb{T}^4$}{T4}]{\boldmath Boundary states in the seed \texorpdfstring{$\mathbb{T}^4$}{T4}}

In the seed $\mathbb{T}^4$ SCFT, we have $4$ free bosons and $4$ free fermions. We group them into two complex pairs as $\p X^{j\pm}$, $\psi^{j\pm}$, where $j=1,2$. They then satisfy the nontrivial OPEs
\begin{align}
\p X^{j+}(x)\p X^{k-}(y) \sim \frac{\delta^{jk}}{(x-y)^2}\,,\qquad \psi^{j+}(x)\psi^{k-}(y) \sim \frac{\delta^{jk}}{x-y}\ .
\end{align}

Let us assume that we impose standard Dp-brane boundary conditions directly on the free bosons and fermions; for example, for a D$0$-brane, they would read 
\begin{subequations}
\begin{align}
(\alpha_n^{j\pm} -\bar{\alpha}_{-n}^{j\pm})| \beta,\varepsilon,s\rangle\!\rangle&=0\ , \label{alphaz}\\
(\psi_r^{j\pm} - i\varepsilon\bar{\psi}_{-r}^{j\pm})| \beta,\varepsilon,s\rangle\!\rangle&=0\ ,
\end{align}
\end{subequations}
where $s\in \{\mathrm{NS},\mathrm{R}\}$ denotes the closed string sector, and we have $n\in\mathbb{Z}$, while $r\in \mathbb{Z}+1/2$ for  $s=\mathrm{NS}$ and $r\in \mathbb{Z}$ for $s=\mathrm{R}$. (Here the $\alpha^{j\pm}_n$ are the modes for the bosons, and the $\psi^{j\pm}_r$ those of the fermions.) 

The construction of the corresponding boundary states is standard, see e.g.\ \cite{Gaberdiel:2000jr}, and we denote them by $\| u,\varepsilon,s\rangle\!\rangle$, where $s\in \{\text{NS}, \text{R}\}$; note that in superstring theory one usually combines the NS-NS and R-R contributions in order to guarantee that the open string will be GSO-projected. However, in the current setup this is not appropriate since the symmetric orbifold of $\mathbb{T}^4$ is not the worldsheet theory of a string theory. Instead it describes the dual CFT which does not have a GSO-projection. In any case, we expand these boundary states in terms of the Ishibashi states as 
\be
\| u,\varepsilon,s \rangle\!\rangle = \sum_\beta B_\beta(u,s) \, | \beta,\varepsilon,s\rangle\!\rangle \ , 
\ee
where here $\beta$ runs over those momentum/winding sectors that are compatible with the gluing conditions (\ref{alphaz}) with $n=0$; for example, for a D$0$ brane, only the pure momentum sectors contribute. Their relative overlaps are then given by --- as in the main part we choose opposite values of $\varepsilon$ for the two boundary states, and hence denote them by $\tilde{\rm NS}$ and $\tilde{\rm R}$
\be
	\hat{Z}_{u|v,\tilde{\mathrm{NS}}}^{\mathbb{T}^4}(\hat{\zeta};\hat{t}\,) =\frac{\theta_4(\frac{\hat{\zeta}}{2};\hat{t}\,)
		\theta_4(-\frac{\hat{\zeta}}{2};\hat{t}\,)
	}{\eta(\hat{t}\,)^6}\, \hat{\Theta}_{u|v}^{\mathbb{T}^4}(\hat{t}\,)\ ,\qquad 
	\hat{Z}_{u|v,\widetilde{\mathrm{R}}}^{\mathbb{T}^4}(\hat{\zeta};\hat{t}\,) =      \frac{\theta_1(\frac{\hat{\zeta}}{2};\hat{t}\,)
		\theta_1(-\frac{\hat{\zeta}}{2};\hat{t}\,)
	}{\eta(\hat{t}\,)^6}\, \hat{\Theta}_{u|v}^{\mathbb{T}^4}(\hat{t}\,)\ ,
\ee
where $\hat{\Theta}_{u|v}^{\mathbb{T}^4}(\hat{t}\,)$ depends on which momentum/winding sectors contribute, and $\hat{\zeta}$ is the $\mathfrak{su}(2)$ chemical potential.

Given that we have chosen opposite values of $\varepsilon$ for the two boundary states, the corresponding open string is then in the ${\rm R}$ or $\tilde{\rm R}$ sector, and we find from the $S$-modular transformation, writing $\hat{t} = -1/t$ and $\hat{\zeta}=-\hat{t}\zeta$, the two expressions
%
\be
 	\label{eq:ZT4b}
	{Z}_{u|v,\mathrm{R}}^{\mathbb{T}^4}
	(\zeta; {t}) = \frac{\theta_2(\frac{\zeta}{2}; {t})
		\theta_2(-\frac{\zeta}{2}; {t})
	}{\eta( {t})^6}\,  {\Theta}_{u|v}^{\mathbb{T}^4}({t})\qquad 
	{Z}_{u|v,\widetilde{\mathrm{R}}}^{\mathbb{T}^4}
	( {t}\,) = -\frac{\theta_1(\frac{\zeta}{2}; {t})
		\theta_1(-\frac{\zeta}{2}; {t})
	}{\eta( {t})^6} \, {\Theta}_{u|v}^{\mathbb{T}^4}( {t})\ ,
\ee
where ${\Theta}_{u|v}^{\mathbb{T}^4}({t})$ describes the momentum/winding modes from the open string perspective; they are determined via the relation 
\begin{align}
\frac{\Theta_{u|v}^{\mathbb{T}^4}(t)}{\eta(t)^4} = \frac{\hat{\Theta}_{u|v}^{\mathbb{T}^4}(\hat{t}\,)}{\eta(\hat{t}\,)^4} \ .
\end{align}

\subsection[\texorpdfstring{$\mathrm{Sym}(\mathbb{T}^4)$}{Sym(T4)} boundary states]{\boldmath \texorpdfstring{$\mathrm{Sym}(\mathbb{T}^4)$}{Sym(T4)} boundary states}

Next we want to generalise the symmetric orbifold construction of Section~\ref{sec:symorb} to incorporate correctly the signs from eq.~\eqref{eq:BoundaryTraceSCFT}. We shall only concentrate on the boundary states that come from the $s={\rm NS}$ sector; recall that the $s={\rm R}$ sector has central charge proportional to $N$, and hence does not correspond to perturbative string degrees of freedom from the AdS perspective. Following \eqref{maxfracbrane}, we make the ansatz 
\be
\| u , \rho,\varepsilon,\mathrm{NS} \rangle\!\rangle = \sum_{(\underline{\beta},[\sigma])} B_{(\underline{\beta},[\sigma])}(u,\rho,\mathrm{NS})\, 
|  {\underline{\beta}},\varepsilon ,\mathrm{NS}\rangle \!\rangle_{[\sigma]} \,,
\ee
where the boundary state coefficients $B_{(\underline{\beta},[\sigma])}(u,\rho,\mathrm{NS})$ can be written as
\begin{align}
B_{(\underline{\beta},[\sigma])}(u,\rho,\mathrm{NS}) = \Bigl( \frac{|[\sigma]|}{N!} \Bigr)^{\frac{1}{2}}\, \chi_\rho([\sigma]) \, \prod_{j=1}^{r}  B_{\beta_j}(u,s(l_j)) \ ,
\end{align}
with $s(l_j)=\mathrm{NS}$ when $l_j\in 2\mathbb{Z}+1$, while $s(l_j)=\mathrm{R}$ when $l_j\in 2\mathbb{Z}$. Correspondingly, the Ishibashi states $|  {\underline{\beta}},\varepsilon,\mathrm{NS}\rangle \!\rangle_{[\sigma]}$ satisfy the fermionic conditions (for $j=1,\ldots,r$)
\be
\Bigl( \psi^{[j]}_p + i\varepsilon( \bar{\psi}^{[j]})_{-p} \Bigr) \, | {\underline{\beta}},\varepsilon,\mathrm{NS} \rangle \!\rangle_{[\sigma]} = 0 \ ,
\ee
where $p \in \tfrac{1}{l_j} \mathbb{Z}$ for $l_j\in 2\mathbb{Z}$, and $p \in \tfrac{1}{l_j} (\mathbb{Z}+1/2)$ for $l_j\in 2\mathbb{Z}+1$. Thus the individual single-cycle twisted sectors look as though they are in the Ramond sector for even cycle-length.
Choosing opposite spin structures as before, the corresponding overlaps are then 
\begin{align}
\hat{\mathcal{Z}}^{S_N}_{(u,\rho_1)|(v,\rho_2),\tilde{\mathrm{NS}}}(\hat{\zeta};\hat{t}\,)=\frac{1}{|S_N|}\sum_{\sigma\in S_N}\bar{\chi}_{\rho_1}([\sigma]){\chi}_{\rho_2}([\sigma]) \prod_{j=1}^{r} \hat{Z}_{u|v,\tilde{s}(l_j)}^{\mathbb{T}^4}\big(\hat{\zeta};\tfrac{\hat{t}}{l_j}\big)\ . 
\end{align}
Upon the $S$-modular transformation, going to the open string description, $\tilde{\rm NS}\rightarrow {\rm R}$, while $\tilde{\rm R}\rightarrow \tilde{\rm R}$, and hence the open string is always in the Ramond sector, but there is a sign depending on whether $l_j$ is even or odd, reflecting precisely the signs in eq.~(\ref{eq:BoundaryTraceSCFT}). Thus the open string is in the ${\rm R}^{\otimes N}$ sector --- this is because we took the two $\varepsilon$ values of the boundary states opposite --- and the group factors imply again that we project onto those states that transform in the representation $\rho_1 \otimes \rho_2^\ast$ with respect to $S_N$. 

For the simple case  $\rho_1=\rho_2=\mathrm{id}$, so that $\bar{\chi}_{\rho_1}([\sigma])={\chi}_{\rho_2}([\sigma])=1$ for all $\sigma\in S_N$, it is again convenient to go to the grand canonical ensemble 
	\begin{align}
	\hat{\mathfrak{Z}}_{u|v,\tilde{\mathrm{NS}}}(p,\hat{\zeta};\hat{t}\,)&=\sum_{N=1}^\infty p^N  \hat{\mathcal{Z}}^{S_N}_{(u,\mathrm{id})|(v,\mathrm{id}),\tilde{\mathrm{NS}}}(\hat{\zeta};\hat{t}\,)\ , 
	\end{align}
which can be rewritten, as in  \eqref{symorbopen}, as 
	\begin{align}
		\hat{\mathfrak{Z}}_{u|v,\tilde{\mathrm{NS}}}(p,\hat{\zeta};\hat{t}\,)&=
	\exp\bigg(
	\sum_{\substack{w=1\\ \text{$w$ odd}}}^\infty \frac{p^w}{w} \hat{Z}_{u|v,\tilde{\mathrm{NS}}}^{\mathbb{T}^4}
	(\hat{\zeta};\tfrac{\hat{t}}{w})+	\sum_{\substack{w=1\\ \text{$w$ even}}}^\infty \frac{p^w}{w} \hat{Z}_{u|v,\tilde{\mathrm{R}}}^{\mathbb{T}^4}
	(\hat{\zeta};\tfrac{\hat{t}}{w})
	\bigg)
	\,.\label{eq:ZT4hat}
	\end{align}
Finally, modular transforming into the open string channel and keeping track of the elliptic prefactors (which we did not include in \eqref{eq:ZT4b}), we obtain the grand canonical Ramond boundary partition function
	\begin{align}
{\mathfrak{Z}}_{u|v,{\mathrm{R}}}(p,{\zeta};{t})&=
\exp\bigg(
\sum_{\substack{k=1\\ \text{$k$ odd}}}^\infty \frac{p^k}{k} e^{\frac{\pi i k\zeta^2}{2t}}{Z}_{u|v,{\mathrm{R}}}^{\mathbb{T}^4}
(k{\zeta};k{{t}})+	\sum_{\substack{k=1\\ \text{$k$ even}}}^\infty \frac{p^k}{k}e^{\frac{\pi i k\zeta^2}{2t}} {Z}_{u|v,\widetilde{\mathrm{R}}}^{\mathbb{T}^4}
(k{\zeta};k{{t}})
\bigg)
\,,\label{eq:ZT4}
\end{align}
where we have relabelled $w\mapsto k$.

We should mention that also the spin structures in \eqref{eq:ZT4hat} arise naturally from the interpretation of the grand canonical ensemble as a sum over covering spaces of the cylinder. Indeed, given a covering map of degree $w$, a fermion which picks up a $(-1)$ monodromy on the base cylinder will pick up a $(-1)^w$ monodromy on the covering cylinder --- intuitively, the compact cycle on the covering cylinder maps to $w$ copies of the compact cycle on the base cylinder. That is, if we have a fermion in the ${\text{NS}}$ sector on the base space, then it will be in the ${\text{NS}}$ sector on the covering space if $w$ is odd, but in the ${\text{R}}$ sector if $w$ is even.

\section{Planar coverings of the disk}\label{sec:covering-maps}

In Section \ref{sec:correlators} we argued that correlation functions of spectrally flowed highest-weight states with a a spherical brane boundary condition are calculable in terms of holomorphic covering maps $\Gamma:\overline{\mathbb{D}}\to\overline{\mathbb{D}}$, where we denote by $\overline{\mathbb{D}}$ be the closed unit disk $\{z\in\mathbb{C}:|z|^2\leq 1\}$. Such a map should have the following properties:
\begin{itemize}

	\item On the interior $\mathbb{D}$ of the unit disk, $\Gamma$ should be a holomorphic function taking values on $\mathbb{D}$. That is,
	\begin{equation}
	\overline{\partial}\Gamma(z)=0\ ,\quad|\Gamma(z)|<1\ ,\quad z\in\mathbb{D}\ .
	\end{equation}

	\item $\Gamma$ should map the boundary of the unit disk to itself. That is,
	\begin{equation}
	|\Gamma(z)|^2=1\ ,\quad |z|^2=1\ .
	\end{equation}

	\item Finally, for $\Gamma$ to be a branched covering, there should exist $n$ marked points $z_i$ such that, near $z=z_i$, $\Gamma$ has a critical point of order $w_i$. That is,
	\begin{equation}
	\Gamma(z)\sim x_i+\mathcal{O}\bigl((z-z_i)^{w_i}\bigr)\ ,\qquad z\to z_i\ .
	\end{equation}

\end{itemize}
Below, we detail the algebraic constraints arising from these properties, and demonstrate that such a covering map only exists given that the insertion points $z_i$ are chosen appropriately.

\subsection{Algebraic constraints}

Given any branched covering between two surfaces $X$ and $Y$, the degree $N$ of the covering map $\Gamma:X\to Y$ (i.e.\ the number of preimages of $\Gamma$ at a generic point) is determined by the Riemann-Huwritz formula 
\begin{equation}
\chi(Y)\, N=\chi(X)+\sum_{i=1}^{n}(w_i-1)\ .
\end{equation}
Since the disk has Euler characteristic $\chi(\overline{\mathbb{D}})=1$, the degree of $\Gamma$ thus turns out to be
\begin{equation}
N=1+\sum_{i=1}^{n}(w_i-1)\ ,
\end{equation}
where $w_i$ is the degree of the critical point at $z_i\in\mathbb{D}$. Note that we do not allow critical points at the boundary.

Now, it is well known that all holomorphic functions from the disk to itself are rational. Thus, we can assume that $\Gamma$ takes the form $\Gamma(z)=Q_N(z)/P_N(z)$, where $Q_N$ and $P_N$ are polynomials of order $N$. The requirement that $\Gamma$ maps the circle to itself can be rephrased as
\begin{equation}
|Q_N(e^{i\phi})|^2=|P_N(e^{i\phi})|^2\ .
\end{equation}
If we write
\begin{equation}
Q_N(z)=\sum_{a=0}^{N}q_az^a\ ,\qquad P_N(z)=\sum_{a=0}^{N}p_az^a\ ,
\end{equation}
then we have
\begin{equation}
\begin{split}
|Q_N(e^{i\phi})|^2&=\sum_{a,b=0}^{N}q_aq_b^*e^{i(a-b)\phi} =\sum_{a=0}^{N}|q_a|^2+2\sum_{a<b}\Re\left(q_aq^*_be^{i(a-b)\phi}\right)\ , 
\end{split}
\end{equation}
and similarly for $|P_N|^2$. Requiring $|Q_N|^2=|P_N|^2$ along the unit circle means matching coefficients of $e^{i(a-b)\phi}$ for each value of $a-b$. This in the end gives $2N+1$ real constraints. Since the covering map $\Gamma$ had $2N+1$ complex degrees of freedom originally (the coefficients of $Q_N$ and $P_N$), we see that requiring $\Gamma$ to map the unit circle to itself reduces the number to $2N+1$ \textit{real} degrees of freedom. This reflects the fact that an open string has only half the chiral degrees of freedom of a closed string.

Next, we demand that $\Gamma$ has an appropriate critical point at each $z_i$. That is, we demand
\begin{equation}
\Gamma(z)-x_i\sim\mathcal{O} \bigl((z-z_i)^{w_i}\bigr)\ .
\end{equation}
In terms of the polynomials $Q_N$ and $P_N$, this constraint takes the form
\begin{equation}
Q_N(z)-x_iP_N(z)\sim\mathcal{O} \bigl((z-z_i)^{w_i} \bigr)\ .
\end{equation}
This introduces $w_i$ (complex) constraints at each critical point, and so overall there are $2\sum_{i}w_i$ \textit{real} constraints. Thus, the moduli space of such maps has dimension 
\begin{equation}
\text{dim}_{\mathbb{R}}(\Gamma:\mathbb{D}\to\mathbb{D},\text{ }z_i,x_i\text{ fixed})=2N+1-2\sum_{i=1}^{n}w_i=-2n+3\ .
\end{equation}
Note that $2n-3$ is exactly the dimension of the moduli space of a disk with $n$ marked points. Indeed, there are $2n$ (real) moduli for the $n$ points $z_i$, while we have three real conformal Killing vectors which can be used to fix $z_1=0$ and $z_2\in(0,1)$. Thus, 
\begin{equation}\label{eq:rigid}
\text{dim}_{\mathbb{R}}(\Gamma:\mathbb{D}\to\mathbb{D},\text{ }x_i\text{ fixed})=0\ ,
\end{equation}
and we conclude that constructing such a map is a rigid problem.

\subsubsection*{\boldmath Example: $w_1=w_2=1$}

The simplest example of such a covering map is to consider the case where $w_1=w_2=1$. The degree of such a map is $N=1$, i.e.\ $\Gamma$ is a rational linear function. Requiring that $\Gamma$ maps the unit circle onto itself restricts it to be of the form
\begin{equation}
\Gamma(z)=e^{i\phi}\frac{z-a}{1-\bar{a}z}\ ,
\end{equation}
where $a\in\mathbb{D}$ and $\phi\in\mathbb{R}$.

In order to simplify our life, we can use the three real moduli of the disk, in order to fix $z_i$ and $x_i$ to be of the form $z_1=x_1=0$ and $z_2,x_2\in(0,1)$. The resulting covering map is simply the identity function $\Gamma(z)=z$, and we find that this only maps $z_2$ to $x_2$ if $z_2=x_2$, which gives one real constraint on the insertion points $z_i$. This constraint, in turn, corresponds to the one real modulus of a disk with two marked points.

\subsection{The doubling trick}

We can also characterise $\Gamma$ in a different way by analytically continuing it. That is, instead of treating $\Gamma$ as a map from the disk to itself, we can instead treat $\Gamma$ as a holomorphic map on the full Riemann sphere. Indeed, if we define
\begin{equation}
\Gamma(z)=\frac{1}{\overline{\Gamma(1/\bar{z})}}\ ,\qquad |z|>1\ ,
\end{equation}
then the induced map $\Gamma:\mathbb{CP}^1\to\mathbb{CP}^1$ satisfies $|\Gamma(z)|^2=1$ along the unit circle by construction. Furthermore, $\Gamma$ now has double the critical points as its restriction to the disk. If we expand around $z=1/\bar{z}_i$, we find that $\Gamma$ has the critical behaviour
\begin{equation}
\Gamma\left(\frac{1}{\bar{z}_i}+\varepsilon\right)\sim\frac{1}{\bar{x}_i}+\mathcal{O}(\varepsilon^{w_i})\ .
\end{equation}
Thus, $1/\bar{z}_i$ is also a critical point of order $w_i$ with $\Gamma(1/\bar{z}_i)=1/\bar{x}_i$.

The doubling trick allows us to treat our covering map as a map on the sphere, as opposed to a covering map on the disk. The resulting map has critical points at $z_i$ and $1/\bar{z}_i$. We can thus calculate the resulting degree from Riemann-Hurwitz for sphere coverings, and we find
\begin{equation}
N=1+\sum_{i=1}^{n}\frac{w_i-1}{2}+\sum_{i=1}^{n}\frac{w_i-1}{2}=1+\sum_{i=1}^{n}(w_i-1)\ ,
\end{equation}
which agrees with the argument from the covering of the disk. The fact that $\chi(\mathbb{CP}^1)=2\chi(\mathbb{D})$ is compensated by the fact that $\Gamma:\mathbb{CP}^1\to\mathbb{CP}^1$ has twice as many critical points as $\Gamma:\mathbb{D}\to\mathbb{D}$. Furthermore, just as on the disk, since any holomorphic map $\Gamma:\mathbb{CP}^1\to\mathbb{CP}^1$ is rational, we can again write $\Gamma(z)=Q_N(z)/P_N(z)$ for some polynomials $Q_N$ and $P_N$ of degree $N$.

The moduli counting argument also works in the context of the doubling trick. Note that each pole $\lambda_a$ of $\Gamma$ is accompanied by a corresponding zero at $1/\bar{\lambda}_a$, by definition of the analytic continuation of $\Gamma$. Thus, we can write $\Gamma$ as
\begin{equation}
\Gamma(z)=C\prod_{a=1}^{N}\left(z-\lambda_a\right)^{-1}\left(z-1/\bar{\lambda}_a\right)\ .
\end{equation}
This parametrisation gives us $N+1$ complex degrees of freedom. Furthermore, the requirement that $|\Gamma(z)|^2=1$ along the unit circle removes one real degree of freedom. Indeed, we have
\begin{equation}
|\Gamma(e^{i\phi})|^2=|C|^2\prod_{a=1}^{N}\frac{|e^{i\phi}-1/\bar{\lambda}_a|^2}{|e^{i\phi}-\lambda_a|^2}=|C|^2\prod_{a=1}^{N}\frac{1}{|\lambda_a|^2}\ ,
\end{equation}
and so requiring $\Gamma$ to map the unit circle to itself simply fixes the norm of $C$. All of the remaining constraints come from simply demanding that $\Gamma$ has critical points at $z=z_i$. The criticality of the points at $z=1/\bar{z}_i$ is then automatic.

\subsection{Covering by the UHP}

In the context of boundary conformal field theory, it is often more convenient to consider the worldsheet as the upper half-plane (UHP) instead of the disk, so that the gluing conditions are implemented along the real line. In this case, the relevant covering maps are of the form $\Gamma:\mathbb{H}\to\mathbb{D}$ or $\Gamma:\mathbb{H}\to\mathbb{H}$. The UHP (with a point at infinity) and the unit disk are conformally equivalent under, for example, the map $f:\mathbb{H}\to\mathbb{D}\cup\{\infty\}$ given by
\begin{equation}
f(z)=\frac{z-i}{z+i}\ .
\end{equation}
Thus, by composing with $f$, one can easily turn a covering map $\Gamma:\mathbb{D}\to\mathbb{D}$ into $\Gamma\circ f:\mathbb{H}\to\mathbb{D}$ or $f^{-1}\circ\Gamma\circ f:\mathbb{H}\to\mathbb{H}$. In this way, the theory of branched coverings between the disk and UHP, and between the UHP and itself, is entirely equivalent to that of coverings from the disk to the disk. That said, it is convenient to review the algebraic properties of these types of maps.

\subsubsection*{UHP to disk}

A map $\Gamma:\mathbb{H}\to\mathbb{D}$ can be analytically extended to a branched covering of the Riemann sphere by the functional equation
\begin{equation}
\Gamma(z)=\frac{1}{\overline{\Gamma(\bar{z})}}\ .
\end{equation}
The resulting function is rational of order $N$ and has the form
\begin{equation}
\Gamma(z)=C\prod_{a=1}^{N}(z-\lambda_a)^{-1}(z-\bar{\lambda}_a)\ ,
\end{equation}
where $|C|^2=1$. Finally, $\Gamma$ satisfies the algebraic conditions
\begin{equation}
\Gamma(z)-x_i\sim\mathcal{O}((z-z_i)^{w_i})\,,\qquad\Gamma(z)-\frac{1}{\bar{x}_i}\sim\mathcal{O}\left((z-\bar{z}_i)^{w_i}\right)\ ,
\end{equation}
near $z=z_i$ and $z=\bar{z}_i$, respectively, where $x_i$ are marked points on the image disk.

\subsubsection*{UHP to UHP}

A branched covering of the form $\Gamma:\mathbb{H}\to\mathbb{H}$ can similarly be extended to a branched covering of the Riemann sphere via the analytic continuation
\begin{equation}\label{eq:uhp-uhp-continuation}
\Gamma(z)=\overline{\Gamma(\bar{z})}\ .
\end{equation}
The resulting function is rational of order $N$ and so generically has the form
\begin{equation}
\Gamma(z)=C\, \frac{\prod_{a=1}^{N}(z-Q_a)}{\prod_{a=1}^{N}(z-P_a)}\ .
\end{equation}
The functional equation \eqref{eq:uhp-uhp-continuation}, along with the fact that the boundary of the UHP is $\mathbb{R}\cup\{\infty\}$, tells us that the normalisation $C$, the zeroes $Q_a$, and the poles $P_a$ all have to be real. Finally, $\Gamma$ satisfies the algebraic conditions
\begin{equation}
\Gamma(z_i)-x_i\sim\mathcal{O}\left((z-z_i)^{w_i}\right)\,,\qquad \Gamma(z)-\bar{x}_i\sim\mathcal{O}\left((z-\bar{z}_i)^{w_i}\right)\ ,
\end{equation}
near $z=z_i$ and $z=\bar{z}_i$, where $x_i$ are marked points on the UHP.

\endgroup

\begin{thebibliography}{99}


\bibitem{Berkovits:1999im}
N.~Berkovits, C.~Vafa and E.~Witten,
``Conformal field theory of AdS background with Ramond-Ramond flux,''
JHEP \textbf{03} (1999), 018
{\tt  [\href{https://arxiv.org/abs/hep-th/9902098}{hep-th/9902098}]}.


\bibitem{Eberhardt:2018ouy}
L.~Eberhardt, M.R.~Gaberdiel and R.~Gopakumar,
``The Worldsheet Dual of the Symmetric Product CFT,''
JHEP \textbf{04} (2019), 103
{\tt [\href{https://arxiv.org/abs/1812.01007}{arXiv:1812.01007 [hep-th]}]}.

\bibitem{Dei:2020zui}
A.~Dei, M.R.~Gaberdiel, R.~Gopakumar and B.~Knighton,
``Free field world-sheet correlators for ${\rm AdS}_3$,''
JHEP \textbf{02} (2021), 081
{\tt [\href{https://arxiv.org/abs/2009.11306}{arXiv:2009.11306 [hep-th]}]}.

\bibitem{Gaberdiel:2018rqv}
M.R.~Gaberdiel and R.~Gopakumar,
``Tensionless string spectra on AdS$_{3}$,''
JHEP \textbf{05} (2018), 085
{\tt [\href{https://arxiv.org/abs/1803.04423}{arXiv:1803.04423 [hep-th]}]}.

\bibitem{Giribet:2018ada}
G.~Giribet, C.~Hull, M.~Kleban, M.~Porrati and E.~Rabinovici,
``Superstrings on AdS$_{3}$ at $k =1$,''
JHEP \textbf{08} (2018), 204
{\tt [\href{https://arxiv.org/abs/1803.04420}{arXiv:1803.04420 [hep-th]}]}.

\bibitem{Eberhardt:2019ywk}
L.~Eberhardt, M.R.~Gaberdiel and R.~Gopakumar,
``Deriving the AdS$_{3}$/CFT$_{2}$ correspondence,''
JHEP \textbf{02} (2020), 136
{\tt [\href{https://arxiv.org/abs/1911.00378}{arXiv:1911.00378 [hep-th]}]}.

\bibitem{Eberhardt:2020akk}
L.~Eberhardt,
``AdS$_{3}$/CFT$_{2}$ at higher genus,''
JHEP \textbf{05} (2020), 150
{\tt [\href{https://arxiv.org/abs/2002.11729}{arXiv:2002.11729 [hep-th]}]}.

\bibitem{Knighton:2020kuh}
B.~Knighton,
``Higher genus correlators for tensionless AdS$_{3}$ strings,''
JHEP \textbf{04} (2021), 211
{\tt [\href{https://arxiv.org/abs/2012.01445}{arXiv:2012.01445 [hep-th]}]}.

\bibitem{Maldacena:2000hw}
J.M.~Maldacena and H.~Ooguri,
``Strings in AdS$_3$ and SL(2,$\mathds{R}$) WZW model 1.: The Spectrum,''
J.\ Math.\ Phys.\ \textbf{42} (2001), 2929-2960
{\tt  [\href{https://arxiv.org/abs/hep-th/0001053}{hep-th/0001053}]}.

\bibitem{Bachas:2000fr}
C.~Bachas and M.~Petropoulos,
``Anti-de Sitter D-branes,''
JHEP \textbf{02} (2001), 025
{\tt  [\href{https://arxiv.org/abs/hep-th/0012234}{hep-th/0012234}]}.

\bibitem{Giveon:2001uq}
A.~Giveon, D.~Kutasov and A.~Schwimmer,
``Comments on D-branes in AdS$_3$,''
Nucl.\ Phys.\ B \textbf{615} (2001), 133-168
{\tt  [\href{https://arxiv.org/abs/hep-th/0106005}{hep-th/0106005}]}.

\bibitem{Petropoulos:2001qu}
P.M.~Petropoulos and S.~Ribault,
``Some remarks on anti-de Sitter D-branes,''
JHEP \textbf{07} (2001), 036
{\tt  [\href{https://arxiv.org/abs/hep-th/0105252}{hep-th/0105252}]}.

\bibitem{Lee:2001xe}
P.~Lee, H.~Ooguri, J.W.~Park and J.~Tannenhauser,
``Open strings on AdS$_2$ branes,''
Nucl.\ Phys.\ B \textbf{610} (2001), 3-48
{\tt  [\href{https://arxiv.org/abs/hep-th/0106129}{hep-th/0106129}]}.

\bibitem{Hikida:2001yi}
Y.~Hikida and Y.~Sugawara,
``Boundary states of D branes in AdS$_3$ based on discrete series,''
Prog.\ Theor.\ Phys.\ \textbf{107} (2002), 1245-1266
{\tt  [\href{https://arxiv.org/abs/hep-th/0107189}{hep-th/0107189}]}.

\bibitem{Rajaraman:2001cr}
A.~Rajaraman and M.~Rozali,
``Boundary states for D-branes in AdS$_3$,''
Phys.\ Rev.\ D \textbf{66} (2002), 026006
{\tt  [\href{https://arxiv.org/abs/hep-th/0108001}{hep-th/0108001}]}.

\bibitem{Lee:2001gh}
P.~Lee, H.~Ooguri and J.w.~Park,
``Boundary states for AdS$_2$ branes in AdS$_3$,''
Nucl.\ Phys.\ B \textbf{632} (2002), 283-302
{\tt  [\href{https://arxiv.org/abs/hep-th/0112188}{hep-th/0112188}]}.

\bibitem{Ponsot:2001gt}
B.~Ponsot, V.~Schomerus and J.~Teschner,
JHEP \textbf{02} (2002), 016
{\tt  [\href{https://arxiv.org/abs/hep-th/0112198}{hep-th/0112198}]}.

\bibitem{Takayanagi:2011zk}
T.~Takayanagi,
``Holographic Dual of BCFT,''
Phys. Rev. Lett. \textbf{107} (2011), 101602
{\tt [\href{https://arxiv.org/abs/1105.5165}{arXiv:1105.5165 [hep-th]}]}.

\bibitem{Fujita:2011fp}
M.~Fujita, T.~Takayanagi and E.~Tonni,
``Aspects of AdS/BCFT,''
JHEP \textbf{11} (2011), 043
{\tt [\href{https://arxiv.org/abs/1108.5152}{arXiv:1108.5152 [hep-th]}]}.

\bibitem{Karch:2000gx}
A.~Karch and L.~Randall,
``Open and closed string interpretation of SUSY CFT's on branes with boundaries,''
JHEP \textbf{06} (2001), 063
{\tt  [\href{https://arxiv.org/abs/hep-th/0105132}{hep-th/0105132}]}.

\bibitem{Gaberdiel:2021njm}
M.R.~Gaberdiel and K.~Naderi,
``The physical states of the Hybrid Formalism,''
{\tt [\href{https://arxiv.org/abs/2106.06476}{arXiv:2106.06476 [hep-th]}]}.

\bibitem{Alex}
A.~Belin, S.~Biswas and J.~Sully, to appear.

\bibitem{Eberhardt:2020bgq}
L.~Eberhardt,
``Partition functions of the tensionless string,''
JHEP \textbf{03} (2021), 176
{\tt [\href{https://arxiv.org/abs/2008.07533}{arXiv:2008.07533 [hep-th]}]}.

\bibitem{Ferreira:2017pgt}
K.~Ferreira, M.R.~Gaberdiel and J.I.~Jottar,
``Higher spins on AdS$_{3}$ from the worldsheet,''
JHEP \textbf{07} (2017), 131
{\tt [\href{https://arxiv.org/abs/1704.08667}{arXiv:1704.08667 [hep-th]}]}.

\bibitem{Giribet:2007wp}
G.~Giribet, A.~Pakman and L.~Rastelli,
``Spectral Flow in AdS$_3$/CFT$_2$,''
JHEP \textbf{06} (2008), 013
{\tt [\href{https://arxiv.org/abs/0712.3046}{arXiv:0712.3046 [hep-th]}]}.

\bibitem{Cardy:1989ir}
J.L.~Cardy,
``Boundary Conditions, Fusion Rules and the Verlinde Formula,''
Nucl.\ Phys.\ B \textbf{324} (1989), 581-596.

\bibitem{Recknagel:2013uja}
A.~Recknagel and V.~Schomerus,
``Boundary Conformal Field Theory and the Worldsheet Approach to D-Branes,''
Cambridge University Press (2013). 

\bibitem{Gaberdiel:2002my}
M.R.~Gaberdiel,
``D-branes from conformal field theory,''
Fortsch.\ Phys.\ \textbf{50} (2002), 783-801
{\tt  [\href{https://arxiv.org/abs/hep-th/0201113}{hep-th/0201113}]}.

\bibitem{Ishibashi:1988kg}
N.~Ishibashi,
``The Boundary and Crosscap States in Conformal Field Theories,''
Mod. Phys. Lett. A \textbf{4} (1989), 251-264.

\bibitem{Dijkgraaf:1996xw}
R.~Dijkgraaf, G.W.~Moore, E.P.~Verlinde and H.L.~Verlinde,
``Elliptic genera of symmetric products and second quantized strings,''
Commun.\ Math.\ Phys.\ \textbf{185} (1997), 197-209
{\tt  [\href{https://arxiv.org/abs/hep-th/9608096}{hep-th/9608096}]}.

\bibitem{Bantay:1997ek}
P.~Bantay,
``Characters and modular properties of permutation orbifolds,''
Phys.\ Lett.\ B \textbf{419} (1998), 175-178
{\tt  [\href{https://arxiv.org/abs/hep-th/9708120}{hep-th/9708120}]}.

\bibitem{Maldacena:2000kv}
J.M.~Maldacena, H.~Ooguri and J.~Son,
``Strings in AdS$_3$ and the SL(2,$\mathds{R}$) WZW model. Part 2. Euclidean black hole,''
J.\ Math.\ Phys.\ \textbf{42} (2001), 2961-2977
{\tt  [\href{https://arxiv.org/abs/hep-th/0005183}{hep-th/0005183}]}.

\bibitem{Recknagel:2002qq}
A.~Recknagel,
``Permutation branes,''
JHEP \textbf{04} (2003), 041
{\tt  [\href{https://arxiv.org/abs/hep-th/0208119}{hep-th/0208119}]}.

\bibitem{Lunin:2001ew}
O.~Lunin and S. D.~Mathur,
``Correlation functions for $M_N/S_N$ orbifolds,''
Comm. Math. Phys. \textbf{219} (2001), 399
{\tt  [\href{https://arxiv.org/abs/hep-th/0006196}{hep-th/0006196}]}.

\bibitem{Dei:2019iym}
A.~Dei and L.~Eberhardt,
``Correlators of the symmetric product orbifold,''
JHEP \textbf{01} (2020), 108
{\tt [\href{https://arxiv.org/abs/1911.08485}{arXiv:1911.08485 [hep-th]}]}.

\bibitem{Pakman:2009zz}
A.~Pakman, L.~Rastelli, S.S.~Schlomo,
``Diagrams for Symmetric Product Orbifolds,''
JHEP \textbf{10} (2009), 034
{\tt [\href{https://arxiv.org/abs/0905.3448}{arXiv:0905.3448 [hep-th]}]}.

\bibitem{Sen:2020cef}
A.~Sen,
``D-instanton Perturbation Theory,''
JHEP \textbf{08} (2020), 075
{\tt [\href{https://arxiv.org/abs/2002.04043}{arXiv:2002.04043 [hep-th]}]}.

\bibitem{Brunner:2005fv}
I.~Brunner and M.R.~Gaberdiel,
``Matrix factorisations and permutation branes,''
JHEP \textbf{07} (2005), 012
{\tt  [\href{https://arxiv.org/abs/hep-th/0503207}{hep-th/0503207}]}.

\bibitem{Gaberdiel:2000jr}
M.R.~Gaberdiel,
``Lectures on nonBPS Dirichlet branes,''
Class.\ Quant.\ Grav.\ \textbf{17} (2000), 3483-3520
{\tt  [\href{https://arxiv.org/abs/hep-th/0005029}{hep-th/0005029}]}.


 \end{thebibliography}
\end{document}